\documentclass[12pt]{article}

\usepackage{axodraw4j}
\usepackage{color}

\usepackage{jheppub}
\usepackage{amsmath,amssymb,euscript,array,mathrsfs}

\usepackage{epsfig}

\def\XXint#1#2#3{{\setbox0=\hbox{$#1{#2#3}{\int}$}
     \vcenter{\hbox{$#2#3$}}\kern-.5\wd0}}

\def\a{\alpha}
\def\b{\beta}
\def\c{\gamma}
\def\d{\delta}
\def\e{\epsilon}

\def\k{\kappa}
\def\l{\lambda}
\def\m{\mu}
\def\n{\nu}

\def\r{\rho}
\def\s{\sigma}
\def\t{\tau}

\def\w{\omega}

\def\D{\Delta}
\def\L{\Lambda}

\def\BH{{\boldsymbol H}}

\def\BF{\boldsymbol{F}}
\def\BG{\boldsymbol{G}}

\def\Dbarslash{\,\,{\raise.15ex\hbox{/}\mkern-12mu {\bar D}}}
\def\Dslash{\,\,{\raise.15ex\hbox{/}\mkern-12mu D}}
\def\delslash{\,\,{\raise.15ex\hbox{/}\mkern-9mu \partial}}
\def\delbarslash{\,\,{\raise.15ex\hbox{/}\mkern-9mu {\bar\partial}}}

\def\half{\frac{1}{2}}

\def\rta{\rightarrow}

\title{\begin{center}The c and a-theorems \\
and the Local Renormalisation Group 
\end{center}}

\author{Graham M. Shore}

\affiliation{Department of Physics,\\ 
Swansea University,\\
Swansea,\\ 
SA2 8PP, UK.}

\emailAdd{g.m.shore@swansea.ac.uk}

\abstract{
The Zamolodchikov $c$-theorem has led to important new insights in our understanding of the 
renormalisation group and the geometry of the space of QFTs. Here, we review the parallel developments of 
the search for a higher-dimensional generalisation of the $c$-theorem and of the Local Renormalisation Group.

The idea of renormalisation with position-dependent couplings, running under local Weyl scaling, is traced 
from its early realisations to the elegant modern formalism of the local renormalisation group. 
The key r\^ole of the associated Weyl consistency conditions in establishing RG flow equations for 
the coefficients of the trace anomaly in curved spacetime, and their relation to the $c$-theorem and
four-dimensional $a$-theorem, is explained in detail.

A number of different derivations of the $c$-theorem in two dimensions are presented -- using spectral 
functions, RG analysis of Green functions of the energy-momentum tensor $T_{\m\n}$, and dispersion relations 
-- and are generalised to four dimensions. The obstruction to establishing monotonic $C$-functions related 
to the $\b_c$ and $\b_b$ trace anomaly coefficients in four dimensions is discussed. 
The possibility of deriving an $a$-theorem, involving the coefficient $\b_a$  of the Euler-Gauss-Bonnet density 
in the trace anomaly, is explored initially by formulating the QFT on maximally symmetric spaces.
Then the formulation of the weak $a$-theorem
using a dispersion relation for four-point functions of $T^\m{}_\m$ is presented. 

Finally, we describe the application of the local renormalisation group to the issue of limit cycles in theories 
with a global symmetry and it is shown how this sheds new light on the geometry
of the space of couplings in QFT.

\vskip1cm}

\begin{document}

\maketitle

\setlength{\parskip}{10pt}


\section{Introduction}\label{sect 1}

The Zamolodchikov $c$-theorem \cite{Zamolodchikov:1986gt, Zamolodchikov:1987ti}, originally formulated in 1986,
has proved to be one of the most influential results in the recent development of the renormalisation group
in quantum field theory, and has led to important new insights in our modern perspective on the geometry 
and topology of the space of QFTs.

The theorem identifies a $C$-function constructed from the two-point Green functions of the energy-momentum tensor
in two-dimensional QFTs, which satisfies a RG flow equation of the form
\begin{equation}
\b^I \frac{\partial}{\partial g^I}\, C ~=~ \tfrac{3}{2}\,G_{IJ} \b^I \b^J \ ,
\label{z1}
\end{equation}
where $\b^I$ are the beta functions corresponding to the marginal operators $O_I$ characterising the theory. 
At a fixed point, $C(g^*)$ reduces to the central charge $c$ of the associated conformal field theory.
This can also be identified as the coefficient of the Ricci scalar in the conformal anomaly for the trace of the
energy-momentum tensor in curved spacetime, {\it i.e.}
\begin{equation}
T^\m{}_\m ~=~ \tfrac{1}{2}\b_c\, R \ ,
\label{z2}
\end{equation}
where $\b_c = c/12\pi$.
As the function $G_{IJ}$, similarly defined from the Green functions $\langle O_I ~ O_J\rangle$, is positive definite,
it may be interpreted as a metric on the space of couplings $\{g^I\}$. 
The relation (\ref{z1}) then implies that the RG flow of $C(g)$ is monotonic. This is the `$c$-theorem' and 
is a powerful constraint on the nature of RG flows, apparently ruling out, for example, the existence of limit cycles. 
It immediately follows that for a RG flow from an UV to an IR fixed point, the `weak $c$-theorem' holds, {\it viz.}
\begin{equation}
c^{UV} \, - \, c^{IR} ~ \ge ~ 0 \ ,
\label{z3}
\end{equation}
and since $c = n_s + \tfrac{1}{2}n_f$, this relation gives information on the spectrum of the CFTs related by the RG flow.
Moreover, since all the key ingredients of the proof of the $c$-theorem -- the existence of a conserved energy-momentum 
tensor, the trace anomaly and unitarity -- are very general properties of QFTs, it seems natural that a generalisation 
of the $c$-theorem beyond two dimensions should exist.

The search for a higher-dimensional version of the $c$-theorem has been a catalyst in the development of the 
second main topic of this paper, the Local Renormalisation Group. The essential idea underlying the local RGE is 
to promote the elementary couplings of the QFT to position-dependent functions $g^I(x)$ so that they act as 
sources for the operators $O_I$ appearing in the Lagrangian. The RG flow of these local couplings, generated by the 
corresponding beta functions $\b\left(g^I(x)\right)$, is then related to local Weyl rescalings of the metric, 
$g^{\m\n} \rta \Omega(x)^2 g^{\m\n}$. Since the metric is the source for the
energy-momentum tensor, the resulting local RGE \cite{Drummond:1977dg,Shore:1986hk},
\begin{equation}
\left(\Omega(x)\frac{\d}{\d \Omega(x)} - \b^I \frac{\d}{\d g^I(x)} - ~\ldots ~\right)\, W ~=~ {\mathcal A} \ ,
\label{z4}
\end{equation}
immediately describes the RG properties of Green functions involving the energy-moemntum tensor. 
The rhs of (\ref{z5}) is the curvature-dependent conformal anomaly, where
\begin{equation}
T^\m{}_\m ~=~ \b^I O_I ~+~ {\mathcal A} \ .
\label{z5}
\end{equation}
The introduction of local couplings has, however, important implications for the renormalisation of the theory
and, as shown by Osborn in a series of key papers 
\cite{Osborn:1987au,Osborn:1988wu,Osborn:1989td,Jack:1990eb,Osborn:1991gm},  
in order to achieve a complete, finite renormalisation it is necessary to introduce a new type of counterterm,
depending entirely on the local couplings $g^I(x)$ and their derivatives. As we shall see, these new counterterms and 
their corresponding RG functions add vital new contributions to the anomaly ${\mathcal A}$ in (\ref{z4}) and play 
a very powerful r\^ole in uncovering the full implications of anomalous Weyl symmetry in quantum field theory.

In four dimensions, there are more anomalous curvature contributions to the anomaly ${\mathcal A}$ 
and we have \cite{Capper:1974ic} (see also \cite{Deser:1993yx} for a classification in arbitrary dimensions)
\begin{equation}
T^\m{}_\m = \b^I O_I + \b_c {\BF} - \b_a {\BG} - \b_b {\BH} + \ldots 
\label{z6}
\end{equation}
where ${\BF}$, ${\BG}$ and ${\BH}$ are curvature-squared terms determined respectively 
by the Weyl tensor, the Euler-Gauss-Bonnet density and the square of the Ricci scalar. 
At first sight, it therefore appears there are three candidates 
for a four-dimensional Zamolodchikov-like theorem. Indeed, noting the relation between the central charge and 
the anomaly (\ref{z2}), and that in two dimensions the Ricci scalar is the Euler density, Cardy{\cite{Cardy:1988cwa}
conjectured that the analogue of the $c$-theorem in four dimensions should involve the coefficient $\b_a$
of the Euler-Gauss-Bonnet density ${\BG}$. For free theories, we have 
$\b_a = \tfrac{1}{(4\pi)^2}\tfrac{1}{360}\left(n_s + 11n_f + 62n_v\right)$.

The local renormalisation group places important and highly non-trivial constraints on the evolution of the
coefficients $\b_a$ and $\b_b$ under Weyl scalings. These are special cases of an intricate web of identities
which are elegantly determined by Weyl consistency conditions. 
Most important is the remarkable result \cite{Osborn:1989td,Jack:1990eb}
discovered by Osborn on the RG flow of $\b_a$: 
\begin{equation}
\b^I\partial_I \tilde{\b}_a = \tfrac{1}{8} \chi^g_{IJ} \b^I \b^J \ ,
\label{z7}
\end{equation}
where $\tilde{\b}_a = \b_a + O(\b^I)$ and $\chi^g_{IJ}$ is the RG function corresponding to a new counterterm
propertional to $G^{\m\n}\partial_\m g^I \partial_\n g^J$ ($G_{\m\n}$ is the Einstein tensor).
A very similar relation holds in two dimensions for the RG flow of $\b_c$ in (\ref{z2}). However, unlike the case in
two dimensions where the equivalent metric function $\chi_{IJ}$ is directly related to the two-point function 
$\langle O_I ~ O_J\rangle$, the corresponding identification of $\chi^g_{IJ}$ is more subtle and it is not
similarly guaranteed to be positive definite. Nevertheless, it can be shown to be positive in the neighbourhood of
fixed points where the beta functions $\b^I$ are sufficiently small and it has been found to be positive in all
known perturbative calculations. The relation (\ref{z7}) is of course tantalisingly close to realising the 
anticipated $a$-theorem in four dimensions.

A direct approach to looking for a four-dimensional theorem may be made by following the original Zamolodchikov
construction to find the analogues of the two-dimensional $C$-function in terms of two-point functions of the 
energy-momentum tensor and relating them to the RG coefficients in the anomaly (\ref{z6}).
This was carried out in \cite{Shore:1990wq,Shore:1990ng} following the spectral function method
first applied to the $c$-theorem in \cite{Friedan:1990,Cappelli:1990yc}. 
The result is two RG flow equations
for functions $C^{(s)}$,
\begin{equation}
\b^I\partial_I C^{(s)} = 24 G^{(s)} - 2 C^{(s)} \ ,
\label{z8}
\end{equation}
where the index $s=0,2$ indicates the contribution from spin 0 and spin 2 intermediate states in 
$\langle T_{\m\n} ~ T_{\r\s}\rangle$. 
For spin 0, up to normalisations, we find $C^{(0)} = \b_b$ at a fixed point, while $G^{(0)} = \tfrac{1}{9} G_{IJ} \b^I \b^J$
can be written in terms of a positive definite metric as in the original theorem. However, the presence of the 
extra term on the rhs of (\ref{z8}) (in $n$ dimensions, the coefficient is $n-2$)
means that the flow of $C^{(0)}$ is not monotonic. Moreover, the Weyl consistency conditions confirm that
$\b_b$ is itself of $O(\b^I)$ and so vanishes at a fixed point.
For spin 2, the $C^{(2)}$ function reduces to $\b_c$ at a fixed point. However, although $G^{(2)}$ is positive definite
it cannot be written entirely in terms of the metric $G_{IJ}$ and again the extra term proportional to $C^{(2)}$ in
(\ref{z8}) means its RG flow is not necessarily monotonic.

The remaining, and most interesting, anomaly coefficient does not however enter into the local RGEs for the 
two-point functions of the energy-momentum tensor, and so to find the corresponding $a$-theorem we 
need to look further. One possibility is to consider QFT on a spacetime of constant curvature, where the 
Euler index is non-vanishing and the the local RGEs for the two-point functions of $T_{\m\n}$ involve $\b_a$.
This approach was pursued in depth in \cite{Osborn:1999az} (see also \cite{Forte:1998dx})
and summarised here in section \ref{sect 9}.
While this essentially reproduces the consistency condition (\ref{z7})
it turns out that this method is not able to establish the existence of an exact, monotonic $a$-theorem,
for reasons discussed briefly in sections \ref{sect 9} and \ref{sect 8}. 

Eventually, a major breakthrough was made comparatively recently with the realisation by Komargodski and 
Schwimmer \cite{Komargodski:2011vj} (see also \cite{Komargodski:2011xv, Luty:2012ww})
that $\b_a$ can be isolated in the local RGEs for the four-point Green function of the trace of the energy-momentum 
tensor $T^\m{}_\m$ and, crucially, that these four-point functions do exhibit a positivity property through the
optical theorem. A dispersion relation argument then yields the weak $a$-theorem,
\begin{equation}
\b_a^{UV} - \b_a^{IR} \,=\, \frac{1}{\pi}\, \int_0^\infty \frac{ds}{s^3}\,{\rm Im}\,{\mathcal A}(s) \,\ge\, 0 \ ,
\label{z9}
\end{equation}
where ${\mathcal A}(s)$ is the four-point function written as a forward amplitude in terms of the 
Mandelstam variable $s$. Even so, a complete $a$-theorem establishing monotonicity of a well-defined
$C$-function along an entire RG trajectory not only in the vicinity of fixed points, and connecting 
to the Weyl consistency relation (\ref{z7}), remains to be found.

In this paper, we review and trace the common threads running through this line of research in the 
development of the local RGE and the search for a four-dimensional Zamolodchikov-like theorem 
constraining RG flows. It should be emphasised that this is not in any way intended as a comprehensive review
of the whole subject. Rather, the presentation is focused simply on the chain of ideas and technical 
advances outlined above.\footnote{For the same reason, the reference list here is not intended to reflect the
extensive literature on this subject, but is limited for the most part to those papers that were used more or less
directly in preparing this review.  For a more general review, see, for example, \cite{Nakayama:2013is} 
and references therein.}
Many fascinating topics and important work in the field, especially on supersymmetric theories, are therefore 
not touched on here, including explicit model calculations, $a$-maximisation, holography, applications to
entanglement entropy, and many more.

The discussion is organised as follows. The remainder of this section sets the scene with a brief review of the 
essentials of scale, conformal and Weyl invariance, a first look at the derivation of the $c$-theorem as originally 
presented by Zamolodchikov, and a brief account of the early development of the local renormalisation group.
Our approach is then to start with two dimensions and give an in-depth explanation of how the local RG
is applied in this comparatively simple context. Then, having established the general principles, we describe
the key points in the four-dimensional theory in a relatively condensed form, without over-burdening
the presentation with excessive detail.

We therefore begin in section \ref{sect 2} with a discussion of the theory of renormalisation with local couplings
and how the divergent contact terms appearing in the Green functions of composite operators, specifically
the energy-momentum tensor, are renormalised. Some of these derivations are worked through in different ways
to illustrate the self-consistency of the formalism and to build experience in the general theory, with
the more intricate details relegated to appendices \ref{sect A} and \ref{sect B}.

In section \ref{sect 3}, the more elegant methods of the modern local RG and anomalous Ward identities 
are introduced, together with the all-important Weyl consistency conditions. Then, in section \ref{sect 4}
we present a number of different derivations of the $c$-theorem in two dimensions, with the aim of
gaining insight into which may be generalised to four dimensions, and where features specific to 
the dimension arise.

The analysis is then repeated for four-dimensional QFTs, with section \ref{sect 5} describing the local RG
and Weyl consistency conditions while section \ref{sect 6} discusses the generalisation of the various 
derivations of the $c$-theorem described in section \ref{sect 4} -- the spectral function approach yielding
flow equations related to $\b_c$ and $\b_b$, a direct RG analysis of the two-point functions in momentum 
and position space and the potential role of the Weyl consistency condition for $\b_a$. 

Section \ref{sect 9} considers the anomalous Ward identities and local RG on spacetimes
of constant curvature, where a Weyl consistency condition involving both $\b_b$ and $\b_a$ may be
derived directly from two-point Green functions. The possibility of deriving the $c$ or $a$-theorems on 
homogeneous spaces is explored. 

Then, in section \ref{sect 10}, we describe the recent dispersion relation derivation of the weak $a$-theorem from 
four-point Green functions of the energy-momentum tensor in flat spacetime, comparing with the equivalent 
dispersion relation derivation of the $c$-theorem in two dimensions given in section \ref{sect 4} and
emphasising the assumptions inherent in the proof.

Finally, in section \ref{sect 7}, we describe a beautiful result due to Fortin, Grinstein and Stergiou 
\cite{Fortin:2012hn} which uses the local RG in QFTs with a global internal symmetry to resolve an 
apparent paradox concerning RG flows which end on a limit cycle, and sheds new light on 
the true nature of the space of couplings on which the renormalisation group acts.
Section \ref{sect 8} contains a summary and some additional reflections.

\subsection{Scale, conformal and Weyl invariance}\label{sect 1.1}
The maximal extension of the Lorentz group for massless particles in four dimensions is the $SO(2,4)$ conformal group.
This has 15 generators:  $J_{\m\n}$ (Lorentz transformations and rotations), $P_\m$ (translations),
$K_\m$ (special conformal transformations) and $D$ (dilatations).
The commutation relations for the generators of the Poincar\'e group $ISO(1,3)$ are 
\begin{align}
[J_{\m\n},J_{\r\s}] &=i\left(\eta_{\m\r}J_{\n\s} - \eta_{\m\s}J_{\n\r} - \eta_{\n\r}J_{\m\s} + \eta_{\n\s}J_{\m\r} \right) \nonumber\\
[J_{\m\n},P_\r] &= i\left(\eta_{\m\r}P_\n - \eta_{\n\r}P_\m \right) \nonumber\\
[P_\m,P_\n]&= 0 \ .
\label{a1}
\end{align}
where $\eta_{\m\n}$ is the Minkowski metric.
For a global scale-invariant theory, we also have the dilatations,
\begin{align}
[J_{\m\n},D] &= 0 \nonumber\\
[P_\r,D] &= iP_\r \ .
\label{a2}
\end{align}
while including special conformal transformations extends the symmetry group to $SO(2,4)$,
\begin{align}
[J_{\m\n},K_\s] &= i\left(\eta_{\m\s}K_\n - \eta_{\n\s}K_\m\right)  \nonumber\\
[P_\r,K_\s] &= 2 i\left(\eta_{\r\s}D + J_{\r\s}\right) \nonumber\\
[D,K_\s] &= i K_\s \nonumber\\
[K_\r,K_\s] &= 0 \ .
\label{a3}
\end{align}
Note that the algebra comprising only $J_{\m\n}$, $P_\r$ and $D$ closes, to the semi-direct product
$ISO(1,3) \rtimes R$, so in principle a theory can be scale invariant 
without also being invariant under the full conformal group $SO(2,4)$.

Translation invariance implies the existence of a conserved, symmetric energy-momentum tensor $T_{\m\n}$,
that is
\begin{equation}
\partial^\m T_{\m\n} = 0 \ .
\label{a4}
\end{equation}

Global scale transformations are defined by $x^\m \rta \l x^\m$, with $\l$ constant.
If the trace of the energy-momentum tensor is a total derivative, 
\begin{equation}
T^\m{}_\m = \partial^\m V_\m \ ,
\label{a5}
\end{equation}
for some `virial current' $V_\mu$, then there exists a conserved scale current, $\partial^\m D_\m = 0$,
given in terms of the energy-momentum tensor by $D_{\m} = x^\n T_{\m\n} - V_\m$.
This is related to the dilatation generator by $D=\int d^3 x D_0$. 

Special conformal transformations are defined by $ x^\m \rta \frac{x^\m + a^\m x^2}{1 + 2 a.x + a^2 x^2}$.
In this case, if 
\begin{equation}
T^\m{}_\m = \partial^\m \partial^\n L_{\m\n} \ ,
\label{a6}
\end{equation}
which follows if $V_\m = \partial^\n L_{\n\m}$, then there exists a conserved special conformal current,  
$\partial^\m K_{\m\r} = 0$, with the corresponding generator $K_\r = \int d^3 x K_{0\r}$, given by
$K_{\m\r} = (x^2 \eta_{\r\n} - 2 x_\r x_\n)T^\n{}_\m -2 x_\r V_\m + 2 L_{\m\r}$.
A suitable redefinition of $T_{\m\n}$ then  gives a conserved, symmetric {\it and} traceless operator. 
These properties of the trace of the energy-momentum tensor therefore provide a criterion to determine whether 
a theory is fully conformal invariant or simply scale
invariant.\footnote{The issue of whether a quantum field theory can be globally scale invariant without being
special conformal invariant has a substantial and still active literature. For recent developments and a review, 
see for example \cite{Fortin:2012cq,Fortin:2012hn,Dymarsky:2013pqa,Nakayama:2013is}} 
This will be important when we discuss scale and conformal invariance
in the context of limit cycles in section \ref{sect 7}.

In curved spacetime, we define the energy-momentum tensor as 
\begin{equation}
T_{\m\n}(x) = \frac{2}{\sqrt{-g}} \frac{\d S}{\d g^{\m\n}(x)} \ .
\label{a7}
\end{equation}
Diffeomorphism invariance of the action $S$ implies the conservation of $T_{\m\n}$,
\begin{equation}
D^\m T_{\m\n} = 0 \ .
\label{a8}
\end{equation}
Weyl transformations are defined as a scaling of the metric, $g^{\m\n} \rta \Omega^2 g^{\m\n}$,
and we write $\Omega(x) = e^{\s(x)}$. Global Weyl transformations (with $\s$ constant) are a symmetry if the action
is invariant, {\it i.e.}~if the Lagrangian transforms as $\d {\mathcal L} = - \s D^\m V_\m$ for some current $V_\m$.
Then,
\begin{equation}
T^\m{}_\m = \frac{2}{\sqrt{-g}} g^{\m\n} \frac{\d S}{\d g^{\m\n}} = D^\m V_\m \ .
\label{a9}
\end{equation}
Local Weyl invariance under $g^{\m\n}\rta e^{2\s(x)} g^{\m\n}$ implies the energy-momentum tensor is
strictly traceless,
\begin{equation}
T^\m{}_\m = 0 \ .
\label{a10}
\end{equation}

However, Weyl invariance is broken in the quantum theory in curved spacetime, even for free field theories,
and there is a Weyl anomaly proportional to the curvature.  In two dimensions,
\begin{equation}
T^\m{}_\m  = \frac{c}{24\pi} R \ ,
\label{a11}
\end{equation}
where $c$ is the central charge of the infinite-dimensional Virasoro algebra which extends the conformal
group in two dimensions. For free theories with $n_f$ fermions and $n_s$ scalars, $c = n_s + \tfrac{1}{2}n_f$.

In four dimensions,
\begin{equation}
T^\m{}_\m = \frac{1}{(4\pi)^2}\left(c F_4 - a E_4 + \tfrac{1}{9} b R^2 + \hat{b} \Box R \right)\ ,
\label{a12}
\end{equation}
where $F_4$ is the square of the Weyl tensor $C_{\m\n\r\s}$ and $E_4$ is the Euler-Gauss-Bonnet density.
The $\Box R$ term can always be removed by adding a finite local counterterm to the action 
and will be neglected from now on. Explicitly,
\begin{align}
F_4 ~&=~ R^{\m\n\r\s}R_{\m\n\r\s} - 2 R^{\m\n}R_{\m\n} + \frac{1}{3} R^2 ~=~ C^{\m\n\r\s}C_{\m\n\r\s} \ , \nonumber\\
E_4 ~&=~ R^{\m\n\r\s}R_{\m\n\r\s} + 4 R^{\m\n}R_{\m\n} + R^2 ~~~=~ 
\tfrac{1}{4} \e^{\m\n\r\s}\e_{\a\b\c\d}R^{\a\b}{}_{\m\n}R^{\c\d}{}_{\r\s} \ .
\label{a13}
\end{align}
For free theories, $a =\frac{1}{360}\left(n_s + 11n_f + 62n_v\right)$ and $c= \frac{1}{120}\left(n_s + 6n_f + 12n_v\right)$.

In addition, for interacting theories in flat or curved spacetime, the energy-momentum tensor has the usual
conformal anomaly proportional to the renormalisation group beta functions. With marginal couplings
${\cal L} \sim g^I O_I$ in the Lagrangian, this is
\begin{equation}
T^\m{}_\m = \b^I O_I \ ,
\label{a14}
\end{equation}
where $O_I$ denotes the corresponding renormalised composite operator. Also note that, as we see later, 
the coefficient $b \sim O(\b^I)$ in (\ref{a12}) and so vanishes at a fixed point.

\subsection{Renormalisation group flows and the Zamolodchikov c-theorem}\label{sect 1.2}

Our key motivation is to determine topological properties of the manifold of quantum field theories
by studying the flow of couplings under the renormalisation group (RG), {\it i.e.}~flows generated by the
beta functions $\b^I$ which may be regarded as vector fields on the space of couplings.
In particular, we are looking for information on the possible UV and IR behaviour of theories, 
the existence of fixed points and limit cycles and the RG flow in their vicinity, constraints on
phase transitions, and so on.

The question of whether RG flows must end on a fixed point, or whether limit cycles or more intricate,
even chaotic, behaviour is possible, is especially interesting. In particular, we would like to know whether
the theory at a fixed point or on a limit cycle is necessarily conformal, or whether it could be scale invariant
but not fully conformal.

An important constraint on RG flows for two-dimensional QFTs was given in 1986 by Zamolodchikov
\cite{Zamolodchikov:1986gt, Zamolodchikov:1987ti},
 who found a function $C(g)$ whose RG flow is monotonically decreasing. 
This appears to rule out limit cycles in two dimensions. Moreover, this `$c$-theorem' has a clear physical 
interpretation, since at a fixed point, $C(g_*) = c$, the central charge of the corresponding Virasoro algebra, 
which counts the (weighted) degrees of freedom. The theorem shows that under a flow from 
UV to IR fixed points, $c_{IR} < c_{UV}$, which we refer to as the `weak $c$-theorem'. 
In this paper, we present a variety of different proofs of the $c$-theorem in two dimensions, each of which
illustrates a different aspect of the underlying physics. This also points the way towards our ultimate goal,
a possible proof of the analogue of the theorem in four dimensions.

We begin with a sketch of the original proof of the $c$-theorem
presented by Zamolodchikov \cite{Zamolodchikov:1986gt}.
We work in 2-dim Euclidean space, and introduce complex spacetime coordinates 
$z,\bar{z}$ where $z = x^1 + i x^2$.
The independent components of the energy-momentum tensor $T_{\m\n}$ are then $T_{zz}$, $T_{\bar{z}\bar{z}}$ 
and $T_{z\bar{z}} = T^\m{}_\m$, and the conservation equation becomes
\begin{equation}
\partial_{\bar{z}} T_{zz} + 4 \partial_z T_{z\bar{z}} = 0 \ .
\label{a15}
\end{equation}
Zamolodchikov now defines a set of dimensionless functions from the two-point correlation functions of the 
energy-momentum tensor:
\begin{align}
F &= (2\pi)^2 \,z^4 ~\langle T_{zz}(x) ~ T_{\bar{z}\bar{z}}(0) \rangle  \nonumber\\
H &= (2\pi)^2 \,z^3 \bar{z} ~\langle T_{z\bar{z}}(x) ~ T_{\bar{z}\bar{z}}(0) \rangle  \nonumber\\
G &= (2\pi)^2 \,z^2 \bar{z}^2 ~\langle T_{z\bar{z}}(x) ~ T_{z\bar{z}}(0) \rangle \ ,
\label{a16}
\end{align}
where $F= F(\m |z|)$, {\it etc.} The pre-factors of $O(z^4)$ remove any short distance singularities in the
two-point functions. The $C$ function is then constructed as the combination,
\begin{equation}
C = 2\left(F - \tfrac{1}{2} H - \tfrac{3}{16} G\right) \ .
\label{a17}
\end{equation}
This has the following crucial property, which follows from the conservation equation (\ref{a15}):
\begin{equation}
\frac{\partial C}{\partial(\log|z|)} = -\tfrac{3}{2} \,G \ ,
\label{a18}
\end{equation}
and using dimensional analysis and the renormalisation group equation, we find 
\begin{equation}
\b^I \frac{\partial}{\partial g^I} C = \tfrac{3}{2} \,G  \ .
\label{a19}
\end{equation}

If we now use the trace anomaly equation (\ref{a14}), {\it viz.}~$T_{z\bar{z}} = \b^I O_I$,
and define the function $G_{IJ}$ as
\begin{equation}
G_{IJ} = (2\pi)^2 z^2 \bar{z}^2 ~\langle O_I(x) ~ O_J(0) \rangle \ ,
\label{a20}
\end{equation}
we can rewrite (\ref{a19}) in the form
\begin{equation}
\b^I \frac{\partial}{\partial g^I} C = \tfrac{3}{2} \,G_{IJ} \b^I \b^J  \le 0\ .
\label{a21}
\end{equation}
This is the Zamolodchikov $c$-theorem.
The final inequality in (\ref{a21}) follows from the fact that the function $G_{IJ}$ defines a positive-definite 
`metric' on the space of couplings. This shows that the function $C(g)$ is monotonically decreasing under a
renormalisation group flow to the IR.

At a fixed point, $\b^I(g_*) = 0$ and so $H(g_*) = G(g_*) = 0$, and $C(g_*)$ is determined purely by the
correlator $z^4 \langle T_{zz}(x) ~ T_{\bar{z}\bar{z}}(0) \rangle$. Since these correlators are determined uniquely by the
central charge, we can deduce $C(g_*) = c$.

\subsection{Local renormalisation group}\label{sect 1.3}

The main methodology for analysing Green functions of the energy-momentum tensor and the associated 
Zamolodchikov-like theorems constraining RG flows is the local renormalisation group.

The key idea of promoting the coupling constants of the original QFT to be functions of position 
and relating their RG flow to local Weyl transformations was introduced by Drummond and Shore 
\cite{Drummond:1977dg} in their study of trace anomalies for interacting QFTs in curved spacetime.
The term `local renormalisation group' itself was initially used in \cite{Shore:1986hk} where a RGE 
involving a position-dependent scale $\m(x)$ was proposed in the context of establishing diffeomeorphism 
invariant conditions for conformal invariance in string-related sigma models.
Since then, largely through the seminal work of Osborn \cite{Osborn:1987au,Osborn:1988wu,Osborn:1989td,
Jack:1990eb,Osborn:1991gm}, the idea of a local RGE has been perfected 
and developed into what is now a mature and comprehensive formalism for analysing anomalous Weyl, 
scale and conformal invariance and the associated RG flows.

From the definition (\ref{a7}), it follows that the expectation value of the trace of the energy-momentum tensor 
is determined by the response of the generating functional $W$ to a Weyl scaling of the metric, where $W$ is  
given by the sum of vacuum Feynman diagrams.
The essential observation of \cite{Drummond:1977dg}  was that in evaluating the dependence of these diagrams
on the Weyl scaling $g^{\m\n} \rta \Omega(x)^2 g^{\m\n}$, under which the propagator transforms as
$\D_{F}(x,y) \rta\Omega(x)^{n/2 -1} \D_F(x,y)\Omega(y)^{n/2-1}$, the Weyl factor $\Omega(x)$ enters at the vertices 
in exactly the same way as the RG scale associated with the anomaly. 
Since in $\phi^4$ theory, for example, the bare and renormalised couplings 
are related by $g_B = \m^{4-n} \left(g + L(g)\right)$, where $L(g)$ is a series of pole terms in dimensional regularisation,
we see that $\Omega(x)$ may be absorbed into a {\it position-dependent} RG scale $\m(x) = \m\,\Omega(x)^{-1}$.
Introducing the corresponding {\it local} renormalised coupling $g(x)$ and associated beta function following the
usual RG prescription\footnote{See section \ref{sect 2.1}. In the original discussion \cite{Drummond:1977dg,
Shore:1986hk}, the bare coupling $g_B$ was considered to be constant, with the entire position-dependence of
the running coupling $g(x)$ depending implicitly on the Weyl scaling $\Omega(x)$. This can of course be relaxed, 
with the couplings regarded completely as position-dependent sources.} with $\m(x)$, it was concluded 
in \cite{Drummond:1977dg}
that
\begin{equation}
\Omega(x)\frac{\d W}{\d\Omega(x)} ~=~ \b(g(x)) \frac{\d W}{\d g(x)}  ~+~ {\mathcal A} \ ,
\label{a22}
\end{equation}
where the first term on the rhs defines the renormalised normal product $[\phi^4(x)]$ while ${\mathcal A}$ 
denotes the purely curvature contributions to the trace anomaly (\ref{a12}). 

This appproach was subsequently formalised in \cite{Shore:1986hk} and compared in detail with the 
complete renormalisation of the trace anomaly in $\phi^4$ in terms of normal products presented
in \cite{Hathrell:1981zb}. Here, the local RGE was expressed in terms of a position-dependent RG scale
$\m(x)$ in the form
\begin{equation}
\left[\m(x)\frac{\d}{\d\m(x)} + \b^I\frac{\d}{\d g^I(x)} + \b^a_\eta\frac{\d}{\d \eta^a(x)} 
+ \b_c\frac{\d}{\d c(x)} + \b^a \frac{\d}{\d a(x)} + \b_b\frac{\d}{\d b(x)} + \ldots \right]\,W ~=~0 \ ,
\label{a23}
\end{equation}
where we have used the notation $O_I$ for the dimension 4 operators, $O_a$ for dimension 2 operators
such as $[\phi^2]$, and a coupling $\eta^a$ for the operators of the form $R O_a$. (Note that terms 
explicitly dependent on $\eta^a$ itself have been omitted in (\ref{a23}), together with the contributions 
from the renormalisation of the dimension 2 operators.)
This may then be converted to an equation for the local Weyl variation using the dimensional relation
\begin{equation}
\Omega(x)\frac{\d W}{\d \Omega(x)} ~=~ \m(x) \frac{\d W}{\d \m(x)} ~+~ \partial_\m {\mathcal W}^\m 
~+~\ldots \ ,
\label{a24}
\end{equation}
where the total divergence $\partial_\m {\mathcal W}^\m$ contains divergent terms such as 
$c_B \partial^2 R$ and $\eta_B^a\partial^2 O_a$ arising from the local variation of the curvature,
which is not wholly reflected in the dependence on the RG scale $\m(x)$. 
(Recall that with $\Omega^2(x) = e^{2\s(x)}$, $R \rta 2\s(x) R + 6 \partial^2\s(x)$ in four dimensions.)

In \cite{Shore:1986hk}, this form of local RGE was used together with the Ward identity for diffeomorphisms
to establish that in string-related sigma models, the condition for conformal invariance is actually the
vanishing of generalised, diffeomorphism invariant, beta functions rather than those occurring in
the standard RGE. We will return to related issues in section \ref{sect 7}.
However, it was already noted in that paper that with this renormalisation of the Lagrangian parameters,
the functional derivatives $\d W/\d g^I$ only defined finite normal products $[O_I]$ up to total divergences.
Moreover, since the Weyl variation on the lhs of (\ref{a24}) is manifestly finite, it follows that the
local RG scale variation $\m(x) \d W/\d \m(x)$,  is itself only finite up to total derivatives.

These problems were resolved by Osborn \cite{Osborn:1987au,Osborn:1988wu,Osborn:1989td}, 
and Jack and Osborn \cite{Jack:1990eb},
who made the crucial step of recognising that a complete renormalisation of the QFT requires the addition
of further counterterms depending on derivatives of the position-dependent couplings.
Examples include $e_I \partial_\m R \partial^\m g^I$,~$g_{IJ}
G_{\m\n}\partial^\m g^I \partial^\n g^J$,~$a_{IJ} D^2 g^I D^2 g^J,$ {\it etc}.
The first provides a necessary extra counterterm to ensure finiteness of the normal product 
$[O_I] = \d W/\d g^I(x)$  in a general background (compare with appendix B of \cite{Shore:1986hk}),
while the latter two renormalise the contact terms arising in the two-point Green functions $\langle O_I ~ O_J\rangle$.
In addition, the local RGE should be formulated directly as an equation for the Weyl variation
$\Omega(x) \d W/\d \Omega(x)$, bypassing the introduction of the local RG scale and defining
running couplings directly in terms of their Weyl scaling, as shown explicitly in \cite{Shore:1986hk}.

In this review, we will build all this up in stages, beginning with the renormalisation of two-dimensional QFTs
with running couplings in section \ref{sect 2} and finally progressing to four-dimensional theories with 
global symmetries in section \ref{sect 7}. In anticipation, however, it will be useful to outline the form
of the local RGE in its modern presentation \cite{Osborn:1991gm} immediately. Since the full RGE comprising 
dimension 4 and 2 operators and symmetry currents is extremely complicated and contains a plethora of terms, 
we refer the reader to the comprehensive recent papers by Jack and Osborn \cite{Jack:2013sha} and by 
Baume, Keren-Zur, Rattazzi and Vitale \cite{Baume:2014rla}. 

The local RGE is written in terms of the Weyl variation operators, which take the most general form constrained 
by power counting and symmetries. For a transformation $g^{\m\n} \rta 2\s(x) g^{\m\n}$, we define
\begin{align}
\D_\s^W &= 2\int d^4 x \, \s\, g^{\m\n}\frac{\d}{\d g^{\m\n}} \nonumber \\
\D_\s^\b &= \int d^4 x\, \left[\s\b^I \frac{\d}{\d g^I} + 
\left(\s \r_I^A D_\m  g^I - \partial_\m\s S^A\right) \frac{\d}{\d A_\m^A} +
\left(\s\left(\c_b^a - 2 \d_b^a\right)m^b + \eta^a R + \ldots \right)\frac{\d}{\d m^a}   \right]  
\label{a25}
\end{align}
where as before the couplings $g^I$ are sources for $O^I$, and we have defined $m^a$ (which have dim 2)
as sources for $O_a$ and $A_\m^A$ as sources for $G_F$ global symmetry currents $J_\m^A$. $D_\m g^I$ denotes a 
$G_F$ covariant derivative. The local RGE can then be presented as
\begin{equation}
\left(\D_\s^W -\D_\s^\b\right) \, W ~=~ \int d^4 x \sqrt{-g}\, {\mathcal A} \ ,
\label{a26}
\end{equation}
where the anomaly is 
\begin{align}
&\int d^4 x \sqrt{-g} \,{\mathcal A} \nonumber\\
&=\int d^4 x \sqrt{-g} \left[\s\left(\b_c\,{\BF} - \b_a\,{\BG} - \b_b\,{\BH} 
- \tfrac{1}{3} \chi_I^e \partial^\m R D_\m g^I 
- \tfrac{1}{2} \chi_{IJ}^a D^2 g^I D^2 g^J + \ldots  \right)
-\partial_\m \s \,Z^\m  \right] 
\label{a27}
\end{align}
where the nature of the total derivative term $Z^\m$ is discussed below.
The local RGE in its conventional form follows from differentiation w.r.t.~the position-dependent scale parameter 
$\s(x)$, giving
\begin{equation}
\left(\Omega(x)\frac{\d}{\d \Omega(x)} - \b^I \frac{\d}{\d g^I(x)} - ~\ldots ~\right)\, W 
= \b_c\, {\BF} - \b_a\,{\BG} - \b_b\, {\BH} + \ldots + D_\m Z^\m  \ .
\label{a28}
\end{equation}

The standard RGE is recovered from (\ref{a26}) by considering a global Weyl scaling with $\s = {\rm constant}$,
in which case the novel total derivative terms do not contribute. The dimensional relation is simply
\begin{equation}
\left(\int d^4 x\,\left[\Omega(x)\frac{\d}{\d\Omega(x)} + 2 m^a(x) \frac{\d}{\d m^a(x)} \right] \,+\, 
\m\frac{\partial}{\partial \m} \right)\,W ~=~0 \ ,
\label{a29}
\end{equation}
and the RGE follows (see section \ref{sect 3.3}).

With this overview, we are ready to build up our picture of the local renormalisation group in two
and then four dimensions, and apply the results to obtain a deeper understanding of the
Zamolodchikov $c$-theorem and its possible generalisations in higher dimensions.

\section{Renormalisation and the Conformal Anomaly}\label{sect 2}

In order to study the conformal anomaly and related Green functions of
the energy-momentum tensor, we need to renormalise the quantum 
field theory in curved spacetime including sources for all the
operators of interest. In this section, which follows \cite{Osborn:1989td},
we restrict to two dimensions 
and consider an interaction Lagrangian of the form ${\cal L} = g^I O_I$, where the 
$O_I$ are a complete set of dimension 2 operators.  The corresponding 
sources are simply the position-dependent couplings $g^I(x)$. 
In addition to the conventional 
renormalisation of the bare parameters, there are then further
counterterms depending on the spacetime curvature and on derivatives 
of the couplings. The latter provide the extra counterterms required
to define renormalised Green functions of the composite operators
$O_I$, removing the contact-term divergences which arise even in 
flat spacetime.

\subsection{Renormalisation and local couplings}\label{sect 2.1}

In dimensional regularisation, the action is therefore
\begin{equation}
S = S_0 + \int d^n x \sqrt{-g}\Big(g_B^I O_I + \tfrac{1}{2(n-1)} c_B R
- \L_B\Big) \ ,
  \label{b1}
\end{equation}
where the free action $S_0$ is invariant under Weyl transformations.
The bare couplings are defined as \footnote{The power $\m^{k_I \e}$ in $g_B^I$ 
fixes the dimension of the bare operator 
$O_{BI}$ away from $n=2$. This allows for composite operators with different combinations
of the elementary fields, whose dimension is set by the dimensionality/Weyl invariance
of the free action. The renormalised couplings are dimensionless, while with 
the definition in (\ref{b14}) the renormalised operators $O_I$ 
all have dimension $n$ before the physical limit $n\rightarrow 2$ is taken. Note that
no sum over the index on $k_I$ is implied here or in subsequent equations. }
 
\begin{align}
g_B^I &= \m^{k_I\e}(g^I + L^I(g)) \nonumber \\
c_B &= \m^{-\e}(c + L_c(g)) \nonumber \\
\L_B &= \m^{-\e}(\L + L_\L(g))  \ ,
  \label{b2} 
\end{align}
where 
\begin{equation}
L_\L = \tfrac{1}{2} A_{IJ}(g) \partial_\m g^I \partial^\m g^J \ .
  \label{b3}
\end{equation}
Here, $\e = 2-n$ and $L^I, L_c, A_{IJ}$ denote series of poles in $1/\e$ as
usual. The renormalised couplings $c$ and $\L$ will normally be taken
as zero.

Beta functions $\b_c$ and $\b_\L$ are defined from these couplings 
in the same way as the standard beta function $\b^i$.
Explicitly,
\begin{equation}
\hat{\b}^I \equiv \m \frac{d}{d\m} g^I = -k_I\e g^I + \b^I \ ,
  \label{b3a}
\end{equation}
with 
\begin{equation}
\b^I = -k_I\e L^I - \m \frac{d}{d\m} L^I \ .
  \label{b3aa}
\end{equation}
Note that this is a mass-independent renormalisation scheme so the
counterterms have no explicit $\m$-dependence, {\it i.e.}
$\m\, \partial L^I/\partial \m  = 0$.
In the same way,
\begin{equation}
\hat\b_c \equiv \m\frac{d}{d\m}c = \e c + \b_c \ ,
  \label{b4}
\end{equation}
with 
\begin{equation}
\b_c = \e L_c - \m \frac{d}{d\m}L_c \ ,
  \label{b5}
\end{equation}
and
\begin{equation}
\hat\b_\L = \e \L + \b_\L \ ,
  \label{b6}
\end{equation}
with
\begin{equation}
\b_\L = \e L_\L  - \m \frac{d}{d\m}L_\L \ .
  \label{b7}
\end{equation}
To evaluate the final term, we use
\begin{equation}
\m \frac{d}{d\m} = \m\frac{\partial}{\partial \m}
+ \int d^n x \left( 
\hat\b^I \frac{\d}{\d g^I(x)}
+ \hat\b_c \frac{\d}{\d c(x)}
+ \hat \b_\L \frac{\d}{\d \L(x)} \right) \ ,
  \label{b7a}
\end{equation}
and all couplings are
ultimately set to their constant physical values.
In what follows, we frequently simply write 
$\hat \b^I \frac{\partial}{\partial g^I}$ as shorthand for 
$\int d^n x \hat\b^I \frac{\d}{\d g^I(x)}$ and also abbreviate
$\frac{\partial}{\partial g^I} \equiv \partial_I$.
It then follows that 
\begin{equation}
\b_\L = \tfrac{1}{2} \chi_{IJ} \partial_\m g^I \partial^\m g^J
  \label{b8}
\end{equation}
with
\begin{equation}
\chi_{IJ} = \e A_{IJ} - {\cal L}_{\hat\b}A_{IJ} \ ,
  \label{b9}
\end{equation}
where the Lie derivative is given by
\begin{equation}
{\cal L}_{\hat\b} A_{IJ} = \hat\b^K \partial_K A_{IJ} 
+ \partial_I\hat\b^K A_{KJ}  + \partial_J \hat\b^K A_{IK} \ .
  \label{b10}
\end{equation}

\subsection{Trace anomaly}\label{sect 2.2}

Recall that the energy-momentum tensor is derived as the variation of
the curved-spacetime action with respect to the metric:
\begin{equation}
T_{\m\n}(x) = \frac{2}{\sqrt{-g}} \frac {\d S}{\d g^{\m\n}(x)} \ .
  \label{b11}
\end{equation}
Since the metric is itself a finite parameter, this defines a
renormalised operator.  To evaluate this, we need the variation of the
curvature counterterms, especially
\begin{equation}
\frac{\d}{\d g^{\m\n}(x)} R(y) = \sqrt{-g}\left(R_{\m\n} + \D_{\m\n} \right) \d(x,y) \ ,
\label{b12}  
\end{equation}
where we define $\D_{\m\n} = g_{\m\n}D^2 - D_\m D_\n$. After contracting with the metric, 
this gives the following expression in terms of bare quantities 
for the trace of the energy-momentum tensor:
\begin{equation}
T^\m{}_\m = \e \left(-k_I g_B^I O_{BI} + \tfrac{1}{2(n-1)} c_B R - \L_B \right)
+ D^2 c_B + 2 \m^{-\e}\L \ .
  \label{b13} 
\end{equation}

The renormalised operators $O_I$ are defined as the functional
derivatives of the action with respect to the couplings, viz.
\begin{equation}
O_i = \frac{1}{\sqrt{-g}} \frac{\d S}{\d g^I} \ .
  \label{b14}  
\end{equation}
The curvature and derivative counterterms therefore contribute:
\begin{align}
O_i &= \frac{1}{\sqrt{-g}}\,\frac{\d}{\d g^I} \int d^n x 
\sqrt{-g} \,\left(g_B^J O_{BJ} + 
\tfrac{1}{2(n-1)} c_B R -\L_B \right) \nonumber \\
&= Z_I{}^J O_{BJ} + \m^{-\e}\left( \tfrac{1}{2(n-1)} \partial_i L_c\, R
- \half \partial_I A_{JK} \partial g^J \partial g^K
+ D_\m(A_{IJ}\partial^\m g^J) \right)   \ .
  \label{b15} 
\end{align}
with
\begin{equation}
Z_{IJ} = \m^{k_I\e}(\d_{IJ} + \partial_I L_J) \ .
  \label{b16} 
\end{equation}
This can be re-expressed, after a short but intricate calculation (see
appendix \ref{sect A}), in terms of the
beta functions:
\begin{equation}
T^\m{}_\m ~=~ \hat{\b}^I O_I + \m^{-\e} \left(\tfrac{1}{2(n-1)} \hat{\b}_c R - \hat{\b}_\L \right) 
+ \m^{-\e} (D^2 c + 2 \L) + \m^{-\e} D_\m Z^\m \ ,
  \label{b17} 
\end{equation}
with
\begin{equation}
Z_\m = \partial_\m L_c - \hat\b^I A_{IJ} \partial_\m g^J \ .
  \label{b18} 
\end{equation}
Finally, we define the key function $w_i$ \cite{Osborn:1989td} through 
\begin{equation}
Z_\m = w_I \partial_\m g^I
  \label{b19}  
\end{equation}
and reducing to $n=2$ and setting the couplings $c=\L=0$, 
we find our final result for the energy-momentum tensor trace:
\begin{equation}
T^\m{}_\m = \b^I O_I + \tfrac{1}{2} \b_c R -\b_\L + D_\m Z^\m \ .
  \label{b20}  
\end{equation}

The r.h.s.~is the conformal, or trace, anomaly.  As well as the familiar
operator and curvature contributions, this expression includes the
dependence on derivatives of the position-dependent couplings.
These are crucial in renormalising higher-point Green functions
involving $T^\m{}_\m$.  
Evaluating with the couplings set to constants, we recover the 
familiar expression for the conformal anomaly:
\begin{equation}
\langle T^\m{}_\m\rangle  - \langle \Theta\rangle =   
\frac{1}{2} \b_c R\ ,
  \label{b21}  
\end{equation}
where $\Theta \equiv \b^I O_I$ is the usual operator anomaly and
$\frac{1}{2} \b_c R$ is the curvature contribution.

Since all the other terms in \eqref{b20} are finite, it follows that 
$Z^\m$ itself must also be finite. This gives rise to a consistency
condition which plays an important role in the analysis of the
$c$-theorem. Indeed, as we see later, this is equivalent to the
Wess-Zumino consistency condition which follow from the abelian
nature of the Weyl anomaly. 
To derive this, we start from the
definition \eqref{b18} of $Z_\m$ and apply the operator 
$\e - \hat\b^K \partial_K$
to both sides, or equivalently act with $\m \frac{d}{d\m}$ on 
$\m^{-\e} Z_\m$.   We have
\begin{equation}
\left(\e - \hat\b^K \partial_K\right)\left(\partial_\m L_c - \hat\b^I
A_{IJ}\partial_\m g^J\right) = 
\left(\partial_I\b_c  - \chi_{IJ}\hat\b^J\right) \partial_\m g^I
  \label{b22} 
\end{equation}
and
\begin{equation}
\left(\e - \hat\b^K \partial_K\right)\left(w_I \partial_\m g^I\right)
= -\left( \e g^J \partial_J w_I + \b^J \partial_J w_I +
\partial_I \b^J w_J \right) \partial_\m g^I \ .
  \label{b23}  
\end{equation}
Equating the terms of $O(1)$, we isolate the crucial identity
\begin{equation}
\partial_I\b_c = \chi_{IJ}\b^J - {\cal L}_\b w_I \ .
  \label{b24}
\end{equation}
If we now define 
\begin{equation}
\tilde\b_c = \b_c + \b^I w_I
  \label{b25} 
\end{equation}
we can rewrite \eqref{b24} in the form
\begin{equation}
\partial_I \tilde\b_c = \chi_{IJ}\b^J + \left(\partial_I w_J
  - \partial_J w_I\right) \b^J \ .
  \label{b26}  
\end{equation}
Notice the crucial role played here by the function $w_I$, introduced
by Osborn \cite{Osborn:1989td} in his analysis of renormalisation 
in the presence of position-dependent couplings.

Finally, contracting with the beta function, we derive the Weyl
consistency condition for the RG flow of the modified coefficient
$\tilde \b_c$ of the Euler density in the trace anomaly:\footnote{Notice that there is an
arbitrariness in the choice of RG scheme defining the various functions involved here.
This can be parametrised by the following changes:
\begin{equation*}
\d \b_c = {\mathcal L}_\b b \equiv \b^I\partial_I b\ , \qquad \qquad  \d\chi_{IJ} = {\mathcal L}_\b a_{IJ} \ ,
\end{equation*} 
which implies
\begin{equation*}
\d \tilde{\b}_c = a_{IJ}\b^I \b^J \ , \qquad \qquad   \d w_I = - \partial_I b + a_{IJ}\b^J \ .
\end{equation*}
This leaves the form of the consistency relation  (\ref{b27}) invariant.}
\begin{equation}
\b^I \partial_I \tilde \b_c = \chi_{IJ} \b^I \b^J \ .
  \label{b27} 
\end{equation}
Clearly, this condition has a close resemblance to the form of the
$c$-theorem in two dimensions \cite{Osborn:1989td}. 
We will explore this connection further in what follows.

\subsection{Renormalisation of two-point Green functions}\label{sect 2.3}

As noted above, the position-dependent couplings $g^I(x)$ act as
sources for the composite operators $O_I$, while the counterterms
proportional to $\partial_\m g^I(x)$ provide the contact terms
required to renormalise their higher-point Green functions.

We first define the renormalised two-point functions of the energy-momentum
tensor in curved spacetime so as to be symmetric, 
{\it i.e.}\footnote{We further define the two-point functions of the
  trace $T^\m{}_\m(x)$ so that 
\begin{align*}
i\langle T^\m{}_\m(x) ~ T^\r{}_\r(y)\rangle 
&\equiv g^{\m\n} g^{\r\s} i\langle T_{\m\n}(x) ~T_{\r\s}(y)\rangle \\
&= \frac{2}{\sqrt{-g}}g^{\m\n}\frac{\d}{\d g^{\m\n}(x)} 
\left(\frac{2}{\sqrt{-g}}g^{\r\s}\frac{\d W}{\d g^{\r\s}(y)}\right)
- (n+2) \langle T^\m{}_\m(x)\rangle \d(x,y) \ ,
\end{align*}
where the ambiguity in
the definition of the renormalised Green function is reflected in the 
presence of the VEV of $T^\m{}_\m(x)$, itself defined by
\begin{equation*}
\langle T^\m{}_\m(x)\rangle \equiv 
g^{\m\n} \langle T_{\m\n}(x)\rangle \ .
\end{equation*} 
With Minkowski signature, these two-point functions are 
the usual time-ordered products.
Note that throughout this paper $\d(x,y)$ is the delta function density,
where $\d(x-y) = \sqrt{-g} \d(x,y)$, 
and here we have left the dimension $n$
general for future convenience. 

Similarly, letting
$\Theta = \b^I O_I$ as the operator part of the 
trace anomaly, we define
\begin{align*}
i\langle \Theta(x) ~\Theta(y)\rangle &\equiv \b^I \b^J 
i\langle O_I(x) ~ O_J(y)\rangle \\
&= \frac{1}{\sqrt{-g}}\b^I\frac{\d}{\d g^I(x)} 
\left(\frac{1}{\sqrt{-g}} \b^J \frac{\d W}{\d g^J(y)}
\right)
- \b^I \partial_I\beta^J \langle O_J(x)\rangle \d(x,y) \ .
\end{align*}
}
\begin{equation}
i\langle T_{\m\n}(x)~~T_{\r\s}(y)\rangle = 
\frac{2}{\sqrt{-g(x)}}\frac{2}{\sqrt{-g(y)}} \frac{\d}{\d g^{\m\n}(x)} 
\frac{\d}{\d g^{\r\s}(y)} W
  \label{b28}
\end{equation}
and similarly
\begin{equation}
i\langle O_I(x)~~O_J(y)\rangle = \frac{1}{\sqrt{-g(x)}} 
\frac{1}{\sqrt{-g(y)}}\frac{\d}{\d g^I(x)} 
\frac{\d}{\d g^J(y)} W \ .
  \label{b29}
\end{equation}

With these definitions, we readily find the following relations between
the bare and renormalised two-point functions:
\begin{multline}
i\langle T^\m{}_\m(x)~T^\r{}_\r(y)\rangle  ~-~
i\langle T^\m{}_\m(x)~T^\r{}_\r(y)\rangle_B  ~+~
2 \langle T^\m{}_\m(x)\rangle \d(x,y) ~ \\ ~=~ 
\frac{1}{\sqrt{-g(x)}} \frac{1}{\sqrt{-g(y)}} ~
\big\langle \frac{\d^2 S}{\d\s(x) \d\s(y)} \big\rangle 
\label{b30}
\end{multline}
and similarly
\begin{align}
i\langle T^\m{}_\m(x)~O_I(y)\rangle  ~-~
i\langle T^\m{}_\m(x)~O_I(y)\rangle_B   ~&=~ 
\frac{1}{\sqrt{-g(x)}} \frac{1}{\sqrt{-g(y)}} ~
\big\langle \frac{\d^2 S}{\d\s(x) \d g^J(y)} \big\rangle 
\label{b31} \\
i\langle O_I(x)~O_J(y)\rangle ~~-~~
i\langle O_I(x)~O_J(y)\rangle_B  ~~&=~ 
\frac{1}{\sqrt{-g(x)}} \frac{1}{\sqrt{-g(y)}} ~
\big\langle \frac{\d^2 S}{\d g^I(x) \d g^I(y)} \big\rangle 
\label{b32}
\end{align}
The evaluation of the counterterms on the rhs of (\ref{b30}) - 
(\ref{b32}) by differentiation of the action (\ref{b1}) is again very
intricate and details are given in appendix \ref{sect A}. For the Green functions
of interest here, we eventually find:
\begin{multline} 
i\langle T^\m{}_\m(x)~T^\r{}_\r(y)\rangle  ~-~
i\langle T^\m{}_\m(x)~T^\r{}_\r(y)\rangle_B  ~+~
(2-\e) \langle T^\m{}_\m(x)\rangle \,\d(x,y) \\ 
~=~   \e \m^{-\e}\, L_c \,D^2\d(x,y) 
\label{b33} 
\end{multline}
\begin{equation}
i\langle T^\m{}_\m(x)~O_I(y)\rangle  ~-~
i\langle T^\m{}_\m(x)~O_I(y)\rangle_B   ~=~ 
\m^{-\e} \,\partial_I L_c\, D^2 \d(x,y) ~~~~~~~~~~~~~~~~~~~~~
\label{b34} 
\end{equation}
\begin{multline}
i\langle \hat{\Theta}(x)~O_I(y)\rangle  ~-~
i\langle \hat{\Theta}(x)~O_I(y)\rangle_B  ~+~
\partial_I \hat{\beta}^J \,\langle O_J(x)\rangle \,\d(x,y) \\
~=~  -\m^{-\e} \left[\tfrac{1}{2(n-1)} \partial_I \beta_c\, R\, \d(x,y) ~-~
A_{IJ}\hat{\beta^J} \,D^2 \d(x,y)\right] 
\label{b35}
\end{multline}
where $\hat{\Theta} = \hat{\beta}^I O_I$ and as usual we have 
set $c = \Lambda = 0$ and omitted terms
of $O(\partial_\m g^I)$.

We can derive important identities amongst the renormalised 
connected Green functions by using the operator relation
$T^\m{}_\m \sim \b^I O_I$ inside the {\it bare} Green functions.
This immediately gives, for $n=2$,
\begin{equation}
i\langle T^\m{}_\m(x)~T^\r{}_\r(y)\rangle  ~-~
i\langle T^\m{}_\m(x)~\Theta(y)\rangle   ~+~
2\langle T^\m{}_\m(x)\rangle \,\d(x,y) ~=~ 
\b_c\, D^2 \d(x,y) ~~~~~~~~~~~~~~~~~~~
\label{b36} 
\end{equation} 
\begin{multline}
i\langle T^\m{}_\m(x)~O_I(y)\rangle  ~-~
i\langle \Theta(x)~O_I(y)\rangle  ~-~
\partial_I \beta^J \,\langle O_J(x)\rangle \,\d(x,y)  ~~~\\
~=~  \tfrac{1}{2}\partial_I \beta_c\, R\, \d(x,y) ~+~
w_I\,D^2 \d(x,y)
\label{b37}
\end{multline}
where we have used the identities 
$\b_c = \e L_c - \hat{\b}^I \partial_I L_c$  
and $w_I = \partial_I L_c - A_{IJ}\hat{\b}^J$.
Notice the appearance here of $\gamma_I{}^J = \partial_I \b^J$,
which is the anomalous dimension matrix for the operators $O_I$.

Combining these, we find an identitiy which will play an important
role in our analysis of the $c$-theorem, {\it viz.}
\begin{multline}
i\langle T^\m{}_\m(x)~T^\r{}_\r(y)\rangle  ~-~
i\langle \Theta(x)~\Theta(y)\rangle ~+~
2\langle T^\m{}_\m(x)\rangle \,\d(x,y) ~-~
\beta^I\partial_I \beta^J \,\langle O_J(x)\rangle \,\d(x,y) \\
~=~ \tfrac{1}{2}\,\beta^I\partial_I \beta_c\, R\, \d(x,y) ~+~
\tilde{\beta}_c\,D^2 \d(x,y) ~~~
\label{b38}
\end{multline}
where $\tilde{\beta}_c = \beta_c + \beta^I w_I$ as defined in 
(\ref{b25}). Notice that requiring finiteness of this identity between
renormalised Green functions implies the same consistency condition
previously derived in (\ref{b26}) from finiteness of the VEV $\langle
T^\m{}_\m\rangle$ itself.

Finally, it is important for what follows to see how the $\tilde{\b}_c$ terms
have arisen here as a result of the Weyl variation of the Ricci scalar (see
appendix \ref{sect A}).  This is despite the fact that the integral of $R$ 
over the whole spacetime is a topological invariant in two dimensions,
which implies that its Weyl variation vanishes.
To see this explicitly, note that under
the Weyl rescaling $\d g^{\m\n} = 2\s g^{\m\n}$,
the Ricci scalar transforms in $n$ dimensions as $\d R = 2\s R 
+ 2(n-1) D^2\s$. Then, using (\ref{b12}),
\begin{equation}
\frac{\d}{\d g^{\m\n}} \int d^n x\sqrt{-g}\,R ~=~\sqrt{-g}\, G_{\m\n} \ ,
\label{b39}
\end{equation}
where $G_{\m\n}$ is the Einstein tensor, and so
\begin{equation}
\frac{\d}{\d\s} \int d^n x\sqrt{-g}\,R ~=~
\left(1 - \frac{n}{2}\right)\sqrt{-g}\,R \ .
\label{b40}
\end{equation}
which indeed vanishes when $n=2$.

\section{The Local Renormalisation Group and Weyl Consistency Conditions}\label{sect 3}

We now come to the renormalisation group itself and derive the RGEs for the two-point Green functions
which play a central role in the analysis of the $c$-theorem. We also develop the more algebraic approach to
the local RGE and anomalous Ward identities for Weyl symmetry anticipated in section \ref{sect 1.3}.
In this section, we present these results specialised to two dimensions, though the same principles,
and many of the general formulae, apply equally to the four-dimensional generalisation.

\subsection{Diffeomeorphism and anomalous Weyl Ward identities}\label{sect 3.2}

We now consider the relations amongst  two-point Green functions which follow from
the (anomalous) symmetries of the theory. First, consider invariance under diffeomorphisms.
From the transformations of the metric tensor $g_{\m\n}$ and the scalar local couplings $g^I(x)$, 
we have the diffeomorphism Ward identity (see \cite{Shore:1986hk, Osborn:1999az}),
\begin{equation}
\Delta_v W = \int d^2 x \left( - \left(D^\m v^\n + D^\n v^\m\right) \frac{\d}{\d g^{\m\n}} 
+ v^\m \partial_\m g^I \frac{\d}{\d g^I} \right) W = 0.
\label{cc1}
\end{equation} 
This gives the Green function identities which express the conservation of the
energy-momentum tensor. We find the natural result,
\begin{equation}
D^\m \langle T_{\m\n}\rangle = 0\ ,
\label{cc2}
\end{equation}
while, less trivially, for the two-point functions we also determine the contact terms implied by our
definitions in section \ref{sect 2.3} \cite{Osborn:1999az}:
\begin{align}
D^\m i\langle T_{\m\n}(x) ~ T_{\r\s}(y)\rangle &= - \left(\langle T_{\r\s}\rangle D_\n \d(x,y)
+ D_\r\left(\langle T_{\n\s}\rangle \d(x,y)\right)  +  D_\s\left(\langle T_{\r\n}\rangle \d(x,y)\right)~ \right) \ ,
\nonumber\\
D^\m i\langle T_{\m\n}(x) ~ O_I(y)\rangle &= - \left(\langle O_I(x)\rangle D_\n \d(x,y)~ \right)\ .
\label{cc3}
\end{align}

For future use, note that on an $n$-dim homogeneous space (including flat space) where
$\langle T_{\m\n}\rangle =  (1/n)\langle T^\l{}_\l\rangle g_{\m\n}$ with 
$\langle T^\l{}_\l\rangle$ constant, we can define a ``conserved'' two-point function by
\begin{equation}
i\langle T_{\m\n}(x) ~ T_{\r\s}(y)\rangle_{\rm c} = 
i\langle T_{\m\n}(x) ~ T_{\r\s}(y)\rangle 
+ \frac{1}{n} \langle T^\l{}_\l\rangle \left(g_{\m\n} g_{\r\s} + g_{\m\r}g_{\n\s} + g_{\m\s}g_{\n\r}\right) 
\d(x,y) \ ,
\label{cc4}
\end{equation}
such that\footnote{Also note the relation for the traces,
\begin{equation*}
i\langle T^\m{}_\m(x) ~ T^\r{}_\r(y)\rangle_{\rm c} = 
i\langle T^\m{}_\m(x) ~ T^\r{}_\r(y)\rangle 
+ (n+2)\langle T^\l{}_\l\rangle
\end{equation*}
}
\begin{equation}
D^\m i\langle T_{\m\n}(x) ~ T_{\r\s}(y)\rangle_{\rm c} = 0 \ .
\label{cc5}
\end{equation}

The anomalous scale Ward identities for the renormalised Green functions in section \ref{sect 2.3}  
may be derived more directly by differentiation of the corresponding
anomaly equation for the VEV $\langle T^\m{}_\m\rangle$ with respect
to the metric and the position-dependent couplings, which in the local
RG approach act as sources for the dimension 2 operators $O_I(x)$. 
To see this, start from the trace anomaly (\ref{b20}), (\ref{b21}) for 
$\langle T^\m{}_\m\rangle$, but keeping the terms
of $O(\partial_\m g^I)$ which are essential for the renormalisation of
the higher-point Green functions, {\it viz.}
\begin{equation}
\langle T^\m{}_\m\rangle  - \langle \Theta\rangle = {\cal A} \ ,
\label{c15}
\end{equation}
where in two dimensions,
\begin{equation}
{\cal A}  = \tfrac{1}{2} \,\b_c \,R - \b_\L + 
D_\m\left(w_I \partial^\m g^I\right) \ .
\label{c16}
\end{equation}
This reflects the anomalous variation of the generating functional 
under a Weyl rescaling $\d g^{\m\n} = 2\s g^{\m\n}$, for which
(in $2$ dimensions; see appendix \ref{sect A} for the $n$-dimensional formulae)
\begin{equation}
\d R = 2\s R + 2 D^2\s \ , ~~~~~~
\d D^2 = 2\s D^2 \ ,
\label{c17}
\end{equation}
along with the scaling transformation $\d O_I = 2\s O_I$ on the
fundamental fields in the path integral. 

The anomaly equation (\ref{c15}) is equivalent to
\begin{equation}
\frac{\d W}{\d \s(y)} ~-~ \b^J\,\frac{\d W}{\d g^J(y)} ~=~
\sqrt{-g(y)}\, {\cal A}(y) \ .
\label{c18}
\end{equation}
A functional derivative w.r.t.~$g_{\m\n}(x)$ now gives
\begin{equation}
2\frac{\d^2 W}{\d g_{\m\n}(x) \d\s(y)} ~-~ 2\b^I\, \frac{\d^2 W}{\d g_{\m\n}(x) \d g^I(y)} ~=~ 
\sqrt{-g(x)}\sqrt{-g(y)}\,{\cal A}_{\m\n}(x,y) \ ,
\label{c19}
\end{equation}
that is,
\begin{equation}
i\langle T_{\m\n}(x)~T^\r{}_\r(y)\rangle  ~-~
i\langle T_{\m\n}(x)~\Theta(y)\rangle   ~+~
2\langle T_{\m\n}(x)\rangle \,\d(x,y) ~=~ 
{\cal A}_{\m\n}(x,y) \ ,
\label{c20}
\end{equation}
with\footnote{The notation here follows that of \cite{Osborn:1999az},
except that here since we are using Minkowski signature we have 
changed the signs in the definitions of ${\cal A}_{\m\n}$, ${\cal B}_I$ and ${\cal H}$, and
also ${\cal E}$, ${\cal F}$ and ${\cal G}$ in the renormalisation group equations below.
${\cal C}_{\m\n}$ and ${\cal C}_I$ remain the same as in \cite{Osborn:1999az}.} 
\begin{equation}
{\cal A}_{\m\n}(x,y) ~=~ \frac{1}{\sqrt{-g(x)}}\,\frac{1}{\sqrt{-g(y)}}\,2\,
\frac{\d}{\d g^{\m\n}(x)}\,\left(\sqrt{-g(y)} {\cal A}(y)\right) \ .
\label{c21}
\end{equation}
Similarly, differentiating w.r.t.~$g_I(x)$, we find
\begin{equation}
\frac{\d^2 W}{\d g^I(x) \d\s(y)} ~-~
\b^J\,\frac{\d^2 W}{\d g^I(x) \d g^J(y)} ~-~
\partial_I \b^J\,\frac{\d W}{\d g^J(y)}\,\sqrt{-g(y)}\,\d(x,y) ~=~
\sqrt{-g(y)}\,\frac{\d {\cal A}(y)}{\d g^I(x)}
\label{c22}
\end{equation}
that is,
\begin{equation}
i\langle O_I(x)~T^\m{}_\m(y)\rangle ~-~
i\langle O_I(x)~\Theta(y)\rangle -\partial_I\b^J \langle
O_J(x)\rangle\,\d(x,y) ~=~  {\cal B}_I(x,y) \ ,
\label{c23}
\end{equation}
where
\begin{equation}
{\cal B}_I(x,y) ~=~ \frac{1}{\sqrt{-g(x)}}\,
\frac{\d {\cal A}(y)}{\d g^I(x)} \ .
\label{c24}
\end{equation}

Contracting  (\ref{c20}) with $g^{\m\n}(x)$ and (\ref{c23}) with $\b^I$, we therefore recover the
anomalous Ward identity for the two-point Green function of $T^\m{}_\m$, {\it viz.}
\begin{multline}
i\langle T^\m{}_\m(x)~T^\r{}_\r(y)\rangle  ~-~
i\langle \Theta(x)~\Theta(y)\rangle ~+~
2\langle T^\m{}_\m(x)\rangle \,\d(x,y)  \\ ~-~
\beta^I\partial_I \beta^J \,\langle O_J(x)\rangle \,\d(x,y) 
~=~  {\cal H}(x,y) 
\label{c25}
\end{multline}
where
\begin{align}
{\cal H}(x,y) ~&=~ g^{\m\n}{\cal A}_{\m\n}(x,y) ~+~ \b^I {\cal B}_I(x,y) \\
&=~ \frac{1}{\sqrt{-g(x)}}
\frac{1}{\sqrt{-g(y)}}\,\left(\frac{\d}{\d\s(x)} \,+\, \b^I
  \frac{\d}{\d g^I(x)}\right)\,\left(\sqrt{-g(y)}\,{\cal A}(y) \right) 
\label{c26}
\end{align}

Evaluating the anomalous terms ${\cal A}_{\a\b}$ and ${\cal B}_I$ from (\ref{c16}), 
taking the renormalised couplings $c = \L = 0$ and $g^I = {\rm const}.$, we find
\begin{align}
{\cal A}_{\m\n}(x,y) ~&=~ \b_c\,\left(G_{\m\n} + \D_{\m\n}\right)\d(x,y) ~=~ \b_c \,\D_{\m\n}\d(x,y)
\nonumber \\
{\cal B}_I(x,y) ~&=~ \tfrac{1}{2}\,\partial_I \b_c\,R\,\d(x,y) ~+~ w_I\,D^2\d(x,y)
\label{c27}
\end{align}
where we have used the fact that the Einstein tensor vanishes in two dimensions.
Here, $\D_{\m\n} = g_{\m\n}D^2 - D_\m D_\n$.
It follows that
\begin{equation}
{\cal H}(x,y) ~=~ \tfrac{1}{2}\,\b^I\partial_I \b_c\,R\,\d(x,y) ~+~
\tilde{\b}_c \,D^2\d(x,y)  
\label{c28}
\end{equation}
with $\tilde{\b}_c = \b_c + \b^I w_I$ as before. 
This recovers (\ref{b38}).

\subsection{Renormalisation group and Green functions}\label{sect 3.1}

The renormalisation group equations follow from the $\m$ independence of the generating functional,
{\it i.e.} $\m\, dW/d\m = 0$. From (\ref{b7a}), we then have the fundamental RGE,
\begin{equation}
\left[\m\frac{\partial}{\partial\m} + \int d^2 x \left( 
\b^I \frac{\d}{\d g^I(x)}
+ \b_c \frac{\d}{\d c(x)}
+ \b_\L \frac{\d}{\d \L(x)} \right) \right]W = 0 \ .
\label{c1}
\end{equation}
If we now define the reduced RG operator
\begin{equation}
{\mathcal D} = \m \frac{\partial}{\partial\m} + \b^I \frac{\partial}{\partial g^I} \ ,
\label{c2}
\end{equation}
using the same shorthand notation explained after (\ref{b7a}), the fundamental RGE may be written as
\begin{equation}
{\mathcal D}W =  -\int d^2 x\sqrt{-g} \,{\cal A} ~=~ -\int d^2 x \sqrt{-g} \left(\tfrac{1}{2} \b_c R - \b_\L\right) \ ,
\label{c3}
\end{equation}
in two dimensions.
This is the standard RGE for local sources, with the couplings themselves generalised to be local 
functions acting as sources for the corresponding operators.

We derive the RGEs for the various one and two-point Green functions of interest by successive
differentiation of (\ref{c3}) with respect to the relevant sources, in this case the materic $g_{\m\n}$ and 
position-dependent couplings $g^I$. We need the commutators,
\begin{equation}
\left[\frac{\d}{\d g^{\m\n}},{\mathcal D}\right] = 0, \qquad\qquad
\left[\frac{\d}{\d g^I},{\mathcal D}\right] = \partial_I \b^J \frac{\d}{\d g^J} \ ,
\label{c4}
\end{equation}
and recall that the anomalous dimension matrix is 
$\hat{\gamma}_I{}^J = \partial_I \b^J \frac{\partial}{\partial g^J}$.

For the one-point functions, or VEVs, we therefore have
\begin{equation}
{\mathcal D} \frac{2}{\sqrt{-g}}\frac{\d W}{\d g^{\m\n}(x)} = 
-\frac{2}{\sqrt{-g}} \frac{\d}{\d g^{\m\n}(x)}\int d^2 y\sqrt{-g}\, {\cal A} \ ,
\label{c5}
\end{equation}
that is,
\begin{equation}
{\mathcal D} \langle T_{\m\n}\rangle = {\cal C}_{\m\n} \ ,
\label{c7}
\end{equation}
with 
\begin{equation}
{\cal C}_{\m\n}(x) = - \int d^2y \sqrt{-g} \, {\cal A}_{\m\n}(x,y)  = 0\ ,
\label{c55}
\end{equation}
from (\ref{c27}), evaluating in flat spacetime. In the same way, 
\begin{equation}
{\mathcal D} \frac{1}{\sqrt{-g}}\frac{\d W}{\d g^I(x)} = 
- \partial_I \b^J \frac{1}{\sqrt{-g}}\frac{\d W}{\d g^J(x)} 
- \frac{1}{\sqrt{-g}} \frac{\d}{\d g^I(x)}\int d^2 y \sqrt{-g} \,{\cal A}  \ 
\label{c8}
\end{equation}
so 
\begin{equation}
{\mathcal D} \langle O_I\rangle + \gamma_I{}^J \langle O_J\rangle = {\cal C}_I \ ,
\label{c9}
\end{equation}
where, again using
(\ref{c27}), we find, 
\begin{equation}
{\cal C}_I(x) = - \int d^2 y \sqrt{-g}\,{\cal B}_I(x,y)  = -\tfrac{1}{2} \partial_I\b_c\, R\ .
\label{c99}
\end{equation}

Iterating this procedure, we can deduce the RGEs for the two-point functions. These will include contributions
arising from the counterterms introduced in the previous section to remove the new divergences which arise
at coincident points. 

Differentiating twice with respect to the metric, we have
\begin{equation}
{\mathcal D}~ i \langle T_{\m\n}(x) ~ T_{\r\s}(y)\rangle = {\cal E}_{\m\n,\r\s}(x,y) \ , 
\label{c10}
\end{equation}
where
\begin{equation}
{\cal E}_{\m\n,\r\s} (x,y) = -\frac{2}{\sqrt{-g(x)}}  \frac{2}{\sqrt{-g(y)}}
\frac{\d}{\d g^{\m\n}(x)} \frac{\d}{g^{\r\s}(y)}\int d^2 z\sqrt{-g}\,{\cal A}(z) \ .
\label{c1010}
\end{equation}
Evaluating in two dimensions, with the local couplings set to constants, we have 
\begin{equation}
{\cal E}_{\m\n,\r\s} = 0 \ .
\label{c1011}
\end{equation}
Next, we find
\begin{equation}
{\mathcal D}~ i \langle T_{\m\n}(x) ~ O_I(y)\rangle + \c_I{}^J i \langle T_{\m\n}(x) ~ O_J(y)\rangle
= {\cal F}_{\m\n,I}(x,y)\ ,
\label{c12}
\end{equation}
where we define 
\begin{equation}
{\cal F}_{\m\n,I}(x,y) = -\frac{2}{\sqrt{-g(x)}} \frac{1}{\sqrt{-g(y)}} \frac{\d}{\d g^{\m\n}(x)} \frac{\d}{\d g^I(y)}
\int d^2 z \sqrt{-g} \,{\cal A}(z) \ .
\label{c1212}
\end{equation}
Then, using
\begin{equation}
\frac{\d}{\d g^{\m\n}(x)} \frac{\d}{\d g^I(y)} \int d^2x \sqrt{-g} \left(\tfrac{1}{2} \b_c R - \b_\L\right)
= \sqrt{-g(x)}\sqrt{-g(y)} \tfrac{1}{2} \partial_I \b_c \left(G_{\m\n} + \D_{\m\n}\right) \d(x,y) \ ,
\label{c11}
\end{equation}
and setting $G_{\m\n} = 0$, we have
\begin{equation}
{\cal F}_{\m\n,I}(x,y) = -\partial_I\b_c\,\D_{\m\n}\d(x,y) \ .
\label{c1211}
\end{equation} 
Finally, for the two-point functions of the operators $O_I$ we find,
 iterating the commutator (\ref{c4}),
\begin{align}
{\mathcal D}~ i\langle O_I(x) ~ O_J(y)\rangle &+\c_I{}^K i\langle O_K(x) ~ O_J(y)\rangle 
+ \c_J{}^K i\langle O_I(x) ~ O_K(y)\rangle + \partial_I \partial_J \b^K \langle O_K\rangle \d(x,y) \nonumber\\
&= {\cal G}_{IJ}(x,y) \ ,
\label{c14}
\end{align}
with
\begin{equation}
{\cal G}_{IJ}(x,y) = -\frac{1}{\sqrt{-g(x)}} \frac{1}{\sqrt{-g(y)}} \frac{\d}{\d g^I(x)} \frac{\d}{\d g^J(y)}
\int d^2 z \sqrt{-g} \,{\cal A}(z) \ .
\label{c1414}
\end{equation}
Here, using
\begin{equation}
\frac{\d}{dg^I(x)}\frac{\d}{\d g^J(y)} \int d^2x \sqrt{-g} \left(\frac{1}{2} \b_c R - \b_\L\right)
= \sqrt{-g(x)}\sqrt{-g(y)} \left(\tfrac{1}{2} \partial_I \partial_J \b_c ~R + \chi_{IJ} D^2\right) \d(x,y) \ , 
\label{c13}
\end{equation}
we find
\begin{equation}
 {\cal G}_{IJ}(x,y) = -\tfrac{1}{2} \partial_I \partial_J \b_c \,R \,\d(x,y) - \chi_{IJ} D^2 \d(x,y) \ .
\label{c1315}
\end{equation}

\subsection{Weyl consistency conditions} \label{sect 3.4}

Combining these RGEs for the two-point Green functions with the anomalous Ward identities 
(\ref{c20}) and (\ref{c23}) gives rise to important consistency relations amongst the 
renormalisation group functions introduced in section \ref{sect 2}.

First, from (\ref{c20}) we have
\begin{equation}
{\cal D}i\langle T_{\m\n}(x)~T^\r{}_\r(y)\rangle  - {\cal D}i\langle T_{\m\n}(x)~\Theta(y)\rangle   +
2{\cal D}\langle T_{\m\n}(x)\rangle \,\d(x,y) ~=~ {\cal D} {\cal A}_{\m\n}(x,y) \ .
\label{c341}
\end{equation}
Noting that ${\cal A}_{\m\n}$ has no explicit $\m$ dependence, and 
applying the RG equations (\ref{c7}), (\ref{c10}) and (\ref{c12}), we readily find
\begin{equation}
{\cal L}_\b{\cal A}_{\m\n}(x,y) = {\cal E}_{\m\n,\r\s}(x,y)g^{\r\s}(y) - {\cal F}_{\m\n,I}(x,y)\b^I 
+ 2 {\cal C}_{\m\n}(x)\d(x,y) \ .
\label{c342}
\end{equation}
In the same way, from (\ref{c23}) we have
\begin{equation}
{\cal D} i\langle O_I(x)~T^\m{}_\m(y)\rangle - {\cal D} i\langle O_I(x)~\Theta(y)\rangle -
{\cal D} \left(\partial_I\b^J \langle O_J(x)\rangle\right)\,\d(x,y) ~=~  {\cal D} {\cal B}_I(x,y) \ ,
\label{c343}
\end{equation}
from which, using (\ref{c9}), (\ref{c12}) and (\ref{c14}), we derive
\begin{equation}
{\cal L}_\b {\cal B}_I(x,y)  = g^{\m\n}(y) {\cal F}_{\m\n,I}(y,x) - 
{\cal G}_{IJ}(x,y)\b^J -\partial_I\b^J {\cal C}_J(x)\d(x,y) \ ,
\label{c344}
\end{equation}
where ${\cal L}_\b {\cal B}_I \equiv \b^J\partial_J {\cal B}_I + \partial_I \b^J {\cal B}_J$.

Finally, although since it is simply obtained by appropriate contractions with $g_{\m\n}$
and $\b^I$ from (\ref{c20}) and (\ref{c23}), for completeness we can apply the RG operator
directly to the Ward identity (\ref{c25})),
\begin{multline}
{\cal D}i\langle T^\m{}_\m(x)~T^\r{}_\r(y)\rangle  - {\cal D}i\langle \Theta(x)~\Theta(y)\rangle 
+ 2{\cal D} \langle T^\m{}_\m(x)\rangle \,\d(x,y)  \\ ~-~
{\cal D}\left(\beta^I\partial_I \beta^J \,\langle O_J(x)\rangle\right) \,\d(x,y) 
~=~  {\cal D}{\cal H}(x,y) 
\label{c345}
\end{multline}
to find 
\begin{multline}
{\cal L}_\b{\cal H}(x,y) = g^{\m\n}(x) {\cal E}_{\m\n,\r\s}(x,y) g^{\r\s}(y) - {\cal G}_{IJ}(x,y) \b^I \b^J
\\+ 2 g^{\m\n}(x) {\cal C}_{\m\n}(x) \d(x,y) - \b^I\partial_I\b^J {\cal C}_J(x) \d(x,y) \ .
\label{c346}
\end{multline}
This of course follows from the two identities above, given the definition (\ref{c26})
of ${\cal H}$ in terms of ${\cal A}_{\m\n}$ and ${\cal B}_I$. 

The identities (\ref{c342}) and (\ref{c344}) are quite general, and not in an essential way restricted 
to two dimensions as considered in this section. They provide the key Weyl consistency 
conditions which as we shall see are intimately related to the $c$-theorem and its
potential generalisations.

To identify these conditions in two dimensions, we now substitute the explicit expressions
given in sections \ref{sect 3.2} and \ref{sect 3.1} for the functions appearing in (\ref{c342}), 
(\ref{c344}).
In this case, we see immediately that (\ref{c342}) is merely an identity, with no information.
From (\ref{c344}), we have
\begin{multline}
{\cal L}_\b \left(\tfrac{1}{2} \partial_I\b_c R \d(x,y) + \omega_I D^2
  \d(x,y)\right) =
- g^{\m\n} \left(\partial_I\b_c \D_{\m\n}\d(x,y)\right) \\
+\left(\tfrac{1}{2} \partial_I \partial_J \b_c R \d(x,y) + \chi_{IJ} D^2 \d(x,y)\right)\b^J 
 +\partial_I\b^J \left(\tfrac{1}{2}\partial_J\b_c R \d(x,y)\right) \ .
\label{c347}
\end{multline}
Collecting terms, we see that the coefficient of the curvature $R$ vanishes identically. 
From the coefficient of $D^2 \d(x,y)$, however, we find the crucial relation for the RG flow
of the anomaly coefficient $\b_c$,
\begin{equation}
\partial_I\b_c - \chi_{IJ}\b^J + {\cal L}_{\b} w_I = 0 \ ,
\label{c348}
\end{equation}
as already encountered in (\ref{b24}). As for the contracted form,
\begin{equation}
\b^I \partial _ I \tilde{\b_c} = \chi_{IJ} \b^I \b^J \ ,
\label{c349}
\end{equation}
previously derived in (\ref{b27}), this evidently arises directly from the RG relation 
(\ref{c346}) for ${\cal H}(x,y)$.

\subsection{Local RGE and Weyl consistency conditions}\label{sect 3.3}

An elegant way to summarise the essential results of this section \cite{Osborn:1991gm}
is to introduce the following variation operators for Weyl transformations and RG flows: 
\begin{align}
\D_\s^W &= 2 \int d^2 x ~\s(x) g^{\m\n} \frac{\d}{\d g^{\m\n}(x)}  \nonumber \\
\D_\s^\b &= \int d^2 x ~\s(x) \b^I \frac{\d}{\d g^I(x)}  \ .
\label{c32}
\end{align}
where recall a Weyl transformation is $g^{\m\n}\rta \Omega^2 g^{\m\n}$ with $\Omega = e^\s$.
Then,
\begin{equation}
\left(\D_\s^W - \D_\s^\b\right)W = \int d^2 x \sqrt{-g}~ \left[\s(x) \left(
\tfrac{1}{2} \b_c R - \tfrac{1}{2}\chi_{IJ}  \partial_\m g^I \partial^\m g^J \right) -
\partial_\m \s(x) \left(w_I \partial^\m g^I\right)\right] \ .
\label{c33}
\end{equation}

First, consider the restricted case where the parameter $\s$ is constant, when the final total derivative
term $D_\m Z^\m = D_\m \left(w_I\partial^\m g^I\right)$ does not contribute. 
In this case, (\ref{c33}) is equivalent to the standard renormalisation group equation (\ref{c1}) with local
sources,
\begin{equation}
\left[\m\frac{\partial}{\partial\m} + \int d^2 x \left( 
\b^I \frac{\d}{\d g^I(x)}
+ \b_c \frac{\d}{\d c(x)}
+ \b_\L \frac{\d}{\d \L(x)} \right) \right]W = 0 \ .
\label{c34}
\end{equation}
together with the dimensional equation (for theories with only marginal operators),
\begin{equation}
\left[\m \frac{\partial}{\partial \m} + \Omega \frac{\partial}{\partial\Omega}\right]W = 0 \ .
\label{c35}
\end{equation}
Recall that for Weyl transformations, 
\begin{equation}
\Omega \frac{\partial}{\partial\Omega} = 2 \int d^2 x~g^{\m\n}\frac{\d}{\d g^{\m\n}} \ .
\label{c36}
\end{equation}

For {\it local} Weyl transformations, corresponding to spacetime-dependent $\s(x)$, the identity 
(\ref{c33}) carries the full content of the local renormalisation group. Taking the functional derivative 
$\d/\d\s(x)$ of (\ref{c33}), we find the local generalisation of (\ref{c34}) and (\ref{c35}):
\begin{equation}
\left[\Omega(x)\frac{\d}{\d \Omega(x)} -  \b^I \frac{\d}{\d g^I(x)} - \b_c \frac{\d}{\d c(x)}
- \b_\L \frac{\d}{\d \L(x)} \right]W = D_\m Z^\m  \ ,
\label{c37}
\end{equation}

Evaluating, we recover the now familiar expression for the trace of the energy-momentum tensor:
\begin{equation}
T^\m{}_\m = \b^I O_I + \tfrac{1}{2}\b_c R - \tfrac{1}{2} \chi_{IJ} \partial_\m g^I \partial^\m g^J 
+ D_\m\left(w_I \partial^\m g^I\right) \ ,
\label{c38}
\end{equation}
 that is, the anomaly equation (\ref{c18}) of the previous section. It is therefore clear that subsequent
derivatives w.r.t.\,\,$g^{\m\n}$ and $g^I$ generate the identities amongst two-point Green functions involving
the energy-momentum tensor trace found there. With (\ref{c35}), all the RGEs for these Green functions
can also be deduced from (\ref{c33}).

One of the most powerful applications of this formalism, however, is in deriving the Weyl consistency conditions,
which appear to be intimately related to the existence of the $c$ and $a$-theorems.
In two dimensions, this means determining (\ref{b24}), which we first encountered in the context of a 
particular renormalisation scheme. In fact, a deeper understanding reveals these consistency conditions
to be fundamentally a consequence of the abelian nature of Weyl transformations. This method is especially
direct and useful in four dimensions, where the situation is considerably more complicated and there
are many conditions.

Since Weyl transformations are abelian, the commutator of two variation operators vanishes, that is
\begin{equation}
\left[\D_\s^W - \D_\s^\b, \D_{\s'}^W - \D_{\s'}^b\right] W = 0 \ .
\label{c39}
\end{equation}
Evaluating the l.h.s., we find
\begin{equation}
\int d^2 x \sqrt{-g}~\left(\s' \partial_\m\s - \s\partial_\m\s'\right) \left(\partial_\m\b_c - 
\chi_{IJ}\b^J \partial_\m g^I + \b^I \partial_I\left(w_J\partial_\m g^J\right)\right) \ ,
\label{c40}
\end{equation}
and therefore determine the Weyl consistency condition:
\begin{equation}
\partial_I \b_c - \chi_{IJ} \b^J + {\mathcal L}_\b w_I = 0 \ .
\label{c41}
\end{equation}
Then as before, if we define $\tilde{\b}_c = \b_c + \b^I w_I$, we have
\begin{equation}
\partial_I \tilde{\b}_c - \chi_{IJ}\b^J + \left(\partial_I w_J - \partial_J w_I\right)\b^J = 0 \ ,
\label{c42}
\end{equation}
and so 
\begin{equation}
\b^I \partial_I\tilde{\b}_c = \chi_{IJ}\b^I \b^J \ .
\label{c43}
\end{equation}

\section{c-Theorem in Two Dimensions}\label{sect 4}

With this understanding of the renormalisation group and anomalous Weyl symmetry applied to Green functions
of the energy-momentum tensor, we are ready to discuss the derivation of the $c$-theorem in two dimensions.
We present several derivations, each of which brings some fresh insight and suggests possible ways forward
in looking for an equivalent theorem in four-dimensional QFTs.

In two dimensional flat space, conservation of the energy-momentum tensor
implies that the two-point Green function takes the form
\begin{equation}
i \langle T_{\m\n}(x) ~ T_{\r\s}(0)\rangle_{\rm c} = \D_{\m\n} \D_{\r\s} \,\Omega(t) \ ,
\label{d1}
\end{equation}
where $\D_{\m\n} = \partial^2 \eta_{\m\n} -\partial_\m\partial_\n$ and for spacelike separations,
$t = \tfrac{1}{2} \log(-\m^2 x^2)$.
Note that $\Pi_{\m\n\r\s}^{(0)} = \D_{\m\n} \D_{\r\s}$ is a projection operator onto spin 0 intermediate states.
In two dimensions, this is the only possible Lorentz structure. In contrast, in four dimensions there is
a second projection projector $\Pi_{\m\n\r\s}^{(2)}$ onto spin 2 states.  
Also notice that, strictly speaking, we are 
using the conserved two-point function defined in (\ref{cc4}) to justify writing the Green function 
in terms of the projection operator $\D_{\m\n} \D_{\r\s}$. In fact the distinction is unimportant here,
since the difference only involves contact terms proportional to delta functions, which do not contribute
at the point where we use the anomalous Ward identities in (\ref{d16}) below. 

\subsection{c-theorem and the spectral function }\label{sect 4.1}

A very clear way to understand the physical meaning of the $c$-theorem is to use a spectral function
representation of the two-point functions \cite{Friedan:1990,Cappelli:1990yc,Shore:1990wq,Shore:1990ng}. 
Recall that in general the spectral function, or 
K\"allen-Lehmann, representation of a two-point function of fields $\Phi$ in $n$ dimensions 
(note that our normalisation of $\r(\l^2)$ differs from the standard convention 
by a factor of $2\pi$) is,\footnote{In $n$ dimensions, 
the position-space Feynman propagator is
\begin{align*}
D_F(x;m^2) &= \int \frac{d^n k}{(2\pi)^n}~e^{-ik.x}~ \frac{i}{k^2 - m^2 + i\e} \nonumber\\
&=~\theta(x^2) \frac{(-i)^{n-1}}{(2\pi)^{n/2}} \frac{\pi}{2} 
\frac{m^{\frac{n}{2}-1}}{(\sqrt{x^2 -i\e})^{\frac{n}{2} -1}}
H_{\frac{n}{2}-1}^{(2)}(m\sqrt{x^2 - i\e}) \nonumber \\
&~~+\theta(-x^2) \frac{1}{(2\pi)^{n/2}} \frac{m^{\frac{n}{2}-1}}{(\sqrt{x^2 +i\e})^{\frac{n}{2} -1}}
K_{\frac{n}{2}-1}(m\sqrt{-x^2 + i\e}) \ .
\end{align*}}
\begin{equation}
\langle 0|T~\Phi(x) ~ \Phi(0)|0\rangle = \int_0^\infty d\l^2\,\r(\l^2) \, D_F(x;\l^2) \ ,
\label{d2}
\end{equation}
where the (positive-definite) spectral function $\r(\l^2)$ is
\begin{equation}
\r(\l^2) = \sum_n \d(\l^2 - m_n^2) |\langle 0|\Phi(0)|n\rangle|^2 \ ,
\label{d3}
\end{equation}
and $D_F(x;m^2)$ is the Feynman propagator for a spin 0 field of mass
$m$.
In two dimensions, for spacelike separation ($x^2 < 0$), we simply have
\begin{equation}
D_F(x;m^2) = \frac{1}{2\pi} K_0(m |x|)  \ .
\label{d5}
\end{equation}
where we have used the notation $|x|= \sqrt{-x^2}$. 
We now follow the derivation described in \cite{Shore:1990wq,Shore:1990ng}.
The required spectral function representation for the energy-momentum tensor two-point function
can be written in the form
\begin{align}
i\langle T_{\m\n}(x) ~ T_{\r\s}(0) \rangle_{\rm c} &= \D_{\m\n} \D_{\r\s} \,\Omega(t) \nonumber \\
&= \D_{\m\n} \D_{\r\s} \,\frac{i}{2\pi} \int_0^\infty d\l^2\, \r(\l^2) K_0(\l|x|)  \ ,
\label{d6}
\end{align}
where in two dimensions the dimension of $\r(\l^2)$ is $-2$.  

Carrying out the differentiations we have
\begin{align}
\langle T_{\m\n}(x) ~ T_{\r\s}(0)\rangle_{\rm c} &= \frac{1}{2\pi} \int_0^\infty d\l^2\, \l^4\,\r(\l^2)
\left[A(\l|x|) \eta_{\m\n}\eta_{\r\s} +\ldots + E(\l|x|) \frac{1}{x^4} x_\m x_\n x_\r x_\s \right] \nonumber\\
\langle T_{\m\n}(x) ~ T^\r{}_\r(0)\rangle_{\rm c} &= -\frac{1}{2\pi} \int_0^\infty d\l^2\, \l^4\,\r(\l^2)
\left[P(\l|x|) \eta_{\m\n} ~+~ Q(\l|x|) \frac{1}{x^2} x_\m x_\n  \right] \nonumber\\
\langle T^\m{}_\m(x) ~ T^\r{}_\r(0)\rangle_{\rm c} &= \frac{1}{2\pi} \int_0^\infty d\l^2\, \l^4\,\r(\l^2)\, R(\l|x|) \ ,
\label{d7}
\end{align}
where, using the standard Bessel function recursion relations,
\begin{align}
z\frac{d}{dz} K_\n(z) - \n K_\n(z) &= - z K_{\n+1}(z) \nonumber \\
z K_{\n -1}(z) + 2 \n K_\n(z) &= z K_{\n +1} \ ,
\label{d8}
\end{align}
the important coefficient functions are seen to be remarkably simple:
\begin{align}
E(\l|x|) &= K_4(\l|x|) \nonumber \\
Q(\l|x|) &= K_2(\l|x|) \nonumber \\
R(\l|x|) &= K_0(\l|x|) \ .
\label{d9}
\end{align}

The next step is to identify the dimensionless Zamolodchikov $F$, $H$ and $G$ functions introduced 
in section \ref{sect 1.2}. It can be shown that up to normalisations these are simply
\begin{align}
F &= \frac{1}{2\pi} \int_0^\infty d\l^2\,\r(\l^2) \, (\l|x|)^4\, E(\l|x|) \nonumber \\
H &= \frac{1}{2\pi} \int_0^\infty d\l^2\,\r(\l^2) \, (\l|x|)^4\, Q(\l|x|) \nonumber \\
G &= \frac{1}{2\pi} \int_0^\infty d\l^2\,\r(\l^2)\, (\l|x|)^4\,  R(\l|x|) \ .
\label{d10}
\end{align}
To derive the $c$-theorem, we now look for a combination $C = F + \a H + \b G$ with the property
$|x|\partial C/\partial|x| \sim G$. In fact, we can show using the relations (\ref{d8}) for the 
Bessel functions that
\begin{equation}
C = F + 2 H - 3 G
\label{d11}
\end{equation}
satisfies
\begin{equation}
|x| \frac{\partial C}{\partial |x|} = - 12 G \ .
\label{d12}
\end{equation}
Dimensional analysis and the non-renormalisation of the energy-momentum tensor Green functions
then implies
\begin{equation}
|x| \frac{\partial C}{\partial|x|} = \m \frac{\partial C}{\partial \m}  = -\b^I \frac{\partial C}{\partial g^I} \ .
\label{d13}
\end{equation}

A similar spectral representation can be written for the two-point Green function for the operators $O_I$:
\begin{align}
i \langle O_I(x) ~ O_J(0)\rangle &= \partial^4 \Omega_{IJ}(t)
\nonumber \\
&= \partial^4 \frac{i}{2\pi} \int_0^\infty d\l^2\, \r_{IJ}(\l^2) K_0(\l|x|) \ ,
\label{d14}
\end{align}
and we define 
\begin{equation}
G_{IJ} = \frac{1}{2\pi} x^4 \int_0^\infty d\l^2\,\r_{IJ}(\l^2)\, (\l|x|)^4\, K_0(\l|x|) \ .
\label{d15}
\end{equation}
Then from the Ward identity (\ref{c25}), noting that the delta function contributions vanish for the Green functions
rescaled by $x^4$, we have
\begin{equation}
G = G_{IJ} \b^I \b^J \ .
\label{d16}
\end{equation}

To summarise, we have proved the $c$-theorem:
\begin{equation}
\b^I \partial_I C = 12 G_{IJ} \b^I \b^J \ ,
\label{d17}
\end{equation}
where
\begin{align}
C &= \frac{1}{2\pi} \int_0^\infty d\l^2 \,\r(\l^2)\, (\l|x|)^4\,\left[K_4(\l|x|) + 2 K_2(\l|x| - 3 K_0(\l|x|)\right] 
\nonumber\\
G_{IJ} &= \frac{1}{2\pi} \int_0^\infty d\l^2 \,\r_{IJ}(\l^2)\, (\l|x|)^4\, K_0(\l|x|) \ .
\label{d18}
\end{align}
The function $G_{IJ}$ is positive and is interpreted as a metric on the space of couplings.
Also note that, because of the hierarchy $K_{\n+4} > K_{\n+2} >K_{\n}$ of the (positive definite) Bessel
functions $K_\n(\l|x|)$, the $C$-function itself is also positive.

This allows a very clear physical interpretation of the $c$-theorem \cite{Shore:1990wq,Shore:1990ng}.  
First, write
\begin{equation}
C = \frac{1}{2\pi} \int_0^\infty d\l^2\, \r(\l^2)\, f(\l|x|) \ , 
\label{d19}
\end{equation}
and 
\begin{equation}
G_{IJ} = \frac{1}{2\pi} \int_0^\infty d\l^2\, \r_{IJ}(\l^2)\, h(\l|x|) \ ,
\label{d20}
\end{equation}
 where the `sampling functions' $f(\l|x|)$ and $h(\l|x|)$ are given by
\begin{equation}
f(\l|x|) = (\l|x|)^4 \left(K_4(\l|x|) + 2 K_2(\l|x|) - 3 K_0(\l|x|) \right) \ ,
\label{d21}
\end{equation}
and
\begin{equation}
h(\l|x|) = (\l|x|)^4 K_0(\l|x|) \ .
\label{d22}
\end{equation}
These are plotted in figure \ref{Sampling2d}.
Notice especially how the profile of $f(\l|x|)$, which has the shape of a smoothed step-function, moves
to smaller values of $\l$ as $|x|$ is increased.\footnote{Note the normalisation of these sampling functions:
\begin{equation*}
\int_0^\infty d\l^2\,f(\l|x|) = 2^8\,3\, x^{-2} \ , \qquad \qquad  \int_0^\infty d\l^2\,h(\l|x|) = 2^7\, x^{-2} \ .
\end{equation*} }

\begin{figure}[h!]
\centering
\includegraphics[scale=0.8]{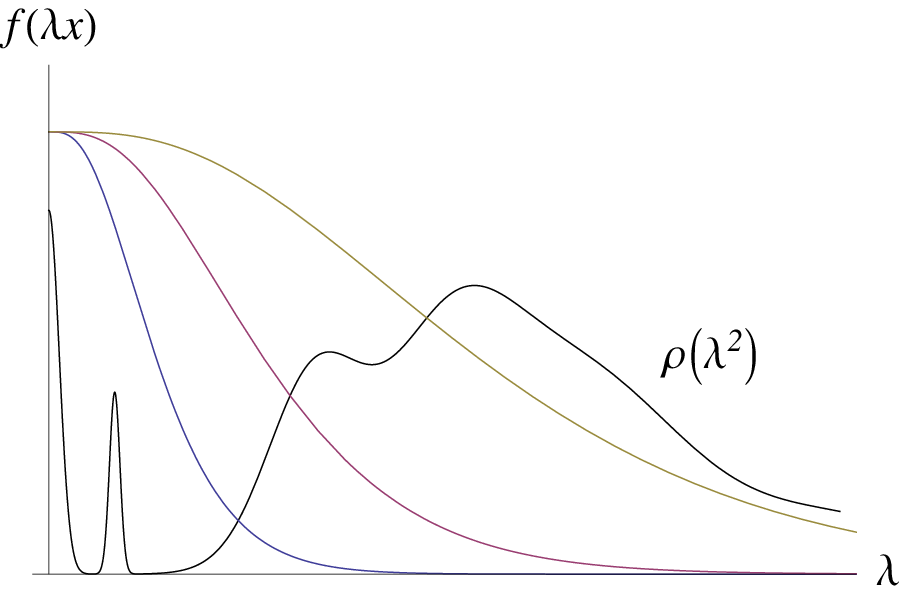}
\includegraphics[scale=0.8]{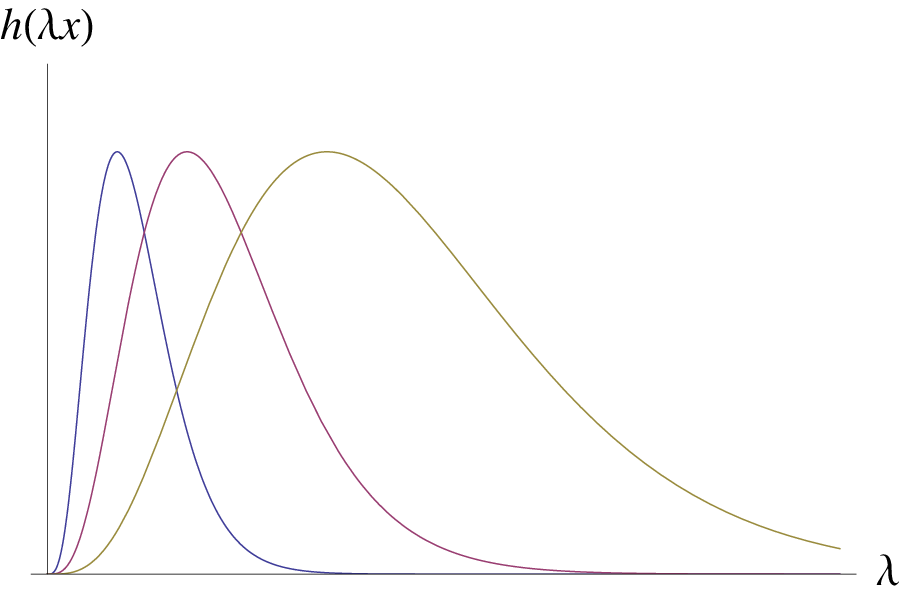}
\caption{The LH figure shows the sampling function $f(\l|x|)$ defining the Zamolodchikov $C$ function
for three values of the scale $|x|$, superimposed on a sketch of the spectral function $\r(\l^2)$. 
As $|x|$ is increased, the sampling function moves towards smaller $\l$ values, progressively sampling the
IR sector of the spectrum. The RH plot shows the corresponding manifestly positive sampling function
$h(\l|x|)$ defining the metric function $G_{IJ}$.}
\label{Sampling2d}
\end{figure}

Now, the spectral function $\r(\l^2)$ measures the density of states at the mass scale $\l$, with a 
typical sketched form superimposed on the sampling function $F(\l|x|)$ in figure \ref{Sampling2d}. 
So, as $|x|$ is increased, the sampling function $f(\l|x|)$ measures less and less
of the high-energy, UV part of the spectrum. We therefore understand why the $C$ function decreases
as we flow into the IR, {\it i.e.}
\begin{equation}
|x| \frac{\partial C}{\partial |x|} < 0 
\label{d23}
\end{equation}
We conclude therefore that with the definition (\ref{d19}), $C(g(|x|)$ is a measure of the degrees of freedom
in the spectrum at scales $\gtrsim |x|$. Evidently, this is a monotonically decreasing function as we flow
into the infra-red.

This construction, with its natural physical interpretation, looks as if it should generalise straightforwardly
to four dimensions. We shall see in section \ref{sect 5} to what extent this is borne out. First though, we 
give an alternative, closely related, derivation which relates $C(g)$ to the RG functions associated with the
Weyl anomaly.

\subsection{c-theorem and renormalisation group flow}\label{sect 4.2}

Once again, we start from the general form (\ref{d1}) for the two-point function of the energy-momentum tensor.
Taking the Fourier transform, we define
\begin{align}
i \langle T^\m{}_\m(x) ~ T^\r{}_\r (0) \rangle_{\rm c} &= \partial^4 \Omega(x) \nonumber \\
&= - \int \frac{d^2 k}{(2\pi)^2}\, e^{-i k.x} \,k^2 \tilde{\Omega}_{TT}(t) \ ,
\label{d24}
\end{align}
where here $t = \tfrac{1}{2} \log(k^2/\m^2)$. Similarly, we define $\tilde{\Omega}_{IJ}$ by
\begin{equation}
i \langle O_I(x)~O_J(0) \rangle = - \int \frac{d^2 k}{(2\pi)^2}\, e^{-i k.x} \,k^2 \tilde{\Omega}_{IJ}(t) \ .
\label{d244}
\end{equation}

Next, we introduce the Ward identities and renormalisation group equations from section \ref{sect 3}.
From the fundamental Ward identity (\ref{c25})
\begin{equation}
i\langle T^\m{}_\m(x)~T^\r{}_\r(0)\rangle  ~-~
i\langle \Theta(x)~\Theta(0)\rangle ~+~\ldots 
~=~  \tilde\b_c \partial^2 \d(x) \ ,
\label{d28}
\end{equation}
where we have omitted the terms involving the VEVs and $\d(x,y)$ with no derivatives,
we deduce the key identity
\begin{equation}
\tilde{\Omega}_{TT} - \b^I \b^J \tilde{\Omega}_{IJ} = \tilde\b_c \ .
\label{d29}
\end{equation}
The RGEs (\ref{c10}) and (\ref{c14}) imply
\begin{align}
&{\mathcal D} \tilde{\Omega}_{TT} =0 \nonumber\\
&{\mathcal D} \tilde{\Omega}_{IJ} + \partial_I\b^K \tilde{\Omega}_{KJ} + \partial_J \b^K \tilde{\Omega}_{IK}
= - \chi_{IJ} \ ,
\label{d30}
\end{align}
that is 
\begin{align}
\frac{\partial}{\partial t} \,\tilde{\Omega}_{TT} &= {\mathcal L}_\b \tilde{\Omega}_{TT}
\equiv \b^I\partial_I \tilde{\Omega}_{TT} \nonumber \\
\frac{\partial}{\partial t} \,\tilde{\Omega}_{IJ} &= \chi_{IJ} + {\mathcal L}_\b \tilde{\Omega}_{IJ} 
\label{d31}
\end{align}

These can be solved as usual to express $\tilde{\Omega}_{TT}(t)$ and $\tilde{\Omega}_{IJ}(t)$ in terms
of the running couplings $g^I(t)$ and anomalous dimensions $\c_I{}^J(t) = \partial_I\b^J$: 
\begin{align}
\tilde{\Omega}_{TT}(t) &= F_{TT}(g(t)) \nonumber \\
\tilde{\Omega}_{IJ}(t) &= e^{\int_0^t dt'\,\c(g(t')} \left[F(g(t)) + 
\int_0^t dt' \chi(g(t')) e^{-\int_0^{t'} dt^{\prime\prime}\,\c(g(t^{\prime\prime})} \right] \ ,
\label{d32}
\end{align}
suppressing indices. Expanding to first order around an arbitrary point on the RG trajectory,
we find (as is also obvious from (\ref{d31}) directly)
\begin{align}
\tilde{\Omega}_{TT}(t) &= F_{TT}(g) + t \b^I\partial_I F_{TT}(g) + O(t^2) \nonumber \\
\tilde{\Omega}_{IJ}(t) &= F_{IJ}(g) + t \left(\chi_{IJ}(g) + {\mathcal L}_\b F_{IJ}(g)\right) + O(t^2) \ ,
\label{d33}
\end{align}
where $g$ is the running coupling at $t=0$, {\it i.e.} at $k^2 = \m^2$.

The anomalous Ward identity (\ref{d29}) then gives \cite{Shore:1990wq}, equating terms 
of $O(1)$ and $O(t)$,
\begin{align}
F_{TT} - \b^I \b^J F_{IJ} &= \tilde{\b}_c \nonumber \\
\b^I\partial_I F_{TT} - \b^I \b^J {\mathcal L}_\b F_{IJ} &= \chi_{IJ} \b^I \b^J \ ,
\label{d34}
\end{align}
where all functions are evaluated at $g(0)$. The second of these can be rewritten as
\cite{Shore:1990wq},
\begin{equation}
\b^K\partial_K\left(F_{TT} - \b^I \b^J F_{IJ} \right) = \chi_{IJ} \b^I \b^J \ ,
\label{d35}
\end{equation}
that is 
\begin{equation}
\b^I \partial_I \tilde{\b}_c = \chi_{IJ} \b^I \b^J \ .
\label{d36}
\end{equation}
This is of course the familiar Weyl consistency condition. 

The Zamolodchikov $F$, $H$ and $G$ functions, and then $C$ and $G_{IJ}$, are defined in terms of the
two-point Green functions through (\ref{d7}) and (\ref{d10}). Writing in terms of $\tilde{\Omega}_{TT}$
and $\tilde{\Omega}_{IJ}$ and inserting the small $t$ expansions (\ref{d33}) under the Fourier integral,
we find \cite{Shore:1990wq,Shore:1990ng} after some calculation that
\begin{align}
C &= 96\pi \left[F_{TT} + \b^I \partial_I F_{TT} \left(-\tfrac{1}{2} \log(- \m^2 x^2) 
+ {\rm const} \right) + \ldots ~\right] \nonumber\\
G_{IJ} &= 8\pi \left[ \chi_{IJ} + {\mathcal L}_\b F_{IJ} + \ldots ~\right]
\label{d38}
\end{align}
Using (\ref{d34}), we can write this as
\begin{equation}
C = 96 \pi \left[\tilde{\b}_c + \b^I \b^J F_{IJ} + \ldots \right] \ .
\label{d39}
\end{equation}

This shows that at a fixed point of the renormalisation group, where the $\b^I = 0$, the $C$ function is simply
given by $\b_c$, {\it i.e.}
\begin{equation}
C(g^*) = 96 \pi \b_c(g^*) \ ,
\label{d40}
\end{equation}
while the RG flow of $C(g)$ itself is given by the $c$-theorem identity
\begin{equation}
\b^I\partial_I C = 12 G_{IJ}\b^I \b^J \ .
\label{d400}
\end{equation}
Moreover, this construction also shows the role of the Weyl consistency condition, which by (\ref{d34}) describes
the flow in a small neighbourhood of an arbitrary point on the RG trajectory. Precisely, (\ref{d400}) reads at
lowest order in this expansion,
\begin{equation}
\b^K\partial_K \left(\tilde{\b}_c + \b^I \b^J F_{IJ} \right) = \b^I \b^J \left(\chi_{IJ} + {\mathcal L}_\b F_{IJ}\right) \ ,
\label{d41}
\end{equation}
which is just the consistency condition with the intrinsic renormalisation scheme arbitrariness (see footnote 5). 
Equivalently, notice that the terms proportional to $F_{IJ}$ cancel, since 
${\cal L}_\b \left(\b^I \b^J F_{IJ}\right) = \b^I \b^J {\cal L}_\b F_{IJ}$. 
The Weyl consistency condition therefore appears as a local approximation to the full $c$-theorem at an arbitrary 
point on the RG trajectory away from the fixed point.

\subsection{c-theorem in position space}\label{sect 4.3}

The derivation of the $c$-theorem above used a momentum space representation for the Green functions.
Although giving less insight into the physical meaning of the $c$-theorem and its relation to the 
renormalisation group and Weyl consistency conditions, we can also give a similar construction 
directly in position space \cite{Osborn:1991gm}.

The key ingredients are, as before, the anomalous Ward identity (\ref{d28}) and the RGEs (\ref{c10})
and (\ref{c14}) for the Green functions
\begin{align}
i\langle T^\m{}_\m(x) ~ T^\r{}_\r(0)\rangle_{\rm c} &= \partial^4 \Omega(t) \nonumber\\
i\langle O_I(x) ~O_J(0)\rangle &= \partial^4 \Omega_{IJ}(t) \ ,
\label{d42}
\end{align}
where $t = \tfrac{1}{2} \log(- \m^2 x^2)$. Since in two dimensions, $\partial^2 t = 2\pi \d(x)$, we can 
rewrite these as
\begin{align}
&\Omega - \b^I \b^J \Omega_{IJ} = - 2\pi \,\tilde{\b}_c\, t \nonumber\\
&{\mathcal D}\Omega = 0 \nonumber \\
&{\mathcal D}\Omega_{IJ} + \partial_I\b^K \Omega_{KJ} + \partial_j\b^K \Omega_{IK} = 2\pi\,\chi_{IJ}\, t \ .
\label{d43}
\end{align}

With the usual definition of the $C$ function, some unilluminating algebra using (\ref{d43}), 
with $\Omega' = d\Omega/dt$, now shows that
\begin{align}
C &=96\pi\left(-\tfrac{1}{2}\Omega' + \tfrac{1}{2}\Omega^{\prime\prime} - \tfrac{1}{8}\Omega^{\prime\prime\prime} \right)
\nonumber\\
&= 96\pi\left(\tilde{\b}_c + a_{IJ} \b^I \b^J \right) \ ,
\label{d44}
\end{align}
with $a_{IJ} = -\tfrac{1}{2} \Omega'_{IJ} + \tfrac{1}{2}\Omega^{\prime\prime}_{IJ} - \tfrac{1}{8}\Omega^{\prime\prime\prime}_{IJ}$.
The metric $G_{IJ}$ is similarly found to be
\begin{align}
G_{IJ} &= 8\pi\left( \tfrac{1}{2}\Omega^{\prime\prime}_{IJ} -  \tfrac{1}{2}\Omega^{\prime\prime\prime}_{IJ} + 
\tfrac{1}{8}\Omega^{\prime\prime\prime\prime}_{IJ} \right)\nonumber\\
&= 8\pi \left(\chi_{IJ} + {\mathcal L}_\b a_{IJ}\right) \ ,
\label{d45}
\end{align}
in clear correspondence with (\ref{d38}).
The $c$-theorem follows:
\begin{equation}
\b^I\partial_I C = 12 G_{IJ}\b^I \b^J \ .
\label{d46}
\end{equation}
Just as in the previous derivation, we see that away from a fixed point the $c$-theorem differs from
the Weyl consistency condition itself, but with the RG functions $\tilde{\b}_c$ and $\chi$ modified by
terms of the form corresponding to the addition of a finite local counterterm in the action.

\subsection{c-theorem and dispersion relations}\label{sect 4.4}

We now present a further derivation of the $c$-theorem using a dispersion relation for the
two-point Green function of $T^\m{}_\m$, which will be of particular interest when we discuss the
generalisation to four dimensions. Note that this only proves the `weak' $c$-theorem,
\begin{equation}
\b_c^{UV} - \b_c^{IR} > 0 \ ,
\label{d47}
\end{equation}
which shows that for an RG flow from a UV to an IR fixed point, $\b_c$ is less in the IR.
Unlike the previous derivations, however, this does not identify a monotonically decreasing $C$ function 
along the flow.

The central idea is to write a dispersion relation for $\Pi(\w)$, defined as
\begin{equation}
i \int d^2 x \,e^{ik.x}\,\langle T^\m{}_\m(x) ~ T^\r{}_\r (0) \rangle = k^2 \Pi(\w) \ ,
\label{d48}
\end{equation}
where, to emphasise the analogy with the Kramers-Kr\"onig dispersion relation familiar from 
QED, we have written the momentum as $k_\m = \w\hat{k}_\m$, where $\hat{k}_\m$ is a unit vector.
A similar definition can be given for $\Pi_{IJ}(\w)$ in terms of $\langle O_I~ O_J\rangle$.

It is important to note that nothing in the following derivation requires there to be a dynamical
`dilaton' field $\t(x)$ coupling to $T^\m{}_\m$ (or indeed a `graviton' coupling to $T_{\m\n}$
if we considered more general Green functions of the energy-momentum tensor) in analogy
with the photon $A_\m(x)$ coupling to the current $J_\m$ in the QED relation.\footnote{In the QED case,
we consider the vacuum polarisation tensor, {\it i.e.} the two-point function of the conserved electromagnetic
current $J_\m$,
\begin{equation*}
\Pi_{\m\n} = \left(k^2 \eta_{\m\n} - k_\m k_\n\right) \Pi(\w) = 
\langle J_\m(x) ~ J_\n(0)\rangle\big|_{\rm F.T.} \ .
\end{equation*}
 The refractive index $n(\w)$ for photons with frequency $\w$ is identified as 
\begin{equation*}
n(\w) -1 = \Pi(\w) \ ,
\end{equation*}
and satifies the Kramers-Kronig dispersion relation 
\begin{equation*}
n(\infty) - n(0) = - \frac{2}{\pi} \int_0^\infty \frac{d\w}{\w}\,{\rm Im}\,n(\w) < 0 \ ,
\end{equation*}
which relates the refractive index in the UV and IR limits. The analogy with (\ref{d48}) is even closer
if we think of the trace operator $T^\m{}_\m \sim \partial^\m D_\m$, where $D_\m$ is the dilatation 
current.}  We also present the derivation in full detail making explicit where all the necessary assumptions
on $\Pi(\w)$ are used, so that we may make a critical assessment of the later generalisation to four
dimensions. 

We now make some (standard) assumptions about $\Pi(\w)$:
\begin{enumerate}
\item $\Pi(\w)$ is analytic in the upper-half complex $\w$-plane, but may have cuts along the real 
$\w$ axis. This is a requirement of causality.

\item $\Pi(\w)$ is bounded at infinity.

\item $\Pi(\w) = \Pi(-\w)$. This is a consequence of translation invariance since the two-point function
$\langle T^\m{}_\m(x) ~ T^\m{}_\m(y)\rangle$ depends only on the magnitude of the separation $|x-y|$.

\item $\Pi(\w^*)= \Pi(\w)^*$, {\it i.e.}~hermitian analyticity. This is a basic axiom of $S$-matrix theory for 
scattering amplitudes (section 1.3 of \cite{Eden:1966}), but see \cite{Hollowood:2008kq} for a critique 
in curved spacetime.

\item ${\rm Im}\,\Pi(\w) > 0$, which is a consequence of unitarity and the key element of the optical theorem 
(but again see \cite{Hollowood:2011yh} for subtleties, especially in curved spacetime).
\end{enumerate}

\begin{figure}[h!]
\centering
\includegraphics[scale=0.6]{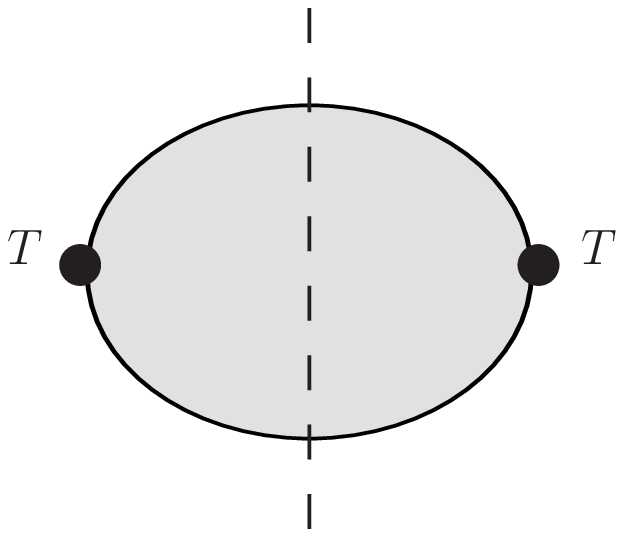}
\qquad \qquad \qquad
\includegraphics[scale=0.6]{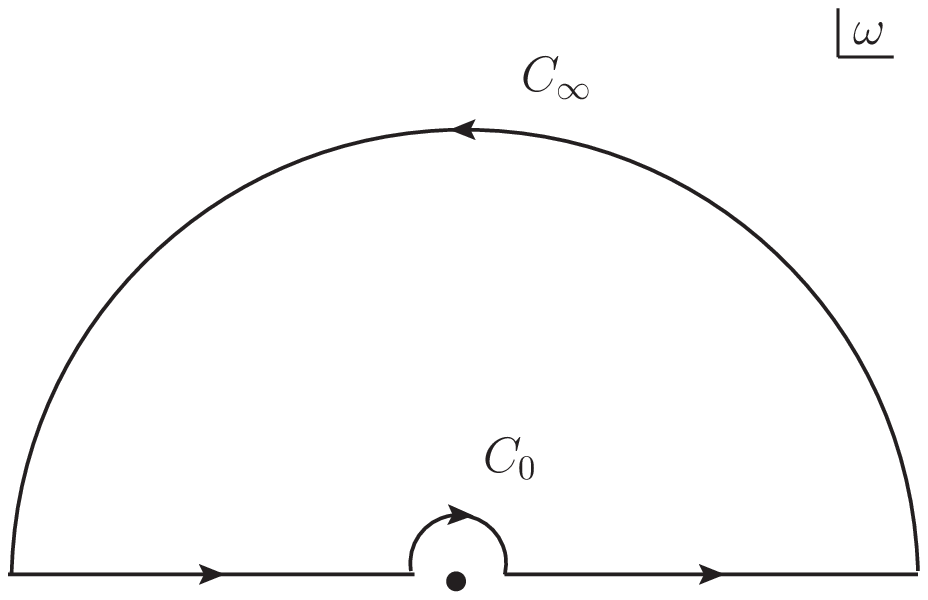}
\caption{The LH figure shows the two-point function $\langle T^\m{}_\m ~ T^\r{}_\r\rangle$ with the cut giving rise 
to ${\rm Im}\,\Pi(\omega)$ in (\ref{d56}). The RH figure shows the contour in the complex $\omega$-plane 
used in the dispersion relation.}
\label{Contour2d}
\end{figure}

Now consider the integral of $\Pi(\w)/\w$ around the contour shown in figure \ref{Contour2d}.
This gives
\begin{equation}
\int_{C_\infty} \frac{d\w}{\w}\,\Pi(\w) + \int_{C_0} \frac{d\w}{\w}\,\Pi(\w) +
 {\mathcal P} \int_{-\infty}^\infty \frac{d\w}{\w}\,\Pi(\w) = 0 \ ,
\label{d49}
\end{equation}
which implies
\begin{equation}
\Pi(\infty) - \Pi(0) = - \frac{1}{i\pi} {\mathcal P} \int_{-\infty}^\infty \frac{d\w}{\w}\,\Pi(\w) \ ,
\label{d50}
\end{equation}
where the principal part integral is
\begin{align}
{\mathcal P} \int_{-\infty}^\infty \frac{d\w}{\w}\,\Pi(\w) &= 
\int_{-\infty}^0 \frac{d\w}{\w}\,\Pi(\w+i\e)  +  \int_0^\infty \frac{d\w}{\w}\,\Pi(\w+i\e) \nonumber\\
&= \int_0^\infty \frac{d\w}{\w}\,\left[\Pi(\w+i\e) - \Pi(-\w + i\e)\right] \ .
\label{d51}
\end{align}
If we now use assumption (3), we have 
\begin{equation}
\left[ ~\ldots ~ \right] = \Pi(\w +i\e) - \Pi(\w -i\e) = {\rm disc}\,\Pi(\w) \ ,
\label{d52}
\end{equation}
with the discontinuity across the cut on the positive real axis. Then using assumption (4),
gives
\begin{equation}
\left[ ~\ldots ~ \right] = \Pi(\w +i\e) - \Pi(\w +i\e)^* = 2 i {\rm Im}\,\Pi(\w) \ ,
\label{d53}
\end{equation}
and we find the required dispersion relation,
\begin{equation}
\Pi(\infty) - \Pi(0)  = - \frac{2}{\pi} \int_0^\infty \frac{d\w}{\w}\,{\rm Im}\,\Pi(\w) < 0 \ .
\label{d54}
\end{equation}
where the final inequality is a consequence of assumption (5).

Finally, from the anomalous Ward identity (\ref{b38}) or (\ref{c25}), we have 
\begin{equation}
k^2 \Pi(\w) = k^2 \left(\b^I \b^J \Pi_{IJ}(\w) - \tilde{\b}_c \right) - 2\b^I \langle O_I\rangle
+ \b^I\partial_I \b^J \langle O_J\rangle \ .
\label{d55}
\end{equation}
Evaluating at the UV and IR fixed points at the ends of a RG trajectory (where $\b^I = 0$), 
we find our final result
\begin{equation}
\b_c^{UV} - \b_c^{IR} = \frac{2}{\pi} \int_0^\infty \frac{d\w}{\w}\,{\rm Im}\,\Pi(\w) >0 \ .
\label{d56}
\end{equation}
This is the weak $c$-theorem.

\section{Local RGE and Weyl Consistency Conditions in Four Dimensions}\label{sect 5}

We now turn to four dimensions and consider the generalisation of the anomalous Ward identities
and renormalisation group equations which underpinned the derivation of the $c$-theorem in two dimensions.
We begin with the renormalisation of the energy-momentum tensor and the Weyl consistency
conditions.

\subsection{Local RGE, trace anomaly and Weyl consistency conditions}\label{sect 5.1}

In four dimensions, exactly the same principles used in the renormalisation of the two-dimensional
theory in section \ref{sect 2.1} apply but now there are many more possible counterterms.
The action (\ref{b1}) generalises to\footnote{For simplicity, we only display dimension 4 operators $O_I$ here,
although a full treatment should also include the dimension 2 operators $O_a$ and global symmetry currents
$J_\m^A$ as discussed in sections \ref{sect 1.3} and \ref{sect 7}. For full details, see 
\cite{Jack:1990eb, Osborn:1991gm, Jack:2013sha, Baume:2014rla}.}
\begin{equation}
S = S_0 + \int d^n x \sqrt{-g}\, \left(g_B^I O_{I} + c_B{\BF} - a_B {\BG} - b_B{\BH} 
- \Lambda_B \right)
\label{e1}
\end{equation}
where ${\BF}$, ${\BG}$ and ${\BH}$ are the curvature terms, which in dimensional
regularisation are taken to be
\begin{align}
{\BF} &= R^{\m\n\r\s} R_{\m\n\r\s} - \frac{4}{n-2} R^{\m\n}R_{\m\n} + \frac{2}{(n-1)(n-2)}R^2 \nonumber\\
{\BG} &= \frac{2}{(n-2)(n-3)}\left(R^{\m\n\r\s} R_{\m\n\r\s} - 4 R^{\m\n} R_{\m\n} + R^2\right) \nonumber\\
{\BH} &= \frac{1}{(n-1)^2} R^2 \ .
\label{e2}
\end{align}
In four dimensions, ${\BF} = F_4 = C^{\m\n\r\s}C_{\m\n\r\s}$ is the square of the Weyl tensor
and ${\BG} = E_4$ is the Euler-Gauss-Bonnet density.

The counterterms depending on derivatives of the local couplings have dimension 4 and so may now also
contain curvature-dependent terms. A complete independent set is
\begin{multline}
\Lambda_B = \m^{-\e} (\Lambda_R + \tfrac{1}{3}e_I\partial_\m R\,\partial^\m g^I 
+\tfrac{1}{6} f_{IJ} R\, \partial_\m g^I \partial^\m g^J + \tfrac{1}{2}g_{IJ} G^{\m\n}\, \partial_\m g^I \partial_\n g^J \\
+ \tfrac{1}{2} a_{IJ} D^2 g^I D^2 g^J + \tfrac{1}{2} b_{IJK} \partial_\m g^I \partial^\m g^J D^2 g^K 
+ \tfrac{1}{2} c_{IJKL} \partial_\m g^I \partial^\m g^J \partial_\n g^K \partial^\n g^L)\ .
\label{e3}
\end{multline}
It follows that these contribute to contact term divergences in Green functions up to the four-point
$\langle O_I ~O_J ~O_K ~O_L\rangle$.
We define RG functions corresponding to these counterterms as described in section \ref{sect 2.1}.

The next step is to repeat the calculation of the trace anomaly $T^\m{}_\m$ following the method 
of section \ref{sect 2.2}. First, we need the variations of the curvature terms:\footnote{
More generally, we have the following metric variations of the curvature squared terms, which will be used
in section \ref{sect 9}:
\begin{align*}
\frac{1}{\sqrt{-g}}\frac{\d}{\d g^{\m\n}(x)} \,\left(R^{\k\l\r\s}R_{\k\l\r\s}\right)(y) &= 
2 R_{\m\l\r\s} R_\n{}^{\l\r\s}\d(x,y) + 4 R_{\m\r\n\s}D^\r D^\s \d(x,y) \nonumber \\
\frac{1}{\sqrt{-g}}\frac{\d}{\d g^{\m\n}(x)} \,\left(R^{\r\s}R_{\r\s}\right)(y) &= 
2 R_{\m\r} R_\n{}^\r \d(x,y) + \left(R_{\m\n} D^2 - 2 R_{\m\r}D^\r D_\n + 
g_{\m\n} R^{\r\s} D_\r D_\s \right)\d(x,y) \nonumber \\
\frac{1}{\sqrt{-g}}\frac{\d}{\d g^{\m\n}(x)} \,R^2(y) &= 2 R \left( R_{\m\n} + \D_{\m\n} \right)\d(x,y) \ .
\end{align*}
}
\begin{align}
\frac{1}{\sqrt{-g}}g^{\m\n} \frac{\d}{\d g^{\m\n}(x)} \,{\BF}(y) &= 2 {\BF} \d(x,y) \nonumber\\
\frac{1}{\sqrt{-g}}g^{\m\n} \frac{\d}{\d g^{\m\n}(x)} \,{\BG}(y) &= 2 {\BG} \d(x,y) 
- \frac{8}{(n-2)} G^{\m\n} D_\m D_\n\d(x,y) \nonumber \\
\frac{1}{\sqrt{-g}}g^{\m\n} \frac{\d}{\d g^{\m\n}(x)} \,{\BH}(y) &= 2 {\BH} \d(x,y) 
+ \frac{2}{(n-1)} R\, D^2 \d(x,y) \ ,
\label{e4}
\end{align}
and
\begin{align}
\frac{1}{\sqrt{-g}}g^{\m\n} \frac{\d}{\d g^{\m\n}(x)} \,G_{\r\s}(y) &= -\left(\frac{n}{2}-1\right)
\left(D^2 g_{\r\s} - D_\r D_\s\right)\d(x,y) \nonumber\\
\frac{1}{\sqrt{-g}}g^{\m\n} \frac{\d}{\d g^{\m\n}(x)} \,D^2 f(y) &= D^2 f \d(x,y) 
- \left(\frac{n}{2}-1\right) D^\m f D_\m \d(x,y) \ .
\label{e5}
\end{align}
for some scalar function $f(y)$.
Eventually we find the analogue of (\ref{b20}) for the trace anomaly:
\begin{equation}
T^\m{}_\m = \b^I O_I + \b_c {\BF} - \b_a {\BG} - \b_b {\BH} 
- \b_{\L} + D_\m Z^\m \ ,
\label{e6}
\end{equation}
with
\begin{equation}
\b_\L = \tfrac{1}{3} \chi^e_I \,\partial^\m R \,\partial_\m g^I 
+ \tfrac{1}{6} \chi^f_{IJ}\, R\, \partial_\m g^I \partial^\m g^J
+ \tfrac{1}{2} \chi^g_{IJ}\, G_{\m\n}\,\partial^\m g^I \partial^\n g^J 
+ \tfrac{1}{2} \chi^a_{IJ} \,D^2 g^I D^2 g^J + \ldots \ ,
\label{e7}
\end{equation}
and 
\begin{equation}
Z_\m = w_I\, G_{\m\n}\, \partial^\n g^I + \tfrac{1}{3} \partial_\m(d R) + \partial_\m\left(U_I D^2 g^I\right) 
+ \tfrac{1}{3} Y_I R \partial_\m g^I + S_{IJ} \partial_\m g^I D^2 g^J + \ldots
\label{e8}
\end{equation}
We have only quoted here the terms which play a role in our discussion of potential generalisations of the
$c$-theorem. For the full set, including a discussion of which of the RG functions may be set to zero using
the freedom to add finite local counterterms to the action, see \cite{Jack:1990eb, Osborn:1991gm}.

As before, we define operators $\D_\s^W$ and $\D_\s^\b$ corresponding to Weyl and RG variations.
Consistent with the trace anomaly (\ref{e6}), these satisfy 
\begin{equation}
\left(\D_\s^W - \D_\s^\b\right)W = \int d^4 x \sqrt{-g}~ \left[\s(x)\left(
\b_c {\BF} - \b_a {\BG} - \b_b {\BH} - \b_\L\right) -\partial_\m \s(x)  Z^\m \right] \ ,
\label{e9}
\end{equation}
generalising (\ref{c33}) to four dimensions. Differentiating wrt sources $g^{\m\n}(x)$ and $g^I(x)$ then gives 
anomalous Ward identities for Green functions involving $T^\m{}_\m$.

First, we derive the Weyl consistency conditions from the fact that the commutator of $\D_\s^W - \D_\s^\b$
vanishes as in (\ref{c39}) due to the abelian property of the Weyl transformations. This gives
\begin{align}
8 \partial_I \b_a - \chi^g_{IJ} \b^J &= - {\mathcal L}_\b w_I \nonumber \\
8 \b_b - \chi^a_{IJ} \b^I \b^J &= {\mathcal L}_\b \left(2d + \b^I U_I\right) 
\equiv {\mathcal L}_\b \left(2\tilde{d}\right) \nonumber \\
2 \chi^e_I + \chi^a_{IJ}\b^J &= - {\mathcal L}_\b U_I \nonumber \\
4 \partial_I\b_b + \left(\chi_{IJ}^f + \chi_{IJ}^a\right) \b^J &= {\cal L}_\b \left(\partial_I d + Y_I - U_I\right) \nonumber \\
\chi^g_{IJ} + 2 \chi^a_{IJ} + 2 \partial_I\b^K \chi^a_{KJ} + \b^K \chi^b_{KIJ} &= {\mathcal L}_\b S_{IJ} \nonumber \\
\ldots 
\label{e10}
\end{align}
where we have only quoted those relevant to the two-point Green functions discussed below.
Again, the complete set may be found in \cite{Jack:1990eb, Osborn:1991gm}.

Clearly the first condition involving the coefficient $\b_a$ of the Euler-Gauss-Bonnet density in the trace anomaly
is a precise analogue of the two-dimensional condition (\ref{c41}) which is implicitly related to 
the $c$-theorem. We will investigate in the next section whether a similar relation also applies
in four dimensions, where it would constitute an `$a$-theorem'.

We can also look at the possibility of theorems related to the RG flow of the coefficients $\b_c$ and
$\b_b$ of the Weyl and Ricci curvatures in the trace anomaly. At first sight, this does not look promising.
The second condition in (\ref{e10}) implies $\b_b = O(\b^I)$ so this function vanishes at a fixed point
so clearly cannot have a monotonic behaviour. Notice also that the Weyl consistency conditions do not
involve $\b_c$. Technically, this is because the Weyl variation of ${\BF}$ vanishes when evaluated
in flat space, {\it i.e.}~there are no derivative terms in the variation (\ref{e4}). This implies that studying
two-point Green functions involving the trace $T^\m{}_\m$ will not give information on the RG flow of $\b_a$.
However, as we shall see, we can still find information on the RG flow of $\b_c$ from two-point functions
involving $T_{\m\n}$ itself rather than its trace.

\subsection{Renormalisation group and anomalous Ward identities for Green functions}\label{sect 5.2}

We now consider the two-point Green functions involving the energy-momentum tensor and discuss the
corresponding Ward identities for anomalous Weyl symmetry and the renormalisation group equations.
The analysis follows that of sections \ref{sect 3.2} and \ref{sect 3.1}, generalising to four dimensions. 
This will allow us to recover the Weyl consistency conditions (\ref{e10}) directly from the Green functions, 
which is important in relating them to potential $a$, $b$ and $c$-theorems in four dimensions.
 
Throughout this section, unlike section \ref{sect 3}, we restrict to flat spacetime and all the expressions
quoted for the Green functions and RG functions are to be understood as evaluated in flat spacetime and
with physical couplings set to constants.

First, diffeomorphism invariance, through the conservation equation $\partial^\m T_{\m\n}= 0$, constrains the 
form of the two-point function:
\begin{equation}
i\langle T_{\m\n}(x) ~ T_{\r\s}(0)\rangle_{\rm c} = \Pi_{\m\n\r\s}^{(0)}\,\Omega^{(0)}(x) ~+~ 
\Pi_{\m\n\r\s}^{(2)}\,\Omega^{(2)}(x) \ ,
\label{e11}
\end{equation}
where $\Pi_{\m\n\r\s}^{(0)}$ and $\Pi_{\m\n\r\s}^{(2)}$ are projection operators onto the spin 0 and spin 2
intermediate states \cite{Cappelli:1990yc,Shore:1990wq,Shore:1990ng}. In $n$ dimensions,
\begin{align}
\Pi_{\m\n\r\s}^{(0)} &= \frac{1}{n-1} \D_{\m\n} \D_{\r\s} \nonumber \\
\Pi_{\m\n\r\s}^{(2)} &= \frac{1}{2(n-1)} \left[-2 \D_{\m\n} \D_{\r\s} + (n-1) \left(\D_{\m\r} \D_{\n\s} +
\D_{\m\s} \D_{\n\r}\right)\right] \ .
\label{e12}
\end{align}
Note that in four dimensions, $\Pi_{\m\n\r\s}^{(2)}$ is transverse and traceless, while it vanishes identically
in two dimensions.

Contracting indices, in four dimensions we have
\begin{equation}
i\langle T^\m{}_\m(x) ~ T^\r{}_\r(0)\rangle_{\rm c} = 3\, \partial^4\, \Omega^{(0)}(x) \ ,
\label{e13}
\end{equation}
which involves only the spin 0 factor. To identify the spin 2 contribution, we need to take a different
contraction \cite{Shore:1990ng}, which will therefore not be determined by the anomalous Ward identities.
In fact, we can show
\begin{equation}
i\langle T^{\m\n}(x) ~ T_{\m\n}(0) \rangle_{\rm c} - \tfrac{1}{3}
i\langle T^\m{}_\m(x) ~ T^\r{}_\r(0)\rangle_{\rm c} 
= 5\, \partial^4\, \Omega^{(2)}(x) \ .
\label{e14}
\end{equation}

The anomalous Ward identities may be derived exactly as in section \ref{sect 3}
by differentiating the Weyl variation equation (\ref{e9}) for $\left(\D_\s^W - \D_\s^\b\right)W$. 
In this case, see (\ref{e6}), (\ref{e7}) and (\ref{e8}), the anomaly is
\begin{equation}
{\cal A} = \b_c {\BF} - \b_a {\BG} - \b_b {\BH}  - \b_{\L} + D_\m Z^\m \ ,
\label{e200}
\end{equation}
and the derived RG functions ${\cal A}_{\m\n}(x,y)$, ${\cal B}_I(x,y)$, \ldots ${\cal H}(x,y)$
are all defined exactly as in section \ref{sect 3}.
We find, with the familiar notation $\Theta = \b^I O_I$ and recalling the identities in
section \ref{sect 5.1} and (\ref{b12}) for the variation of the curvature terms,
\begin{equation}
i\langle T_{\m\n}(x)~T^\r{}_\r(y)\rangle  ~-~ i\langle T_{\m\n}(x)~\Theta(y)\rangle   ~+~
2\langle T_{\m\n}(x)\rangle \,\d(x,y) ~=~ {\cal A}_{\m\n}(x,y) \ ,
\label{e20}
\end{equation}
where now\footnote{Notice that there is no ambiguity in this section in taking all the derivatives
with respect to $x$, since in the flat spacetime limit they are always acting on functions
of $(x-y)$. Later, when we consider curved backgrounds, we need to be careful to act with
covariant derivatives appropriate to the points $x$ or $y$.}
\begin{equation}
{\cal A}_{\m\n} =  \tfrac{2}{3} d\,\D_{\m\n}\partial^2 \d(x,y)  \ .
\label{e201}
\end{equation}
Similarly,
\begin{equation}
i\langle O_I(x) ~T^\r{}_\r(y) \rangle ~-~ i\langle O_I(x)~\Theta(y) \rangle 
-\partial_I\b^J \langle O_J(x)\rangle\,\d(x,y) ~=~  {\cal B}_I(x,y) \ ,
\label{e202}
\end{equation}
with 
\begin{equation}
{\cal B}_I(x,y) =  U_I\,\partial^4 \d(x,y) \ .
\label{e203}
\end{equation}
Notice that these delta-function contributions to the anomalous Ward identities all arise here
from the total divergence $D_\m Z^\m$ term in the trace anomaly ${\cal A}$. In contrast, as evident
from (\ref{e15}) below, such terms do not appear in the RGEs themselves.

As before, we can contract and combine these to find the identity for the two-point function
of the energy -momentum tensor trace,
\begin{multline}
i\langle T^\m{}_\m(x)~T^\r{}_\r(y)\rangle  ~-~ \b^I \b^J i\langle O_I(x)~O_J(y)\rangle ~+~
2\langle T^\m{}_\m(x)\rangle \,\d(x,y)  ~-~
\beta^I\partial_I \beta^J \,\langle O_J(x)\rangle \,\d(x,y)\\
 ~=~ {\cal H}(x,y)\ , 
\label{e21}
\end{multline}
with 
\begin{equation}
{\cal H}(x,y) = 2 \tilde{d}\,\partial^4\d(x,y) \equiv 2 \left(d + \tfrac{1}{2}\b^I U_I\right) \partial^4 \d(x,y) \ .
\label{e211}
\end{equation}
The corresponding relation involving the conserved 
two-point function $i \langle T^\m{}_\m(x)~T^\r{}_\r(y)\rangle_{\rm c}$ follows immediately from the 
footnote following its definition in (\ref{cc4}).

Now consider the renormalisation group equations. In four dimensions, the fundamental RGE generalising
(\ref{c3}) is
\begin{equation}
{\mathcal D} W = - \int d^4 x\sqrt{-g}\,\left(\b_c {\BF} - \b_a {\BG} - \b_b {\BH} 
- \b_{\L} \right) \ .
\label{e15}
\end{equation}
The RGEs for the VEVs are therefore,
\begin{equation}
{\cal D} \langle T_{\m\n}\rangle = {\cal C}_{\m\n}  = 0\ ,
\label{e151}
\end{equation}
and 
\begin{equation}
{\cal D}\langle O_I\rangle + \gamma_I{}^J \langle O_j\rangle = {\cal C}_I = 0 \ ,
\label{e152}
\end{equation}
evaluating in flat spacetime with constant couplings.

To derive the RGE for $\langle T_{\m\n} ~ T_{\r\s}\rangle$, we need the following crucial identities for the 
metric variations of the curvature terms \cite{Shore:1990wq,Shore:1990ng}. Evaluating in the flat-spacetime limit, 
in four dimensions, we find
\begin{align}
\frac{\d}{\d g^{\m\n}(x)} \,\frac{\d}{\d g^{\r\s}(y)} \,\int d^4 x\sqrt{-g}\,{\BF} &=
\Pi_{\m\n\r\s}^{(2)} \d(x,y) \nonumber\\
\frac{\d}{\d g^{\m\n}(x)} \,\frac{\d}{\d g^{\r\s}(y)} \,\int d^4 x\sqrt{-g}\,{\BG} &= 0 \nonumber\\
\frac{\d}{\d g^{\m\n}(x)} \,\frac{\d}{\d g^{\r\s}(y)} \,\int d^4 x\sqrt{-g}\,{\BH} &=
\tfrac{2}{3} \Pi_{\m\n\r\s}^{(0)} \d(x,y) \ .
\label{e16}
\end{align}
It follows immediately that 
\begin{equation}
{\mathcal D} \,i\langle T_{\m\n}(x) ~ T_{\r\s}(y)\rangle = {\cal E}_{\m\n,\r\s}(x,y) \ ,
\label{e17}
\end{equation}
with
\begin{equation}
{\cal E}_{\m\n,\r\s}(x,y) = \tfrac{8}{3} \b_b \,\Pi_{\m\n\r\s}^{(0)} \d(x,y) ~-~ 4\b_c\, \Pi_{\m\n\r\s}^{(2)} \d(x,y) \ ,
\label{e171}
\end{equation}
and so,
\begin{equation}
{\mathcal D}\,\Omega^{(0)} = - \tfrac{8}{3} \b_b \d(x) \ , \qquad \qquad  
{\mathcal D}\,\Omega^{(2)} = - 4 \b_c \d(x) \ .
\label{e18}
\end{equation}

For the remaining Green functions of interest, we find 
\begin{equation}
{\mathcal D}~ i \langle T_{\m\n}(x) ~ O_I(y)\rangle + \c_I{}^J i \langle T_{\m\n}(x) ~ O_J(y)\rangle
= {\cal F}_{\m\n,I}(x,y) \ , 
\label{e19}
\end{equation}
with 
\begin{equation}
{\cal F}_{\m\n,I}(x,y) = -\tfrac{2}{3} \chi^e_I \,\D_{\m\n}\partial^2 \d(x,y) \ ,
\label{e191}
\end{equation}
and
\begin{multline}
{\mathcal D}~ i\langle O_I(x) ~ O_J(y)\rangle +\c_I{}^K i\langle O_K(x) ~ O_J(y)\rangle 
+ \c_J{}^K i\langle O_I(x) ~ O_K(y)\rangle + \partial_I \partial_J \b^K \langle O_K\rangle \d(x,y) \\
= {\cal G}_{IJ}(x,y) \ ,
\label{e192}
\end{multline}
with
\begin{equation}
{\cal G}_{IJ}(x,y) =  \chi^a_{IJ}\, \partial^4 \d(x,y) \ .
\label{e193}
\end{equation}

\subsection{Weyl consistency conditions}\label{sect 5.3}

With these results in hand, we can now establish the consistency conditions implied by the RGEs
for the anomalous Ward identities (\ref{e20}) and (\ref{e202}). These follow exactly as in section \ref{sect 3.3}
where we have the two crucial relations,
\begin{equation}
{\cal L}_\b{\cal A}_{\m\n}(x,y) = {\cal E}_{\m\n,\r\s}(x,y)g^{\r\s}(y) - {\cal F}_{\m\n,I}(x,y)\b^I 
+ 2 {\cal C}_{\m\n}(x)\d(x,y) \ ,
\label{e30}
\end{equation}
and
\begin{equation}
{\cal L}_\b {\cal B}_I(x,y)  = g^{\m\n}(y) {\cal F}_{\m\n,I}(y,x) - 
{\cal G}_{IJ}(x,y)\b^J -\partial_I\b^J {\cal C}_J(x)\d(x,y) \ .
\label{e31}
\end{equation}

Substituting the explicit forms above for the RG functions, we find from (\ref{e30}),
equating coefficients of $\partial^2 \D_{\m\n}\d(x,y)$,
\begin{equation}
 {\cal L}_\b d = 4\b_b + \chi_I^e \b^I \ .
\label{e32}
\end{equation}
Similarly, from (\ref{e31}), equating coefficients of $\partial^4 \d(x,y)$, we find
\begin{equation}
{\cal L}_\b U_I = - \chi_{IJ}^a \b^J - 2 \chi_I^e \ .
\label{e33}
\end{equation}
Combining these and eliminating $\chi_I^e$ leaves
\begin{equation}
8 \b_b = \chi_{IJ}^a \b^I \b^J + {\cal L}_\b \left(2d + U_I \b^I\right) \ ,
\label{e34}
\end{equation}
and together (\ref{e33}) and (\ref{e34}) reproduce two of the Weyl consistency conditions
listed above in (\ref{e10}).

As a final consistency check, we can verify the contracted identity (\ref{c346}), {\it viz.}
\begin{multline}
{\cal L}_\b{\cal H}(x,y) = g^{\m\n}(x) {\cal E}_{\m\n,\r\s}(x,y) g^{\r\s}(y) - {\cal G}_{IJ}(x,y) \b^I \b^J
\\+ 2 g^{\m\n}(x) {\cal C}_{\m\n}(x) \d(x,y) - \b^I\partial_I\b^J {\cal C}_J(x) \d(x,y) \ .
\label{e35}
\end{multline}
With (\ref{e211}) for ${\cal H}(x,y)$, we find directly the relation
\begin{equation}
{\cal L}_\b \tilde{d} \equiv \b^i \partial_I \tilde{d}= 8 \b_b - \chi_{IJ}^a \b^I \b^J \ .
\label{e36}
\end{equation}

We have therefore reproduced some of the Weyl consistency conditions
directly from properties of the two-point Green functions of the energy-momentum
tensor in flat spacetime, opening the way for their use in searching for possible
four-dimensional generalisations of the $c$-theorem. 
As noted above, however, these relations involve only the coefficient of the $R^2$ term in the 
four-dimensional trace anomaly. The Weyl consistency condition for the flow of the
Euler coefficient $\b_a$, which would be of most interest in the search for a four-dimensional
$a$-theorem, is not accessible to the RGEs for two-point Green functions
in flat spacetime. We return to this issue in section \ref{sect 6.3} below.

\section{c, b and a-Theorems in Four Dimensions}\label{sect 6}

We now generalise the various methods that led to the two-dimensional $c$-theorem to four dimensions.
Here, because of the three curvature terms in the trace anomaly, we may look for identities for the 
renormalisation group flow of quantities related to the coefficients $\b_a$, $\b_c$ and $\b_b$ of the
curvature terms ${\BG}$, ${\BF}$ and ${\BH}$ respectively, {\it i.e.}~the Euler
density and squares of the Weyl tensor and Ricci scalar.

However, as we have seen above, there are important differences between two and four dimensions.
It is clear from the decomposition of the Green function $\langle T_{\m\n}(x) ~ T_{\r\s}(0)\rangle$
into spin 0 and spin 2 parts that two-point functions in flat spacetime will only give information 
on $\b_b$ and $\b_c$. 

On the other hand, the Weyl consistency conditions give no constraint on the Weyl coefficient $\b_c$,
while the Ricci coefficient $\b_b$ is $O(\b^I)$ and so vanishes at a fixed point. 
There is, however, a potential $a$-theorem condition \cite{Osborn:1989td, Jack:1990eb},
\begin{equation}
\b^I\partial_I\tilde{\b}_a = \tfrac{1}{8} \chi^g_{IJ} \b^I \b^J \ ,
\label{f1}
\end{equation}
with $\tilde{\b}_a = \b_a + \tfrac{1}{8} \b^I w_I$, but in this case there is no direct positivity constraint
from the two-point functions.

\subsection{Spectral functions and the $c$ and $b$ `theorems'}\label{sect 6.1}

The spin 0 and spin 2 projections of the energy-momentum tensor Green function are defined 
in (\ref{e11}). As in section \ref{sect 4.1}, we can write a spectral representation,
\begin{equation}
\Omega^{(s)}(x) = \frac{i}{(2\pi)^2} \int_0^\infty d\l^2\,\l^2\,\r^{(s)}(\l^2)\,(\l |x|)^{-1} K_1(\l|x|) \ ,
\label{f2}
\end{equation}
for $s = 0, 2$ using the four-dimensional Feynman propagator (for spacelike $x$). The spectral function
$\r(\l^2)$ has dimension 0.

The Green functions $x^8 \langle T_{\m\n}(x) ~ T_{\r\s}(0)\rangle$ and the contractions which isolate the 
functions $\Omega^{(0)}$ and $\Omega^{(2)}$ can be used as before to construct analogues of the Zamolodchikov
$F$, $G$ and $H$ functions. This construction \cite{Shore:1990wq,Shore:1990ng}, which is a straightforward 
generalisation of that in section \ref{sect 4.1},  is described in detail in appendix \ref{sect C}.

We then look for a combination $C = F + \a H + \b G$ which satisfies 
$|x|\frac{\partial C}{\partial |x|} \sim G$ for some constants $\a$ and $\b$. However, it turns out that 
this is only possible in two dimensions. In fact, the best we can do in general is to define
(in $n$ dimensions)
\begin{equation}
C = F + \left(\frac{n}{2}+1\right) H - \left(\frac{n}{2}+ 2\right) G \ ,
\label{f3}
\end{equation}
for which
\begin{equation}
|x|\frac{\partial C}{\partial |x|} = (n-2) C - \tfrac{1}{2} (n+2)(n+4) G \ .
\label{f4}
\end{equation}
This clearly shows the preferred r\^ole of $n=2$.

In four dimensions, therefore, we define the two $C^{(s)}$ functions for $s=0$ and $s=2$
(related to the $\b_b$ and $\b_c$ RG functions respectively) as
\begin{equation}
C^{(s)} = \frac{1}{(2\pi)^2} \int_0^\infty \frac{d\l^2}{\l^2}\,\r^{(s)}(\l^2)\,
\left[ (\l |x|)^7 \left(K_5(\l |x|) + 3 K_3(\l |x|) - 4 K_1(\l |x|) \right) \right] \ ,
\label{f5}
\end{equation}
together with 
\begin{equation}
G^{(s)} = \frac{1}{(2\pi)^2} \int_0^\infty \frac{d\l^2}{\l^2}\,\r^{(s)}(\l^2)\,\left[(\l |x|)^7 K_1(\l |x|)\right] \ .
\label{f6}
\end{equation}
Similarly, defining the corresponding spectral function $\r_{IJ}(\l^2)$ from the Green function 
$x^8 \langle O_I(x) ~ O_J(0)\rangle$, we define
\begin{equation}
G_{IJ} = \frac{1}{(2\pi)^2} \int_0^\infty \frac{d\l^2}{\l^2}\,\r_{IJ}(\l^2)\,\left[(\l x)^7 K_1(\l x)\right] \ .
\label{f7}
\end{equation}
but note that it is only for the spin 0 function that we have the relation
\begin{equation}
G^{(0)} = \tfrac{1}{9} G_{IJ} \b^I \b^J \ .
\label{f8}
\end{equation}

We then have the $b$ and $c$ `theorems' \cite{Shore:1990wq,Shore:1990ng} 
for the spin 0 and spin 2 projections,
\begin{equation}
\b^I \partial_I C^{(s)} = -2 C^{(s)} + 24 G^{(s)} \ ,
\label{f9}
\end{equation}
where $G^{(s)}$, but not necessarily the combination $12 G^{(s)} - C^{(s)}$, is positive definite.
Although (\ref{f9}) determines the RG flow of the functions $C^{(0)}$ and $C^{(2)}$, we cannot
therefore infer that this flow is monotonic.

\begin{figure}[h]
\centering
\includegraphics[scale=0.8]{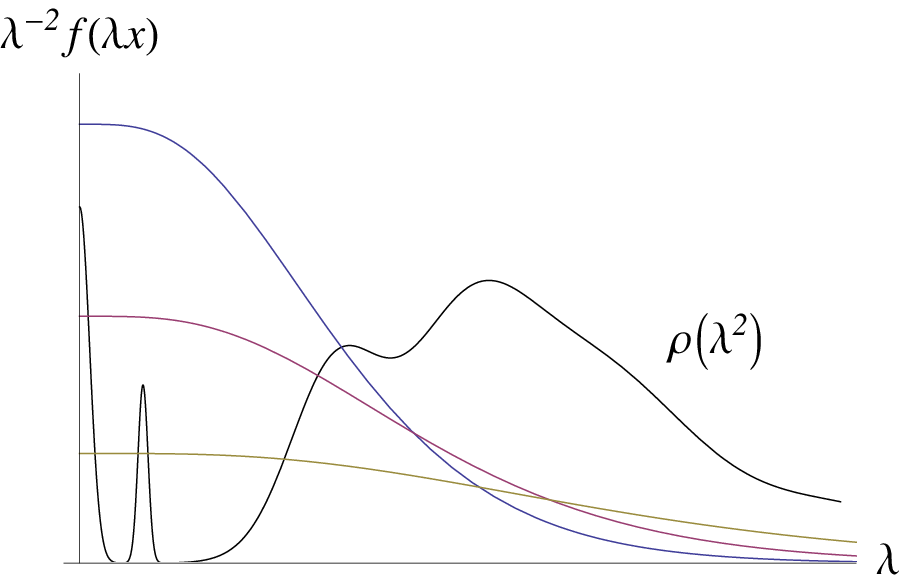}
\includegraphics[scale=0.8]{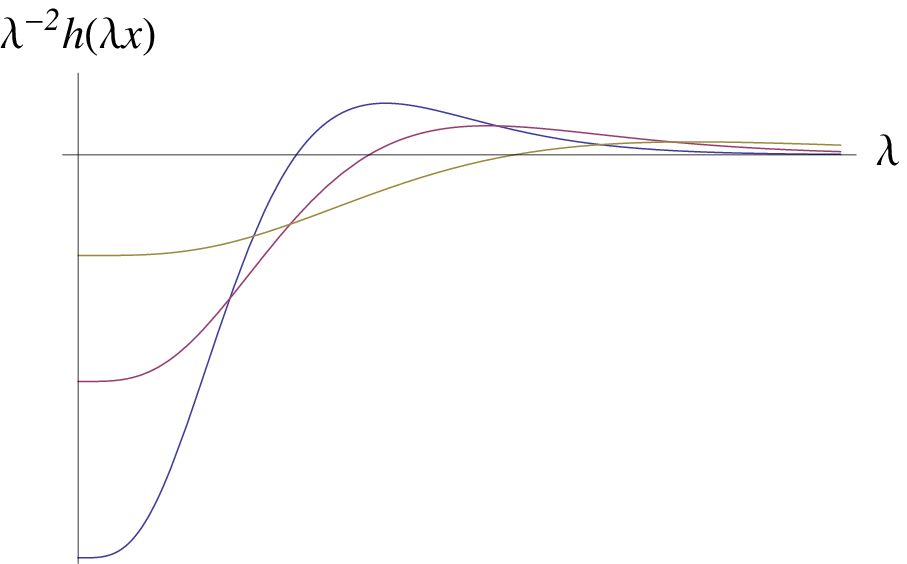}
\caption{The LH figure shows the sampling function $f(\l|x|)$ defining
the four-dimensional Zamolodchikov-like $C^{(s)}$ functions
for three values of the scale $|x|$, superimposed on a sketch of the spectral function $\r(\l^2)$. 
As $|x|$ is increased, the sampling function moves towards smaller
$\l$ values, though monotonicity of $C^{(s)}$ is no longer evident.
The RH plot shows the corresponding sampling function $h(\l|x|)$ defining the functions $G^{(s)}$.}
\label{Sampling4d}
\end{figure}

To understand the physical content of (\ref{f9}), rewrite the $C^{(s)}$ functions in terms of 
the integral of the corresponding spin 0 or spin 2 spectral functions with a sampling
function $f(\l |x|)$ as in (\ref{d19}) and (\ref{d21}), {\it i.e.}
\begin{equation}
C^{(s)} = \frac{1}{(2\pi)^2} \int_0^\infty \frac{d\l^2}{\l^2}\,\r^{(s)}(\l^2)\, f(\l |x|) \ ,
\label{f10}
\end{equation}
with
\begin{equation}
f(\l |x|) = (\l |x|)^7 \left[K_5(\l |x|) + 3 K_3(\ |x|) - 4 K_1(\l |x|) \right] \ .
\label{f11}
\end{equation}
Notice that because of the dimensional dependence, the power of the prefactor $(\l |x|)^7$ is greater 
in this case than the highest order Bessel function $K_5(\l |x|)$ (in two dimensions they were the same). 
As a result, in four dimensions $f(\l |x|)$ itself has a Gaussian shape which moves to smaller $\l$ as $|x|$ 
is increased (similar to the function $h(\l|x|)$ shown in figure \ref{Sampling2d}), 
while the full integrand $\l^{-2}f(\l|x|)$ is plotted in figure \ref{Sampling4d}.  
This shows a clear difference from the corresponding function in the two-dimensional Zamolodchikov
theorem, and in particular monotonicity as $|x|$ increases is no longer evident.\footnote{The integrals
of the sampling functions $f(\l|x|)$ and $h(\l|x|)$ are given here by
\begin{equation*}
\int_0^\infty \frac{d\l}{\l^2}\,f(\l|x|) = 2^{10}\,3^2 \ , \qquad \qquad  
\int_0^\infty \frac{d\l}{\l^2}\,h(\l|x|) = 0 \ .
\end{equation*}
It is a curious feature, whose physical significance is not immediately apparent, that for the particular
combination of $C^{(s)}$ and $G^{(s)}$ appearing on the rhs of the flow equation (\ref{f9}), the integral
over all $\l^2$ of $h(\l|x|)$ vanishes, showing that it gives equal and opposite weight to the UV and IR
regions as defined by the scale $|x|$.}

We can also define the sampling function $h(\l |x|)$ characterising
the rhs of (\ref{f9}). Here, 
\begin{equation}
12 G^{(s)} - C^{(s)} = \frac{1}{(2\pi)^2} \int_0^\infty \frac{d\l^2}{\l^2}\,\r^{(s)}(\l^2)\, h(\l |x|) \ ,
\label{f12}
\end{equation}
where $h(\l |x|)$ is plotted in figure \ref{Sampling4d}. Evidently, this lacks the positivity that would allow
an interpretation (at least for the spin 0 function) as a metric on coupling space.
Nevertheless, although we do not have monotonicity, the relation (\ref{f9}) for the RG flow of the 
Zamolodchikov-like functions $C^{(s)}$ does in principle provide a detailed probe of the spectral 
functions $\r^{(s)}(\l^2)$.  The next step is to show how they are related to the $\b_c$ and $\b_b$ 
anomaly coefficients at fixed points.

Before moving on, we should comment briefly that the construction described here may be straightforwardly 
applied to other operators, particularly (anomalous) symmetry currents $J_\m$, which do not necessarily
appear in the original Lagrangian. The RG formalism with local sources goes through exactly as 
we have described here for the local couplings $g^I$ sourcing the operators $O_I$ (see \cite{Shore:1990wp}
for a formal discussion, including applications to the operator product expansion).
Moreover, we can follow the method described above to define $C$ and $G$ functions from the
corresponding spectral functions characterising the two-point functions $\langle J_\m ~ J_\n\rangle$,
and derive an equation analogous to (\ref{f9}) (including in general an anomalous dimension term 
if $J_\m$ is not conserved)  for the RG flow. These relations would again in principle provide interesting 
spectral information on the theory. Of course, such results lack the universality of the energy-momentum tensor,
whose correlation functions probe the entire spectrum of the QFT and are therefore singularly important
in constraining RG flow.

\subsection{Renormalisation group flow for the $c$ and $b$ `theorems'}\label{sect 6.2}

The renormalisation group analysis of the $C^{(s)}$ functions follows the same pattern described
for the two-dimensional case in section \ref{sect 4.2}, starting from the RGEs
\begin{equation}
{\mathcal D} \Omega^{(0)} = \tfrac{8}{3} \b_b \d(x) \ , \qquad \qquad  
{\mathcal D}\Omega^{(2)} = -4 \b_c \d(x) \ .
\label{f13}
\end{equation}
Full details of the analysis may be found in \cite{Shore:1990wq,Shore:1990ng} 
(see also appendix \ref{sect C}).  
We find, for the spin 2 function, and with the same notation as section \ref{sect 4.2},
\begin{equation}
C^{(2)} =  \frac{2^9\, 3^2}{5}\, \frac{1}{(2\pi)^2}\, \left[ -20\b_c \,-\, \b^I \partial_I F^{(2)}_{TT} \,+\, 
O(\b^I, t) \right]\ ,
\label{f14}
\end{equation}
so this function does indeed reduce to $\b_c$ at a fixed point. Recall that for free theories,
$\b_c = \tfrac{1}{(4\pi^2)}\tfrac{1}{120} \left( n_s + 6n_f + 12n_v\right)$. 

However, although we have the relation (\ref{f9}) controlling the RG flow of $C^{(2)}$ , 
and the result that $C^{(2)}$ is proportional to $\b_c$ at a fixed point, we cannot infer whether $\b_c^{UV}$
is bigger or smaller than $\b_c^{IR}$. In fact, examples with both $\b_c^{UV} > \b_c^{IR}$ and
$\b_c^{UV} < \b_c^{IR}$ are known.\footnote{For example, SUSY QCD with $N_c = n_f$ breaks the chiral
symmetry $SU(n_f)_L \times SU(n_f)_R \times U(1)_V \times U(1)_X$ to
$SU(n_f - m) \times SU(m)_V \times U(1)_X$ with (for any $m$) $n_f^2$ chiral multiplets.
Since the quarks and squarks form $N_c n_f$ chiral multiplets, it is clear that $\b_c^{UV} > \b_c^{IR}$.

SUSY QCD with $\tfrac{3}{2}N_c \le n_f \le 3N_c$ is in the conformal window and, exploiting the fact that 
the conformal anomaly lies in a supermultiplet with the $R$ symmetry current, supersymmetry allows the
calculation of exact results for $\b_c$ and also $\b_a$ in the UV and IR. These show that while 
$\b_a^{UV} \ge \b_a^{IR}$ always, $\D \b_c$ can change sign, in accordance with our `theorem'. }

The results (\ref{f9}) and (\ref{f14}) may also be obtained using the position space representation 
described in section \ref{sect 4.3}. For details of this derivation of the $\b_c$-theorem, see 
the discussion around eq.~(3.22) of \cite {Osborn:1991gm}.

For the $C^{0)}$ function, we can show
\begin{align}
C^{(0)} &= \frac{2^9}{(2\pi)^2} \left(8 \b_b - \b^I\partial_I F^{(0)}_{TT} \,
+\, O(\b^I, t)\,\, \right) \nonumber \\
G^{(0)}_{IJ} &= \frac{2^7\,3}{(2\pi)^2} \left(\chi^a_{IJ} - {\mathcal L}_\b F_{IJ} \,
+\, O(\b^I, t)\,\, \right) \ .
\label{f15}
\end{align}
So at this order, before we reach the $O(t)$ term in the expansion of $C^{0)}$, the RG flow equation
\begin{equation}
-x \frac{\partial C^{(0)}}{\partial x}  ~=~ \b^I\partial_I C^{(0)} = \frac{8}{3} G_{IJ} \b^I \b^J - 2 C^{(0)} \ ,
\label{f16}
\end{equation}
is satisfied by the vanishing of the rhs, that is
\begin{equation}
\b^I\partial_I F^{(0)}_{TT} - \b^I \b^J {\mathcal L}_\b F_{IJ} = -\chi^a_{IJ} \b^I \b^J + 8 \b_b \ ,
\label{f17}
\end{equation}
or equivalently \cite{Shore:1990wq}
\begin{equation}
\b^I\partial_I\left(F^{(0)}_{TT} - F_{IJ} \b^I \b^J\right) = -\chi^a_{IJ} \b^I \b^J + 8 \b_b \ .
\label{f18}
\end{equation}

Compare this with the Weyl consistency condition (\ref{e36})
\begin{equation}
\b^I\partial_I(2\tilde{d}) \equiv {\mathcal L}_\b (2\tilde{d}) = -\chi^a_{IJ} \b^I\b^J + 8\b_b \ .
\label{f19}
\end{equation}
The compatibility of (\ref{f18}) and (\ref{f19}) is assured by the anomalous Ward identity (\ref{e21})
relating the two-point functions $\langle T^\m{}_\m ~ T^\r{}_\r\rangle$ and $\b^I \b^J \langle O_I ~O_J\rangle$.

The `$\b_b$-theorem' is therefore linked to a Weyl consistency condition and the two-point Green functions.
However, once again the $C^{(0)}$ function is not monotonic but exhibits the RG flow determined by (\ref{f9}).
Moreover, since the Weyl consistency conditions imply $\b_b = O(\b^I)$, the function $C^{(0)}$ vanishes at a 
fixed point. Its usefulness therefore appears to be limited although, as for the $\b_c$-theorem, its RG flow
gives dynamical information on the structure of the spin 0 spectrum.

\subsection{Weyl consistency conditions and the $a$-theorem}\label{sect 6.3}

The remaining potential generalisation of the Zamolodchikov theorem to four dimensions involves the
coefficient $\b_a$ of the Euler-Gauss-Bonnet density in the trace anomaly. As we have seen, unlike $\b_c$ and $\b_b$,
this cannot be accessed from the two-point Green functions of the energy-momentum tensor in flat spacetime. 
This means that the most obvious source of positivity, and therefore monotonicity of the RG flow,
is lost. An extensive attempt to use two-point functions in curved spacetime where the Euler characteristic
does not vanish, especially on spaces of constant curvature, was made in \cite{Osborn:1999az}
and is summarised in the following section.

However, $\b_a$ does enter the Weyl consistency conditions, where it satisfies a constraint which is
essentially identical to that for the coefficient of the Euler density in two dimensions. This is, see (\ref{e10}),
\begin{equation}
8 \partial_I \b_a - \chi^g_{IJ} \b^J = - {\mathcal L}_\b w_I \ ,
\label{f20}
\end{equation}
and so, defining 
\begin{equation}
\tilde{\b}_a = \b_a + \tfrac{1}{8}\b^I w_I \ ,
\label{f21}
\end{equation}
we have the key RG flow equation \cite{Osborn:1989td,Jack:1990eb},
\begin{equation}
\b^I\partial_I \tilde{\b_a} = \tfrac{1}{8} \chi^g_{IJ} \b^I \b^J \ .
\label{f22}
\end{equation}

The analogy with the two-dimensional $c$-theorem is clear.
In order to establish (\ref{f22}) as a genuine, monotonic $a$-theorem, however, we need to show 
that the RG function $\chi^g_{IJ}$ is positive definite, as required for it to represent a metric on coupling space.

Now, although $\chi^g$ itself is not related to a two-point function, a further Weyl consistency condition
from (\ref{e10}) relates it to $\chi^a_{IJ}$, which is related to the Green function $\langle O_I ~ O_J\rangle$
and is positive. This identity reads,
\begin{equation}
\chi^g_{IJ} + 2 \chi^a_{IJ} + 2 \partial_I\b^K \chi^a_{KJ} + \b^K \chi^b_{KIJ} = {\mathcal L}_\b S_{IJ} \ .
\label{f23}
\end{equation}
Unfortunately, this identity involves also $\chi^b_{IJK}$, which is related to the three-point functions 
$\langle O_I ~ O_J ~O_K\rangle$ for which there is no obvious positivity constraint. 
However, as emphasised recently in \cite{Baume:2014rla}, these are of $O(\b^I)$ so if they are sufficiently
small, as may be expected in the vicinity of a fixed point, $\chi^g_{IJ}$ coud still inherit the positivity
property of $\chi^a_{IJ}$ in this limited region. 
A general proof of positivity for $\chi^g_{IJ}$ nevertheless appears to be out of reach, although it has been 
calculated to high order in numerous examples and no counter-example has been found.\footnote{The 
importance of the definition
$\tilde{\b}_a$ including the $w_I$ contribution term is evident in pure Yang-Mills theory. In perturbation theory,
we have \cite{Jack:1990eb,Fortin:2012hn},
\begin{equation*}
\b_a(g) = \b_a(0) + \tfrac{1}{8}\frac{\b_1}{(4\pi)^6} n_v g^4 + O(g^6) \ ,
\end{equation*}
where $\b_0$, $\b_1$ are the usual beta functions coefficients, 
and $\b_a(g)$ increases as we flow away from the UV fixed point. However, $\tilde{\b}(g)$ decreases, since 
including the $w_I$ correction,
\begin{equation*}
\tilde{\b}_a(g) = \b_a(0) - \tfrac{1}{4} \frac{\b_0}{(4\pi)^6} n_v g^2 + O(g^4) \ .
\end{equation*} }
It therefore remains a valid conjecture
that $\chi^g_{IJ}$ is positive and (\ref{f22}) is indeed a monotonic $a$-theorem.

\section{Local RGE and Maximally Symmetric Spaces} \label{sect 9}

As we have seen, the two-point Green functions of the energy-momentum tensor in flat spacetime
does not yield information about the coefficient $\b_a$ of the Euler-Gauss-Bonnet density in the trace anomaly
and so does not allow the derivation of a conjectured $a$-theorem in four dimensions.
Since two-point Green functions are the most obvious source of the positivity conditions 
which are necessary to prove monotonicity of the corresponding RG flow, this line of investigation
has clearly reached an impasse. A natural way forward, which was pursued in great detail 
in \cite{Osborn:1999az}, is to generalise the background to a curved spacetime where the 
Euler characteristic is non-vanishing.

In this section, we consider the Green functions of the energy-momentum tensor
on four-dimensional, maximally symmetric spaces of constant curvature. Here, the Riemann tensor takes the
form $R_{\m\n\r\s} = \tfrac{1}{12} R \left(g_{\m\r} g_{\n\s} - g_{\m\s} g_{\n\r}\right)$, where the Ricci scalar 
$R$ is constant. The Ricci tensor is $R_{\m\n} = \tfrac{1}{4} R g_{\m\n}$. 
These spaces are conformally flat, so ${\BF} = 0$, while the Euler-Gauss-Bonnet curvature ${\BG} = \tfrac{1}{6} R^2$
is non-vanishing. Studying Green functions in this background should therefore yield information 
on potential $C$-functions related to the anomaly coefficient $\b_a$.

\subsection{Renormalisation group, Weyl consistency conditions and the $a$-theorem}\label{sect 9.1}

We begin by revisiting the analysis of the anomalous Ward identities and renormalisation group equations
for two-point Green functions presented for flat spacetime in sections \ref{sect 5.2} and \ref{sect 5.3},
this time evaluating the necessary RG functions in a constant curvature background.
In particular, we derive the Weyl consistency conditions which follow from the two-point Green functions
in this case. Naturally, these will involve the coefficient $\b_a$, though since ${\BG}$ is proportional
to $R^2$ it is immediately clear that these identities will mix the coefficients $\b_a$ and $\b_b$.

The anomalous Ward identities and the RGEs take exactly the form already quoted in sections \ref{sect 3.2},
\ref{sect 3.1} and section \ref{sect 5.2}, only the contact term contributions ${\cal A}_{\m\n}, {\cal B}_I, ~\ldots~,
{\cal G}_{IJ}$ being different since they depend on the anomaly ${\cal A}$ and spacetime background.
In order to make this section as self-contained as possible, we repeat the essential identities here, quoting
the contact terms evaluated on a maximally symmetric spacetime. To determine these, we need
the following metric variations, which follow from (\ref{e4}) and the associated footnote, specialised
to the case of non-vanishing, constant curvature:
\begin{align}
\frac{1}{\sqrt{-g}}\frac{\d}{\d g^{\m\n}(x)} \,{\BF} &=   0   \nonumber \\
\frac{1}{\sqrt{-g}}\frac{\d}{\d g^{\m\n}(x)} \,{\BG} &= \tfrac{1}{3} R\left(\tfrac{1}{4}R g_{\m\n} 
+ \D_{\m\n} \right) \d(x,y) \nonumber \\
\frac{1}{\sqrt{-g}}\frac{\d}{\d g^{\m\n}(x)} \,{\BH} &= \tfrac{2}{9} R\left(\tfrac{1}{4}R g_{\m\n} 
+ \D_{\m\n} \right) \d(x,y) \ ,
\label{i1}
\end{align}
where as usual $\D_{\m\n} = g_{\m\n} D^2 - D_\m D_\n$.  The trace anomaly is 
\begin{equation}
{\cal A} = \b_c {\BF} - \b_a{\BG} - \b_b{\BH} - \b_\L + D_\m Z^\m \ ,
\label{i2}
\end{equation}
with $\b_\L$ and $Z^\m$ defined in (\ref{e7}) and (\ref{e8}). 

Beginning with the anomalous Ward identities, we have as usual
\begin{equation}
i\langle T_{\m\n}(x)~T^\r{}_\r(y)\rangle  ~-~ i\langle T_{\m\n}(x)~\Theta(y)\rangle   ~+~
2\langle T_{\m\n}(x)\rangle \,\d(x,y) ~=~ {\cal A}_{\m\n}(x,y) \ ,
\label{i3}
\end{equation}
and with the definition of ${\cal A}_{\m\n}$ (and subsequent contact terms) taken from sections 
\ref{sect 3.2} and \ref{sect 3.1}, we find
\begin{equation}
{\cal A}_{\m\n} = -\tfrac{2}{3} \bar{\b}_a R \D_{\m\n}\d(x,y) +
\tfrac{2}{3} d \left(\tfrac{1}{4} R g_{\m\n} + \D_{\m\n}\right) \d(x,y) \overleftarrow{D}^2 \ ,
\label{i4}
\end{equation}
where for brevity we have introduced the temporary notation $\bar{\b}_a = \b_a +\tfrac{2}{3}\b_b$
since, as noted above, they always appear in this combination in the identities for a 
maximally symmetric background.
Note that in curved spacetime, we have to be careful in general to match the connections 
defining the covariant derivatives to the appropriate spacetime point. This is implicit in the 
above notation, where {\it e.g.}~$\D_{\m\n}$ is understood as involving the covariant derivatives 
with respect to $x$.
Next, we have
\begin{equation}
i\langle O_I(x) ~T^\r{}_\r(y) \rangle ~-~ i\langle O_I(x)~\Theta(y) \rangle 
-\partial_I\b^J \langle O_J(x)\rangle\,\d(x,y) ~=~  {\cal B}_I(x,y) \ ,
\label{i5}
\end{equation}
with 
\begin{equation}
{\cal B}_I(x,y) = -\tfrac{1}{6} \partial_I \bar{\b}_a R^2 \d(x,y) -
\left(\tfrac{1}{4} w_I - \tfrac{1}{3}\left(\partial_I d + Y_I\right)\right) R D^2 \d(x,y) +
U_I D^2 \d(x,y) \overleftarrow{D}^2  \ .
\label{i6}
\end{equation}

The contracted identity of course carries no further information, but because of its relation
to the key RG flow equations it is convenient to state directly:
\begin{multline}
i\langle T^\m{}_\m(x)~T^\r{}_\r(y)\rangle  ~-~ \b^I \b^J i\langle O_I(x)~O_J(y)\rangle ~+~
2\langle T^\m{}_\m(x)\rangle \,\d(x,y)  ~-~
\beta^I\partial_I \beta^J \,\langle O_J(x)\rangle \,\d(x,y)\\
 ~=~ {\cal H}(x,y)\ , 
\label{i7}
\end{multline}
with
\begin{multline}
{\cal H}(x,y) = -\tfrac{1}{6}\b^I \partial_I \bar{\b}^a R^2 \d(x,y) -
\left(2 \bar{\b}_a -\tfrac{2}{3} d + \b^I\left(\tfrac{1}{4} w_I - \tfrac{1}{3}\left(\partial_I d + Y_I\right)
\right)\right) R D^2 \d(x,y) \\
+ \left(2d + \b^I U_I\right) D^2 \d(x,y) \overleftarrow{D}^2 \ .
\label{i8}
\end{multline}

Now we come to the renormalisation group equations.
For the VEVs, evaluating on a constant curvature spacetime, we find
\begin{equation}
{\cal D} \langle T_{\m\n}\rangle = {\cal C}_{\m\n}  = 0\ ,
\label{i9}
\end{equation}
and 
\begin{equation}
{\cal D}\langle O_I\rangle + \gamma_I{}^J \langle O_j\rangle = {\cal C}_I = 
\tfrac{1}{6} \partial_I \bar{\b}_a R^2 \ .
\label{i10}
\end{equation}

The RGE for the two-point Green function of the full energy-momentum tensor is
\begin{equation}
{\mathcal D} \,i\langle T_{\m\n}(x) ~ T_{\r\s}(y)\rangle = {\cal E}_{\m\n,\r\s}(x,y) \ .
\label{i11}
\end{equation}
Here, the full expression for $ {\cal E}_{\m\n,\r\s}(x,y)$ is quite complicated \cite{Osborn:1999az}.
However, for the application to the Weyl consistency conditions, we only need the contracted
expression, which is more simply obtained from the metric variations in (\ref{e4}) and (\ref{i1}).
In this case, we find
\begin{equation}
{\cal E}_{\m\n,\r\s}(x,y) g^{\r\s}(y)= \tfrac{8}{3} \b_b \left(\D_{\m\n} + \tfrac{1}{4}R g_{\m\n}\right) 
\d(x,y) \overleftarrow{D}^2 \ .
\label{i12}
\end{equation}
Note that $\b_a$ cancels from this expression, the contribution arising solely from the second term
on the r.h.s. of the contracted variation equation for ${\BH}$ in (\ref{e4}).
The RGEs for the remaining Green functions are straightforwardly derived and we find
\begin{equation}
{\mathcal D}~ i \langle T_{\m\n}(x) ~ O_I(y)\rangle + \c_I{}^J i \langle T_{\m\n}(x) ~ O_J(y)\rangle
= {\cal F}_{\m\n,I}(x,y) \ , 
\label{i13}
\end{equation}
with 
\begin{equation}
{\cal F}_{\m\n,I}(x,y) =  \tfrac{2}{3} \partial_I \bar{\b}_a R \D_{\m\n}\d(x,y) -
\tfrac{2}{3}\chi_I^e \left(\tfrac{1}{4}R g_{\m\n} + \D_{\m\n}\right) \d(x,y) \overleftarrow{D}^2    \ ,
\label{i14}
\end{equation}
and
\begin{multline}
{\mathcal D}~ i\langle O_I(x) ~ O_J(y)\rangle +\c_I{}^K i\langle O_K(x) ~ O_J(y)\rangle 
+ \c_J{}^K i\langle O_I(x) ~ O_K(y)\rangle + \partial_I \partial_J \b^K \langle O_K\rangle \d(x,y) \\
= {\cal G}_{IJ}(x,y) \ ,
\label{i15}
\end{multline}
with
\begin{equation}
{\cal G}_{IJ}(x,y) = \tfrac{1}{6}\partial_I \partial_J \bar{\b}_a R^2 \d(x,y) -
\left(\tfrac{1}{3} \chi^f_{IJ} - \tfrac{1}{4} \chi^g_{IJ} \right) R D^2 \d(x,y) +
\chi^a_{IJ} D^2 \d(x,y)  \overleftarrow{D}^2   \ .
\label{i16}
\end{equation}

We now have everything we need to analyse the Weyl consistency conditions on a maximally
symmetric spacetime. Recall from sections \ref {sect 3.4} and \ref{sect 5.3} the general identities:
\begin{equation}
{\cal L}_\b{\cal A}_{\m\n}(x,y) = {\cal E}_{\m\n,\r\s}(x,y)g^{\r\s}(y) - {\cal F}_{\m\n,I}(x,y)\b^I 
+ 2 {\cal C}_{\m\n}(x)\d(x,y) \ ,
\label{i17}
\end{equation}
and
\begin{equation}
{\cal L}_\b {\cal B}_I(x,y)  = g^{\m\n}(y) {\cal F}_{\m\n,I}(y,x) - 
{\cal G}_{IJ}(x,y)\b^J -\partial_I\b^J {\cal C}_J(x)\d(x,y) \ .
\label{i18}
\end{equation}

From (\ref{i17}), we can separately equate the coefficients of the independent terms involving
$R \D_{\m\n}\d(x,y) $, $R g_{\m\n}\d(x,y)\overleftarrow{D}^2$ and $\D_{\m\n} \d(x,y) \overleftarrow{D}^2$.
The first is simply an identity involving $\bar{\b}_a$ with no non-trivial content.
The second two give the same identity,
\begin{equation}
4\b_b = -\chi_I^e \b^I + {\cal L}_\b d \ .
\label{i19}
\end{equation}
Unsurprisingly, since this arises from the terms involving only derivatives, this reproduces one
of the consistency relations already found in flat spacetime.

Then, from (\ref{i18}), equating coefficients of $R^2\d(x,y)$, $R D^2\d(x,y)$ and
$D^2 \d(x,y)  \overleftarrow{D}^2$, we again find the first is merely an identity involving $\bar{\b}_a$,
while the second and third conditions are respectively
\begin{equation}
{\cal L}_\b \left(\tfrac{1}{4} w_I - \tfrac{1}{3}\left(\partial_I d + Y_I\right) \right) = \tfrac{2}{3} \chi_I^e
-2 \partial_I\bar{\b}_a + \left(\tfrac{1}{4}\chi_{IJ}^g - \tfrac{1}{3}\chi_{IJ}^a\right)\b^J \ ,
\label{i20}
\end{equation}
and
\begin{equation}
{\cal L}_\b U_I = - 2 \chi_I^e - \chi_{IJ}^a \b^J \ .
\label{i21}
\end{equation}
Again, the final relation arising from the purely derivative terms reproduces a flat space identity.

The new relation arising from working in curved spacetime can be simplified by combining
(\ref{i20}) and (\ref{i21}) to eliminate $\chi_I^e$. Rearranging, this gives
\begin{equation}
\partial_I\left(8 \b_a + \tfrac{4}{3} \b_b\right) + {\cal L}_\b \left(w_I  - \tfrac{4}{3}\left(\partial_I d
+ Y_I - U_I\right)\right) =  \left(\chi_{IJ}^g - \tfrac{4}{3}\left(\chi_{IJ}^f + \chi_{IJ}^a\right) \right)\b^J \ .
\label{i22}\end{equation}
We recognise this as a linear combination of the two Weyl consistency conditions in (\ref{e10}) for
$\partial_I\b_a$ and $\partial_I \b_b$. As anticipated, because  ${\BG}= \tfrac{1}{6} R^2$
for a maximally symmetric spacetime, this construction is unable to distinguish the Weyl consistency
conditions involving $\b_a$ and $\b_b$. 

Contracting with $\b^I$, or equivalently using directly the consistency relation 
\begin{multline}
{\cal L}_\b{\cal H}(x,y) = g^{\m\n}(x) {\cal E}_{\m\n,\r\s}(x,y) g^{\r\s}(y) - {\cal G}_{IJ}(x,y) \b^I \b^J
\\+ 2 g^{\m\n}(x) {\cal C}_{\m\n}(x) \d(x,y) - \b^I\partial_I\b^J {\cal C}_J(x) \d(x,y) \ ,
\label{i23}
\end{multline}
we can write the following identity:
\begin{equation}
\b^I \partial_I \tilde{B}_a= \tilde{G}_{IJ} \b^I \b^J \ ,
\label{i24}
\end{equation}
where we define $\tilde{B}_a = \tilde{\b}_a + \tfrac{2}{3} \tilde{\b}_b$ with
\begin{align}
\tilde{\b}_a &= \b_a + \tfrac{1}{8}\b^I w_I \nonumber \\
\tilde{\b}_b &= \b_b -\tfrac{1}{4}\b^I \left(\partial_I d + Y_I - U_I\right)  \ ,
\label{i25}
\end{align}
and
\begin{equation}
\tilde{G}_{IJ} = \tfrac{1}{8} \chi_{IJ}^g - \tfrac{1}{6} \chi_{IJ}^f - \tfrac{1}{6} \chi_{IJ}^a \ .
\label{i26}
\end{equation}

The identity (\ref{i24}) is the main result of this section. It shows clearly how the Weyl consistency
relation for $\b_a$ is obtained from two-point Green functions on a maximally symmetric spacetime,
which would be a key step in deriving a four-dimensional $a$-theorem on such backgrounds.
On the lhs, notice that although the identity necessarily involves both $\tilde{\b}_a$ and $\tilde{\b}_b$,
we see from (\ref{i19}) that $\b_b$ is itself $O(\b^I)$, so at a fixed point $\tilde{B}_a$
reduces simply to $\b_a$, as we would wish for an $a$-theorem. 
On the rhs, $\tilde{G}_{IJ}$ is given by the three terms of $O(\partial g^I \partial g^J)$ in the 
contribution $\b_\L$ to the trace anomaly, which in turn determine the renormalisation group equation 
(\ref{i16}) for the two-point Green function $\langle O_I(x)~ O_J(y)\rangle$. This would of course be
the natural origin of the metric on the space of couplings required for the $a$-theorem
although, as usual for the Weyl consistency conditions, at this stage there is no direct proof
of positivity of $\tilde{G}_{IJ}$ itself.

\subsection{RG flow of $\langle T^\m{}_\m\rangle$ on maximally symmetric spaces}\label{sect 9.2}

An alternative approach to finding a four-dimensional generalisation of Zamolodchikov's
$c$-theorem is to follow the original suggestion by Cardy \cite{Cardy:1988cwa}, developed further by
Forte and Latorre in \cite{Forte:1998dx}, and define a conjectured $C$-function as the trace of the energy-momentum 
tensor on a background Euclidean space of constant curvature, either the sphere $\mathbb{S}_n$
or hyperboloid $\mathbb{H}_n$ in $n$-dimensions. Since on such maximally symmetric spaces,
\begin{equation}
\langle T_{\m\n}(x)\rangle = - \tfrac{1}{n} C(g) \rho^n g_{\m\n}(x) \ ,
\label{i201}
\end{equation}
where $\rho$ is an inverse length scale characterising the curvature $R = \pm n(n-1)\rho^2$,
we can identify the constant, dimensionless $C$-function as $C(g) =
-(1/\rho^n) \langle T^\m{}_\m\rangle$.

Notice also that integrating the trace anomaly on the sphere and evaluating at an
RG fixed point $\b^I = 0$, the coefficient $\b_a$ of the Euler-Gauss-Bonnet density ${\BG}$ is isolated
and we have, in four dimensions:
\begin{equation}
C(g)\Big|_{\b^I = 0} = - \frac{3}{8\pi^2}\int_{\mathbb{S}_4} d^4x\sqrt{g}\, \langle T^\m{}_\m\rangle \Big|_{\b^I=0} 
= 12\b_a \chi(\mathbb{S}_4) \ ,
\label{i2011} 
\end{equation}
where $\chi(\mathbb{S}_n)$ is the Euler characteristic of $\mathbb{S}_n$.

We now use the anomalous Ward identities and RG equations derived in sections \ref{sect 3.2} and \ref{sect 3.1}
to derive the RG flow of this function $C(g)$.
First, we can consider the response of $C(g)$ to a change in the curvature scale $\r$ or equivalently,
since
\begin{equation}
\r \frac{\partial}{\partial\r} = 2 \int d^n y \,g^{\r\s}\frac{\d}{\d g^{\r\s}(y)} \ ,
\label{i202}
\end{equation}
to a Weyl transformation of the metric. Now, according to (\ref{b28}), the metric derivative of $T_{\m\n}(x)$ 
produces the renormalised two-point function $\langle T_{\m\n}(x) T_{\r\s}(y)\rangle$, we find
\begin{align}
\r \frac{\partial C}{\partial \r} &= \frac{1}{\r^n} \int d^n y \sqrt{g}\, \langle T^\m{}_\m(x) ~ T^\r{}_\r(y)\rangle 
+ 2C \nonumber\\
&= \frac{1}{\r^n} \int d^n y \sqrt{g}\, \left(\langle \Theta(x) ~ \Theta(y) \rangle - {\cal H}(x,y) \right)
- \frac{1}{\r^n} \b^I\partial_I\b^J \langle O_J\rangle \ ,
\label{i203}
\end{align} 
where we have used the identity (\ref{c25}) in the second step (remembering that here we are working
in Euclidean space). Recall from (\ref{c28}) that ${\cal H}(x,y) = g^{\m\n} {\cal A}_{\m\n} + \b^I {\cal B}_I$.

Now, since no other scale is present in the VEV itself, unlike the two-point Green functions which
depend on the geodesic distance between the two points, we have $C = C(g;\m/\r)$, where $\m$ 
is the renormalisation scale. Dimensional analysis then implies
\begin{equation}
\left(\m \frac{\partial}{\partial\m} + \r \frac{\partial}{\partial \r} \right) C(g;\m/\r) = 0 \ .
\label{i204}
\end{equation}
The RGE (\ref{c7}) for the VEV $\langle T^\m{}_\m\rangle$ gives
\begin{equation}
{\cal D} C \equiv \left(\m \frac{\partial}{\partial\m} + \b^I \frac{\partial}{\partial g^I}\right) C
= - g^{\m\n} {\cal C}_{\m\n} \ ,
\label{i205}
\end{equation}
where ${\cal C}_{\m\n} = - \int d^n y \sqrt{g}\,{\cal A}_{\m\n}(x,y)$.
Combining (\ref{i204}) and (\ref{i205}) we find
\begin{equation}
\b^I \frac{\partial C} {\partial g^I}  = \r \frac{\partial C }{\partial \r}  - g^{ \m\n} {\cal C}_{\m\n} \ .
\label{i206}
\end{equation}
Finally, using (\ref{i203}) and noting the cancellation of the factors involving the contact term ${\cal A}_{\m\n}$,
we find the following key identity \cite{Osborn:1999az} for the RG flow of $C(g)$:
\begin{equation}
\b^I \frac{\partial C}{\partial g^I} = \frac{1}{\r^n} \int d^n y \sqrt{g}\, \langle \Theta(x) ~ \Theta(y) \rangle
- \frac{1}{\r^n} \int d^n y \sqrt{g}\,\b^I {\cal B}_I(x,y) - \frac{1}{\r^n} \b^I\partial_I\b^J \langle O_J\rangle \ .
\label{i207}
\end{equation}
Notice that the derivative terms in ${\cal B}_I(x,y)$ vanish under the integral, 
so from (\ref{c27}) and (\ref{i6}) we simply have in two dimensions 
\begin{equation}
\int d^n y \sqrt{g}\,\b^I {\cal B}_I(x,y) = \tfrac{1}{2} \b^I\partial_I \b_c R \ ,
\label{i208}
\end{equation}
and in four-dimensional maximally symmetric spaces,
\begin{equation}
\int d^n y \sqrt{g}\,\b^I {\cal B}_I(x,y) = -\tfrac{1}{6}\b^I\partial_I \bar{\b}_a R^2 \ .
\label{i209}
\end{equation}

There are of course many ways of deriving (\ref{i207})\footnote{For example, as a consistency check
on (\ref{i207}), writing $C(g)$ as the sum of its two components,
$C(g) = -(1/\r^n)\left(\b^I \langle O_I\rangle + {\cal A}\right)$, and recalling that 
$\b^I\partial_I = \int d^n y\,\b^I \d/\d g^I(y)$ acting on the VEV $\langle O_I\rangle$ produces the 
two-point function, we see that (\ref{i207}) is actually just the sum of two obvious identities:
\begin{align*}
\b^I\partial_I\left(\b^J \langle O_J\rangle\right) &= 
-\int d^n y \sqrt{g} \, \b^I \b^J \langle O_I(x)~O_J(y)\rangle + \b^I \partial_I\b^J \langle O_J\rangle \ ,\\
\b^I \partial_I {\cal A} &= \int d^ny \sqrt{g}\,\b^I {\cal B}_I(x,y) \ .
\end{align*}
The second is just the definition of ${\cal B}_I(x,y)$ and has no dynamical content, while the first
is simply the definition of the two-point Green function $\langle O_I(x)~O_J(y)\rangle$.}
and many equivalent expressions for the r.h.s.~which can be obtained using the identities in section \ref{sect 3}.
For example, in terms of the conserved two-point function $\langle T_{\m\n}(x)~T_{\r\s}(y)\rangle$
defined in (\ref{cc4}), we have
\begin{equation}
\b^I \frac{\partial C}{\partial g^I} = \frac{1}{\r^n} \int d^n y \sqrt{g} \, \langle T_{\m\n}(x)~T_{\r\s}(y)\rangle_c 
 - n C - \frac{1}{\r^n} g^{\m\n} {\cal C}_{\m\n} \ ,
\label{i210}
\end{equation}
where, as we have seen in (\ref{c27}) and (\ref{i9}), ${\cal C}_{\m\n} = 0$ in two and four dimensions.

The idea behind interpreting (\ref{i207}) as a $C$-theorem is based on attempting to exploit the positivity of
the two-point Green function $\langle \Theta(x) ~ \Theta(y) \rangle$ for $x\neq y$ to prove monotonicity of 
the RG flow of $C(g)$ \cite{Forte:1998dx}. However, the contact terms 
necessarily present in (\ref{i207}) prevent this conclusion.

It is of course true that the form of the extra terms involving ${\cal B}_I$ and $\langle O_J\rangle$
in (\ref{i207}) depends on the definition (\ref{b29}) of the renormalised two-point function. 
The key point, however, is that whatever prescription is chosen to redistribute contact terms between
the renormalised Green function and the remainder, the r.h.s.~as a whole does contain contact terms
proportional to $\d(x,y)$ appearing under the integral, which necessarily contribute to the
RG flow equation for $C(g)$.
\footnote{To make this explicit, if we refer to (\ref{b35}) where we relate the renormalised
and bare two-point Green functions in two dimensions, we see that (\ref{i207}) can be rewritten as
\begin{equation*}
\b^I \partial_I C = \lim_{\e\rightarrow 0} \int d^2 y \sqrt{g}\,\langle \hat\Theta(x) ~ \hat\Theta(y)\rangle_B \ ,
\end{equation*}
with $\hat\Theta = \hat{\b}^I O_I$. The r.h.s.~is manifestly independent of any renormalisation prescription
for the two-point function, but clearly the bare Green function has contact terms which give a finite, non-singular
contribution under the integral.}

Another difficulty in attempting to establish monotonicity of the flow of $C(g)$ arises if we write a 
spectral expansion of $\langle \Theta(x)~\Theta(y)\rangle$ on the sphere or hyperboloid. 
(For a detailed account see \cite{Osborn:1999az}, following \cite{Cappelli:1990yc}, \cite{Forte:1998dx}.)
The UV behaviour of the Green function requires that its dispersion relation requires a subtraction term, 
whose sign is not constrained by unitarity as is the case for the spectral function itself.

Further discussion of the problems with this approach may be found in \cite{Osborn:1999az}
and here in section \ref{sect 8}.
Ultimately though, we see that even working on a maximally symmetric space where the VEV 
of the energy-momentum tensor is sensitive to the coefficient $\b_a$ of the Euler-Gauss-Bonnet term
in the anomaly, the original Cardy suggestion of identifying a higher-dimensional $C$-function 
with the trace of the energy-momentum tensor is not realised. The corresponding RG flow
equation (\ref{i207}) for $C(g)$ does not exhibit the required monotonicity.

\subsection{Zamolodchikov functions and the search for a $C$-theorem on the sphere}\label{sect 9.3}

The natural next step is therefore to attempt to repeat the Zamolodchikov construction of a $C$-function
 from the two-point Green functions of the energy-momentum tensor, along the lines of sections 
\ref{sect 4} and \ref{sect 6}, this time evaluated on a maximally symmetric curved space background.

\subsubsection{Geometry of constant curvature spaces}

First we need some geometrical preliminaries. (For further details, see \cite{Osborn:1999az}, especially
section 5.) Since we are interested in two-point Green functions, a basic r\^ole is played by the geodesic interval
$\s(x,y) = \tfrac{1}{2} \int_0^1 d\t g_{\m\n} \dot{x}^\m \dot{x}^\n$ along the geodesic $x^\m(\t)$ joining
$x^\m(0) = x^\m$ and $x^\m(1) = y^\m$. This is a bi-scalar ({\it i.e.}~a scalar at both $x$ and $y$) and
satisfies 
\begin{equation}
g_{\m\n}\partial^\m\s \partial^\n\s = 2\s \ .
\label{i220}
\end{equation}
In flat space, $\s(x,y) = \tfrac{1}{2}(x-y)^2$. We also need the bi-vector $I^\m{}_\r(x,y)$ which parallel transports 
vectors from $y$ to $x$ along the geodesic $x^\m(\t)$. This is defined by
\begin{equation}
\s^\n D_\n I^\m{}_\r(x,y) = 0 \ , \hskip2cm  
I^\m{}_\r(x,x) = \d^\m{}_\r \ ,
\label{i221}
\end{equation}
and satisfies notably
\begin{equation}
\partial_\m \s(x,y)I^\m{}_\r(x,y) = - \s(x,y)\overleftarrow{\partial_\r} \ .
\label{i222}
\end{equation}

Now, for spaces of constant curvature we may write $\s(x,y) = \frac{1}{2\r^2}\theta^2(x,y)$, so that
on the sphere,\footnote{For the corresponding results throughout this section for a hyperbolic space,
see \cite{Osborn:1999az}. The only essential change is the substitution of hyperbolic for trigonometric
functions of $\theta$ in all the formulae for the two-point Green functions.} $\theta(x,y)$ 
($0\le\theta\le\pi$) is the angular separation of the points $x$ and $y$ while $\r$ is the inverse radius.
A convenient basis in which to expand bi-tensors, such as the two-point Green function of the 
energy-momentum tensor, is then given by the unit vectors $\hat{x}^\m$ and $\hat{y}^\r$ at $x$
and $y$ respectively, together with the parallel transport bi-vector $I^\m{}_\r(x,y)$, where
\begin{equation}
\hat{x}_\m = \frac{1}{\sqrt{2\s}} \partial_\m\s = \frac{1}{\r}\partial_\m\theta \ , \hskip2cm
\hat{y}_\r = \frac{1}{\sqrt{2\s}} \s\overleftarrow{\partial_\r} = \frac{1}{\r} \theta \overleftarrow{\partial_\r} \ ,
\label{i223}
\end{equation}
with
\begin{equation}
\hat{x}^\m I^\m{}_\r(x,y) = -\hat{y_\r} \ , \hskip2cm 
I_\m{}^\r(x,y) \hat{y}_\r = - \hat{x}_\m \ .
\label{i224}
\end{equation}
We also need their derivatives \cite{Osborn:1999az}, which for the sphere are
\begin{align}
D_\m \hat{x}_\m &= \r \cot\theta\left(g_{\m\n} - \hat{x}_\m \hat{x}_\n\right) \ , \hskip2cm
\partial_\m \hat{y}_\r = -\r \csc\theta \left(I_{\m\r} + \hat{x}_\m \hat{y}_\r\right) \nonumber \\
D_\m I_{\n\r} &= \r \tan\tfrac{1}{2}\theta \left(g_{\m\n}\hat{y}_\r + \hat{x}_\n I_{\m\r}\right) \ .
\label{i225}
\end{align}

\subsubsection{Two-point Green functions and conservation identities}

In this basis, we can expand the two-point Green functions of the energy-momentum tensor in the 
following form:
\begin{multline}
\langle T_{\m\n}(x)~T_{\r\s}(y)\rangle_c = R(\theta) \hat{x}_\m \hat{x}_\n \hat{y}_\r \hat{y}_\s  + 
S(\theta)\left(I_{\m\r}\hat{x}_\n \hat{y}_\s + \m \leftrightarrow\n, \r \leftrightarrow\s\right) \\ +
T(\theta)\left(I_{\m\r} I_{\n\s} + I_{\m\s}I_{\n\r}\right)  + 
U(\theta)\left(\hat{x}_\m \hat{x}_\n g_{\r\s} + \hat{y}_\r \hat{y}_\s g_{\m\n}\right) +
V(\theta) g_{\m\n} g_{\r\s} \ ,
\label{i226}
\end{multline}
where $R, S, \ldots V$ are independent functions of $\theta$. The contracted Green functions are
\begin{equation}
\langle T_{\m\n}(x)~T^\r{}_\r(y)\rangle_c = P(\theta) \hat{x}_{\m} \hat{x}_\n + Q(\theta)g_{\m\n} \ ,
\label{i227}
\end{equation}
where, using the results above in $n$ dimensions, we find 
\begin{equation}
P = R - 4S + nU \ , \hskip2cm  Q = 2T + U + nV \ ,
\label{i228}
\end{equation}
and
\begin{equation}
\langle T^\m{}_\m(x)~T^\r{}_\r(y)\rangle_c = N(\theta) \ ,
\label{i229}
\end{equation}
with
\begin{equation}
N = P + nQ \ .
\label{i230}
\end{equation}
Comparing with the flat space expansions (\ref{d7}) in section \ref{sect 4.1}, we see that the Zamolodchikov functions 
$F$, $H$ and $G$ are to be identified, up to dimension dependent rescalings involving factors of $(n-1)$, 
with the coefficients $R$, $P$ and $N$ respectively.

The next step in the Zamolodchikov programme is to apply the conservation equation,
\begin{equation}
D^\m \langle T_{\m\n}(x)~T_{\r\s}(y)\rangle_c = 0\ .
\label{i231}
\end{equation}
Taking the covariant derivatives using the identities (\ref{i225}), and setting the coefficients of the terms
$\hat{x}_\n g_{\r\s}$, $I_{\n\r}\hat{y}_\s + I_{\n\s}\hat{y}_\r$ and $\hat{x}_\n \hat{y}_\r \hat{y}_\s$ 
to zero, gives three identities:
\begin{align}
\left(R-2S+U\right)' &= -(n-1)\left(\cot\theta\, R + 2 \tan\tfrac{1}{2}\theta \,S\right) 
- 2 \cot\tfrac{1}{2}\theta\, S + 2 \csc\theta \,U \nonumber \\
\left(S-T\right)' &= -n\left(\cot\theta \,S + \tan\tfrac{1}{2}\theta \,T\right) + \csc\theta \,U \nonumber\\
\left(U+V\right)' &= 2\csc\theta \,S - 2 \tan\tfrac{1}{2}\theta \,T - (n-1)\cot\theta \,U \ ,
\label{i232}
\end{align}
where $R'(\theta) = dR/d\theta$, {\it etc.}
The flat space limit ($\s$ fixed with $\r \rightarrow 0$) corresponds to $\theta\rightarrow 0$, and in this limit
it is readily checked that these combinations are precisely those found from the conservation identities
in the original Zamolodchikov construction.

In two dimensions, these identities can be rearranged giving two
equations which involve $R$, $P$ and $N$ only,
{\it viz.}
\begin{align}
\left(R+P\right)' &= -2 \cot\tfrac{1}{2}\theta \,R + 
\left(\cot\tfrac{1}{2}\theta + \tan\tfrac{1}{2}\theta\right) \,P \nonumber \\
\left(P+N\right)' &= - \left(\cot\tfrac{1}{2}\theta - \tan\tfrac{1}{2}\theta\right)\,P \nonumber \\
\left(S-T\right)' &= -\tfrac{1}{2}\csc\theta\,(R-P) + 2 \tan\tfrac{1}{2}\theta\, (S-T) \ ,
\label{i233}
\end{align}
the second of which can also be obtained directly from the conservation equation applied to the 
once-traced Green function (\ref{i227}).

\subsubsection{The Zamolodchikov $C$-function on the sphere}

We now focus on two dimensions and look for an extension of the flat spacetime $c$-theorem to spaces
of constant curvature.

The first step is to define rescaled Zamolodchikov functions. As we shall see, it is natural to use the chord
distance $\sqrt{s} = \frac{2}{\r} \sin\tfrac{1}{2}\theta$ rather than the geodesic distance 
$\sqrt{2\s} = \frac{1}{\r}\theta$, and in two dimensions we define
\begin{equation}
F = s^2\,R \ , \hskip1.5cm  H = - s^2\,P \ , \hskip1.5cm G = s^2\,N \ ,
\label{i234}
\end{equation}
and make the ansatz
\begin{equation}
C = F+2H-3G \ ,
\label{i235}
\end{equation}
for the $C$-function, as in flat space.

From the conservation identities (\ref{i233}), we readily find
\begin{equation}
C' = - 6 \cot\tfrac{1}{2}\theta \, G + 2 \tan\tfrac{1}{2}\theta \, H  \ ,
\label{i236}
\end{equation}
and
\begin{equation}
\left(G-H\right)' = 2 \cot\tfrac{1}{2}\theta \,G - \left(\cot\tfrac{1}{2}\theta + \tan\tfrac{1}{2}\theta\right)\,H \ .
\label{i237}
\end{equation}

Now, in the flat space limit, only the term involving the manifestly positive definite function $G$ would 
appear on the r.h.s.~of (\ref{i236}). In curved space, however, this term, which appears with coefficient
$\cot\tfrac{1}{2}\theta$, is necessarily accompanied by another term with coefficient $\tan\tfrac{1}{2}\theta$
which does not have the required positivity.
This in fact turns out to be quite generic and is the first indication of problems in attempting to establish
a $C$-theorem in curved space.

To see this directly, note that $d/d\theta = \tfrac{1}{2} \cot\tfrac{1}{2}\theta\, \sqrt{s}\, d/d\sqrt{s}$,
so in terms of the chord distance
\begin{equation}
\sqrt{s} \frac{dC}{d\sqrt{s}} = - 12\, G + \r^2 s\,H \ .
\label{i238}
\end{equation}
This is to be compared with the corresponding flat space formula (\ref{d12}). The key difference, of course,
is the additional contribution proportional to $H$ on the r.h.s., which vanishes in the flat space limit
$\r \rightarrow 0$. This means that away from flat space, the flow of $C$ as the separation scale $\theta$
or $\sqrt{s}$ is increased is no longer guaranteed to be monotonic, since there is no general theorem 
constraining the sign of $H$. Even in two dimensions, therefore, the Zamolodchikov proof of the $C$-theorem
fails in curved space.

Similar results may be proved in $n$ dimensions, and in particular $n=4$.
The analysis here is further complicated because the conservation identities do not close on the functions
$R$, $P$ and $N$ alone. In fact, with the Zamolodchikov combination
\begin{equation}
C= R - \frac{(n+2)}{2(n-1)} P - \frac{(n+4)}{2(n-1)^2} N \ ,
\label{i239}
\end{equation}
(see appendix \ref{sect C}), we can show
\begin{equation}
C'(\theta) = -\cot\tfrac{1}{2}\theta\,\left((n+2)C + \tfrac{(n+2)(n+4)}{2(n-1)^2} N \right)
+ \tan\tfrac{1}{2}\theta\, \left(\tfrac{(n^2 -10n +8)}{4(n-1)} P - \tfrac{n(n-2)}{2} U \right) \ .
\label{i240}
\end{equation}
However, this derivation also requires the use of an extra relation,
\begin{equation}
S' = - \csc\theta\,R - \cot\tfrac{1}{2}\theta\,S + 4 \csc\theta\,S \ ,
\label{i241}
\end{equation}
which can be shown to be be a consequence of the explicit representation \cite{Osborn:1999az} for the
two-point Green function restricted to spin 0 intermediate states only. 
Note that in the flat space limit, the term with coefficient $\tan\tfrac{1}{2}\theta$ vanishes, 
while the coefficient of $C$ becomes $(n-2)$ for the dimensionless $C$ function rescaled with a
factor $s^2$ or $\sin^4\tfrac{1}{2}\theta$. This limit may be compared directly with the flat space
formulae (\ref{f3}), (\ref{f4}) and appendix \ref{sect 6}).

Returning to two dimensions,
one way to try to retrieve the theorem would be to redefine $C(\theta)$, allowing it to be a linear combination 
of $F$, $H$ and $G$ with $\theta$-dependent coefficients, chosen in such a way as to cancel the 
unwanted $\tan\tfrac{1}{2}\theta\,H$ term on the r.h.s.~of (\ref{i236}). This can be done using (\ref{i237}),
though this also means the coefficient of $G$ on the r.h.s.~becomes $\theta$-dependent.
A straightforward calculation shows that redefining
\begin{equation}
C = F + 2H -3G + f(\theta)\left(G-H\right) \ ,
\label{i242}
\end{equation}
with
\begin{equation}
f(\theta) = 2 \left[1 - \cot^2\tfrac{1}{2}\theta \,\log(1 + \tan^2\tfrac{1}{2}\theta)\right] \ ,
\label{i243}
\end{equation}
this new $C$-function satisfies
\begin{equation}
C' = - 6 \cot\tfrac{1}{2}\theta\,\left(1 + g(\theta)\right)\,G \ ,
\label{i244}
\end{equation}
with
\begin{equation}
g(\theta) = -\frac{1}{3} \left[1 + (1-\cot^2\tfrac{1}{2}\theta)\log(1+ \tan^2\tfrac{1}{2}\theta)\right] \ ,
\label{i245}
\end{equation}
which is of $O(\theta^2)$.

Alternatively, in terms of the chord distance, 
\begin{equation}
\sqrt{s} \frac{dC}{d\sqrt{s}} = -12\left(1+g(\sqrt{s}\right)\,G \ ,
\label{i246}
\end{equation}
where
\begin{equation}
C = F + 2H -3G + f(\sqrt{s})\left(G-H\right) \ ,
\label{i247}
\end{equation}
with
\begin{equation}
f(\sqrt{s}) = 2 \left[1 + \frac{4}{\r^2 s}\left(1 - \frac{\r^2 s}{4}\right) 
\log\left(1 - \frac{\r^2 s}{4}\right)\right] \ ,
\label{i248}
\end{equation}
and
\begin{equation}
g(\sqrt{s}) = - \frac{1}{3}\left[1 + \frac{4}{\r^2 s}\left(1 - \frac{\r^2 s}{2}\right) 
\log\left(1 - \frac{\r^2 s}{4}\right)\right] \ ,
\label{i249}
\end{equation}
where both $f$ and $g$ are $O(\r^2 s)$ and so vanish in the flat space limit.

Now even if this modification of the $C$-function may be accommodated, (\ref{i246}) could only be the
basis of a monotonic $C$-theorem if the coefficient $\left(1 + g(\sqrt{s})\right)$ of $G$ is always positive.
This function is plotted in figure \ref{Coefficient}.
\begin{figure}[h!]
\centering
\includegraphics[scale=0.8]{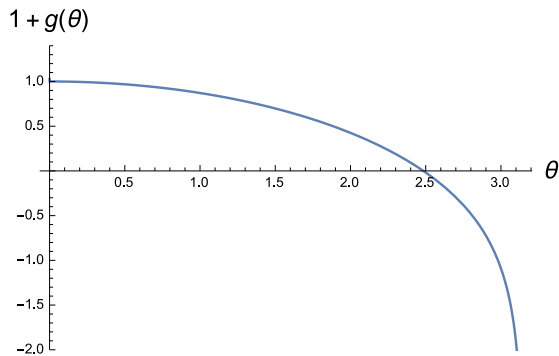}
\caption{The coefficient function $1 + g(\theta)$ governing the scale dependence of $C(\theta)$
in (\ref{i244}) or (\ref{i246}). Note that this is not positive as required over the full range $0\le \theta \le\pi$. }
\label{Coefficient}
\end{figure}
While this is indeed true for small $\theta$ (or $\r^2 s$), {\it i.e.}~in the near flat space limit, unfortunately 
it is not true over the whole required range $0\le \theta \le \pi$ (or $0\le \r\sqrt{s} \le 2$), but 
changes sign for $\theta \gtrsim 2.5$ (or $\r\sqrt{s}\gtrsim 1.8$).
So even with this redefinition of $C$, we do not find a monotonic $C$-function away from flat space.

In fact, there is yet another critical obstruction to establishing a $C$-theorem with monotonic RG flow 
in curved space. We address this below, after first making a final check that we have considered the 
most general potential $C$-theorem on the sphere.

To do this, we now return to the full set of conservation identities (\ref{i233}) and look for the most general
$\theta$-dependent linear combination
\begin{equation}
C = a(\theta) \left(R+P\right) + b(\theta) \left(P+N\right) + c(\theta) \left(S-T\right) \ ,
\label{i250}
\end{equation}
for which $C'$ is proportional to $N$ alone. In fact we immediately see that this requires
\begin{equation}
C'= b'(\theta)\,N \ .
\label{i251}
\end{equation}
Using the conservation identities applied to (\ref{i250}) and imposing that the coefficients of $R$, $P$ and
$(S-T)$ vanish yields the following general solution:
\begin{align}
a(\theta) &= \a \sin^4\tfrac{1}{2}\theta  - \c \cos^4\tfrac{1}{2}\theta  \nonumber \\
b(\theta) &= \a \left(-\sin^4\tfrac{1}{2}\theta + \sin^2\theta \log\cos\tfrac{1}{2}\theta \right)
+ \b \sin^2\theta + \c\left(\cos^4\tfrac{1}{2}\theta - \sin^2\theta \log\sin\tfrac{1}{2}\theta \right)
\nonumber \\
c(\theta) &= \c \cos^4\tfrac{1}{2}\theta \ ,
\label{i252}
\end{align}
with
\begin{multline}
b'(\theta) = \a\left(\sin\theta(\cos\theta -1) + \sin2\theta \log\cos\tfrac{1}{2}\theta \right)
+ \b \sin2\theta \\
+ \c\left(-\sin\theta(\cos\theta +1) - \sin2\theta \log\sin\tfrac{1}{2}\theta \right) \ ,
\label{i253}
\end{multline}
where $\a$, $\b$ and $\c$ are arbitrary constants.

Inspection of the coefficients (\ref{i252}) shows that only the solution with $\b = \c= 0$ reduces to the
Zamolodchikov solution in the flat space limit $\theta\rightarrow 0$.  Normalising with $\a =1$, we then
recognise the overall $\sin^4\tfrac{1}{2}\theta$ factor in $C$ as the familiar scaling factor $s^2$
involving the chord distance already used in defining the Zamolodchikov $F$, $H$ and $G$ functions.
Keeping only this solution, we can rewrite (\ref{i250}) as
\begin{equation}
C = \sin^4\tfrac{1}{2}\theta \left(R-2P-3N + f(\theta) (P + N)\right) \ ,
\label{i254}
\end{equation}
with
\begin{equation}
f(\theta) = 2 \left[1 + 2 \cot^2\tfrac{1}{2}\theta\,\log\cos\tfrac{1}{2}\theta \right] \ ,
\label{i255}
\end{equation}
which we recognise as the same function found above in (\ref{i242}) and (\ref{i243}).
This shows directly that it is indeed the chord distance which gives the natural rescaling for the
Zamolodchikov functions and reassures us that we have exploited the full potential of the
conservation identities in looking for a curved space extension of the $C$-function.

\subsubsection{Dimensional analysis and RG flow of the $C$-function}

Even if (\ref{i246}) had yielded the desired result $\sqrt{s}\, dC/d\sqrt{s} < 0$ for all $\sqrt{s}$, 
we would still not have been able to show that the RG flow of the $C$-function is monotonic.
The reason is straightforward. In order to relate the RG flow $\b^I \partial C/\partial g^I$ 
to $|x|\, \partial C/\partial |x|$ in flat space, we require the intermediate step (\ref{d13}) which
is simply dimensional analysis. The dimensionless function $C$ can only depend on the
combination $\m|x|$, {\it i.e.} $C = C(g;\m|x|)$.

In a constant curvature space, however, there is a further scale $\r$. The dimensional 
analysis identity is now
\begin{equation}
\left(\m \frac{\partial}{\partial \m} + \r \frac{\partial}{\partial \r} - 
\sqrt{s} \frac{\partial}{\partial\sqrt{s}} \right)\,C = 0 \ .
\label{i256}
\end{equation}
(We could of course use $\s = \tfrac{1}{2\r^2}\theta^2$ instead of 
$s = \tfrac{4}{\r^2}\sin^2\tfrac{1}{2}\theta$ here.) 
The direct relation between $\sqrt{s}\, \partial C/\partial\sqrt{s}$ and $\m\,\partial C/\partial\m$
is therefore lost in curved space. Equivalently, $C$ is now a function of {\it two} dimensionless
variables, so here we may write for example $C = C(g;\theta,\m/\r)$.
Only in the flat space limit, where $C \rightarrow C(g;\m\theta/\r)$ as $\r \rightarrow 0$,
does (\ref{i246}) imply a relation for the flow $\b^I\, \partial C/\partial g^I$ through the RGE.

In the end, therefore, we see that this approach to finding a $C$-theorem by working on spaces
of constant curvature is not successful, despite the promising results of section \ref{sect 9.1}
on the Weyl consistency conditions. It is probable that this is not simply a technical obstruction,
but that the $C$-theorem does not in fact hold in curved space. We discuss further evidence in 
support of this conjecture in section \ref{sect 8}.

\section{The Weak $a$-theorem and 4-point Green Functions}\label{sect 10}

Since $\b_a$ does not occur in the anomalous Ward identities or renormalisation group equations for the
two-point Green functions of $T_{\m\n}$ in flat spacetime and, as we have just seen, a $C$-theorem related
to $\b_a$ cannot be proved from one and two-point functions of $T_{\m\n}$ on maximally symmetric, 
curved spaces, we have to look further to try to relate the Weyl consistency condition
(\ref{f22}) to Green functions and find the long sought non-perturbative, monotonic $a$-theorem.

The crucial breakthrough made in \cite{Komargodski:2011vj,Komargodski:2011xv,Luty:2012ww}
comes from the realisation that in fact $\b_a$ does contribute 
to the {\it four-point} Green functions of the energy-momentum tensor trace $T^\m{}_\m$. Moreover, four-point
functions also have an implicit positive property related to the optical theorem, itself a consequence of unitarity.
These properties are then used in a dispersion relation derivation of the {\it weak} $\b_a$ theorem, analogous 
to the proof of the two-dimensional $c$-theorem given in section \ref{sect 4.4}. 

The four-point function of $T^\m{}_\m$ is found by taking four derivatives wrt $\s(x)$ of the local renormalisation 
group identity (\ref{e9}),
\begin{equation}
\left(\D_\s^W - \D_\s^\b\right)W = \int d^4 x \sqrt{-g}~ \big[\s(x)\left(
\b_c {\BF} - \b_a {\BG} - \b_b {\BH} - \b_\L\right)  - \partial_\m \s(x) Z^\m \big] \ .
\label{f24}
\end{equation}
The first derivative gives the trace anomaly,
\begin{equation}
T^\m{}_\m = \b^I O_I + \b_c {\BF} - \b_a {\BG} - \b_b {\BH} - \b_\L + D_\m Z^\m \ .
\label{f25}
\end{equation}
Taking further derivatives, and evaluating in flat space with $\partial_\m g^I = 0$, the only contributions to survive 
are either $O(\b^I)$, and so vanish at a fixed point, or come from the curvature squared terms. We therefore need the
variations of ${\BF}$, ${\BG}$ and ${\BH}$ for $g_{\m\n} \rta e^{2\s(x)}g_{\m\n}$ up to $O(\s^3)$.
This extends the formulae (\ref{e4}) and ({\ref{e16}) for the first and second variations in section \ref{sect 5}. 
These follow from the expansions (in four dimensions)
\begin{align}
\sqrt{-g} {\BF} &= 0  \nonumber \\
\sqrt{-g} {\BG} &= 8 \left(\partial^2\s\right)^2 - \s_{,\m\n} \s^{,\m\n} + 16 \s_{,\m} \s_{,\m} \s^{,\m\n}
+ 8 \s_{,\m} \s^{,\m} \partial^2\s + O(\s^4)  \nonumber \\
\sqrt{-g} {\BH} &= 4 \left(\partial^2\s\right)^2 - 8 \s_{,\m} \s^{,\m} \partial^2\s + O(\s^4) \ ,
\label{f26}
\end{align}
with notation $\s,\m = \partial_\m\s$.
The second is crucial, showing that unlike the case of two-point functions considered previously,
three and higher-point Green functions of the energy-momentum tensor trace
are sensitive to the RG function $\b_a$.

The four-point is then given by (abbreviating $T^\m{}_\m(x) = T(x)$ for simplicity),
\begin{multline}
i\langle T(x_1) ~ T(x_2) ~ T(x_3) ~ T(x_4) \rangle \\= \frac{\d^4}{\d\s(x_1) \d\s(x_2) \d\s(x_3) \d\s(x_4)} 
\int d^4 x \sqrt{-g}\, \s(x)\, \left(\b_a {\BG} + \b_b {\BH}\right)\Big|_{\s = 0}  + O(\b^I) \ ,
\label{f27}
\end{multline}
with ${\BG}$ and ${\BH}$ expanded in $\s(x)$ as above.

The next step is to take the Fourier transform with momenta $p_1, \ldots p_4$, then write the Green function
in terms of Mandelstam variables $s = -(p_1+p_2)^2$, $t = -(p_1-p_3)^2$ and $u= -(p_1-p_4)^2$.
This gives
\begin{multline}
{\mathcal A}(s,t,u) = {\rm f.t.}~i\langle T(x_1) ~ T(x_2) ~ T(x_3) ~ T(x_4) \rangle \\
= {\rm f.t.}~ \frac{\d}{\d\s\,\d\s\,\d\s\,\d\s} \, \int d^4 x \sqrt{-g} \,
\left[ \b_a\left(-\tfrac{1}{2} \s_{,\m} \s_{,\n} \s^{,\m} \s^{,\n}\right)
+ \b_b\left(-\tfrac{1}{2} \s \s_{,\m} \s^{,\m} \partial^2\s\right) \right] + O(\b^I)\ .
\label{f28}
\end{multline}
To isolate the contribution from $\b_a$, we choose the sources $\s(x_i)$ to be null, {\it i.e.} $\partial^2 \s = 0$.
Equivalently, we set the external momenta $p_i^2 = 0$, which implies $s+t+u = 0$.
Then,
\begin{align}
{\mathcal A}(s,t,u) &= -4 \b_a \left(p_1 . p_2 \, p_3 . p_4 \,+\, {\rm perms} \right) + O(\b^I) \nonumber \\
&=  -\b_a \left(s^2 + t^2 + u^2\right)  + O(\b^I) \ .
\label{f29}
\end{align}

\begin{figure}[h!]
\centering
\includegraphics[scale=0.6]{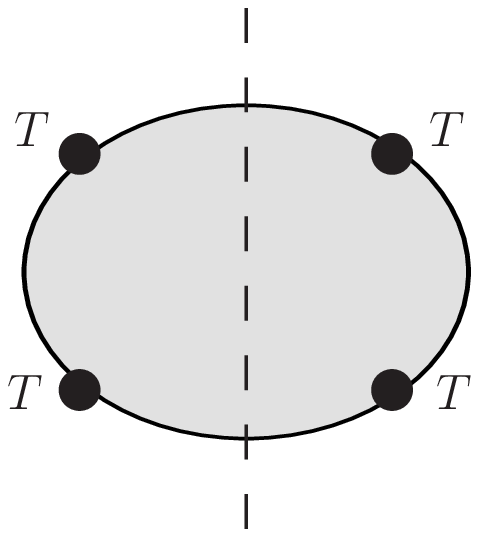}
\qquad \qquad \qquad
\includegraphics[scale=0.6]{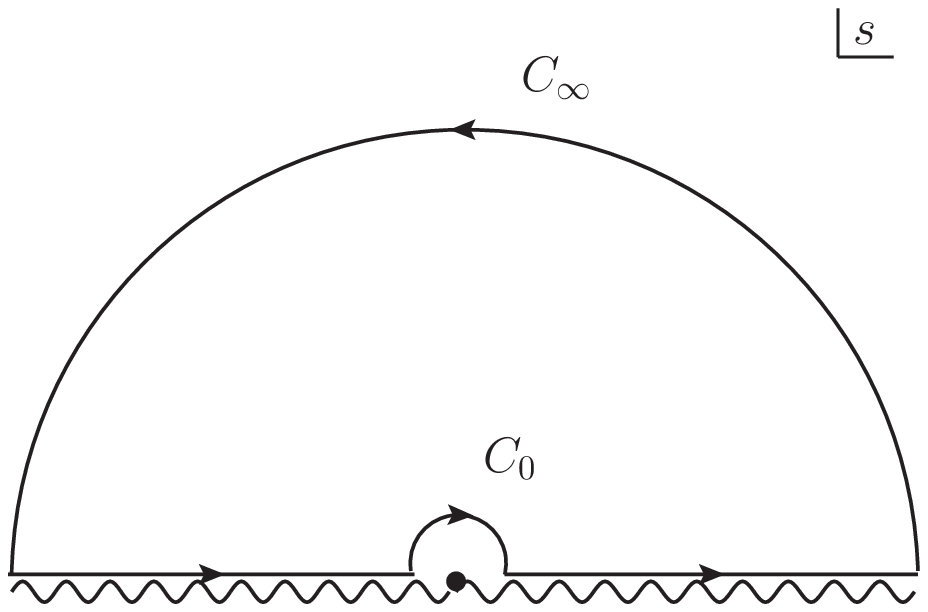}
\caption{The LH figure shows the four-point function $\langle T~ T~ T~ T\rangle$ with the cut giving rise 
to ${\rm Im}\,{\mathcal A}(s)$ in (\ref{f32}). The RH figure shows the contour in the complex $s$-plane 
used in the dispersion relation, showing the cuts on the real $s$-axis. The figure illustrates the case in 
which these may run to the origin.}
\label{Contour4d}
\end{figure}

We now write a dispersion relation for ${\mathcal A}(s,t,u)$ in the `forward scattering' limit $t=0$,
for which
\begin{equation}
{\mathcal A}(s,t,u)\Big|_{t=0} ~\rta~  {\mathcal A}(s) = -2 \b_a s^2 \  .
\label{f30}
\end{equation}
Including a factor $s^{-3}$ to ensure convergence, we integrate ${\mathcal A})s)/s^3$ over the familiar contour 
shown in figure \ref{Contour4d}. Notice that if the theory does not have a mass gap, the cuts on the 
positive and negative real $s$-axis will extend all the way to $s=0$. This is the situation illustrated in the figure.
As in section \ref{sect 4.4}, the derivation of the dispersion depends on a number of critical assumptions:
\begin{enumerate}
\item ${\mathcal A}(s)$ is analytic in the upper-half complex $s$ plane.

\item ${\mathcal A}(-s) = {\mathcal A}(s)$.  This follows from crossing symmetry, which  implies
${\mathcal A}(s,t,u) = {\mathcal A}(u,t,s)$, where $u = -s$ for $t=0$.

\item ${\mathcal A}(s^*) = {\mathcal A}(s)^*$, {\it i.e.}~hermitian analyticity.

\item ${\mathcal A}(s)/s^3$ is bounded for large $s$.

\item ${\rm Im}\,{\mathcal A}(s) \ge 0$, motivated by unitarity and the optical theorem for $2\rta 2$ scattering.
\end{enumerate}

We then have,
\begin{equation}
\int_{C_0} \frac{ds}{s^3} \,{\mathcal A}(s)  + \int_{C_\infty} \frac{ds}{s^3}\, {\mathcal A}(s) 
+ \int_{-\infty}^0 \frac{ds}{s^3} \,{\mathcal A}(s+ i\e)  + \int_0^\infty\frac{ds}{s^3}\, {\mathcal A}(s+i\e)  = 0 \ .
\label{f31}
\end{equation}
Following the same steps detailed in (\ref{d49}) to (\ref{d56}), with ${\mathcal A}(s) = - 2\b_a s^2$
at the UV and IR fixed points, we find
\begin{equation}
\b_a^{UV} - \b_a^{IR} = \frac{1}{\pi} \int_0^\infty \frac{ds}{s^3}\, {\rm Im}\,{\mathcal A}(s) ~ \ge ~ 0 \ .
\label{f32}
\end{equation}
This is the weak $\b_a$-theorem in four dimensions.

Clearly, the proof depends on the validity of the assumptions above, of which the most questionable is
the assumption of hermitian analyticity, ${\mathcal A}(s^*) = {\mathcal A}(s)$. Normally in $S$-matrix theory, 
there is a mass gap between the cuts on the real $s$-axis, allowing ${\mathcal A}(s)$ to be defined as
a single analytic function over the whole complex plane. The assumption of real analyticity is then
assured. However, with the cuts preventing the natural identification of ${\mathcal A}_+(s)$ and
${\mathcal A}_-(s)$ defined on the upper and lower-half planes respectively, assumption (3) above
is not guaranteed to hold. Indeed, this resembles the situation recently found in studies of the
Kramers-Kr\"onig dispersion relation in curved spacetime \cite{Hollowood:2008kq}, where a novel 
geometric-induced analytic structure also produces cuts to the origin in the complex $\omega$ plane, 
and the dispersion relation fails in its usual form.

With this proviso, however, this appears to establish the weak $\b_a$ theorem, showing that
as expected the RG function associated with the Euler-Gauss-Bonnet density in the trace anomaly is less at an
IR fixed point than at the UV fixed point at the origin of the flow. It does not, however, establish
the full desired $a$-theorem analogous to the two-dimensional $c$-theorem, which of course 
requires a monotonically decreasing function all along the RG trajectory.

We can attempt to push this further by studying the origin of the ${\rm Im}\,{\mathcal A}(s)$ contribution 
to the relation (\ref{f32}). Just as for the two-point functions, successive differentiations of (\ref{f24})
yield the anomalous Ward identities relating the four-point functions of $T^\m{}_\m$ to Green functions of the 
operators $O_I$ (and the dimension 2 operators $O_a$, which we are neglecting here). This is the 
generalisation of the Ward identity (\ref{e21}) and involves four, three and two-point functions of the $O_I$,
all terms of course being $O(\b^I)$. In \cite{Baume:2014rla}, it is shown how by taking suitable cuts of
these Green functions we can define a positive definite metric $G_{IJ}$ such that
\begin{equation}
{\rm Im}\,{\mathcal A}(s) =  G_{IJ} \b^I \b^J  \ ,
\label{f33}
\end{equation}
and where for small $\b^I$, $G_{IJ}$ is determined by the two-point function $\langle O_I ~ O_J\rangle$,
bringing (\ref{f32}) closer to the Zamolodchikov form.

\section{Global Symmetries and Limit Cycles}\label{sect 7}

In this final section, we consider the application of these ideas to theories exhibiting limit cycles.
This requires an extension of the formalism of Weyl consistency conditions and the local renormalisation
group to theories with global symmetries. It leads to important consequences for both the issue
of whether scale invariant theories are necessarily conformal and for the nature of the space
of couplings.

\subsection{Global symmetries and Weyl consistency conditions}\label {sect 7.1}

In a theory with several fundamental fields, there may be a global symmetry $G_F$ of the free part of the action.
(For example, a theory with $N$ real scalar fields may have an $O(N)$ global, or flavour, symmetry.)
This modifies the Weyl consistency conditions.

In brief (for a full discussion, see \cite{Jack:1990eb,Osborn:1991gm,Fortin:2012hn} and the more recent
discussions in \cite{Jack:2013sha,Baume:2014rla}), 
if the symmetry transformation acting on fields $\phi$ 
takes the form $\d\phi = - \omega\phi$, where $\omega$ is an element of the Lie algebra of $G_F$,
then the couplings $g^I(x)$ of operators $O_I$, viewed as sources, inherit a compensating
transformation $\d g^I = -(\omega g)^I$ so the combination $g^I O_I$ is invariant.\footnote{It is understood 
here that $\omega$ is always taken in the appropriate 
representation, without explicit notation. Similarly for other elements of $G_F$, in particular $A_\m$.}
Associated with this global symmetry there is a current $J_\m(x)$. Introducing the corresponding source 
$A_\m(x)$, itself an element of $G_F$, so that $J_\m = -\d S/\d A^\m$, we may write the Ward identity 
\begin{equation}
\int d^4 x
\sqrt{-g} \left(D_\m\omega \frac{\d}{\d A_\m} - (\omega g)^I \frac{\d}{\d g^I}\right) W = 0 \ .
\label{g1}
\end{equation}
The consistency conditions follow from considering the extended variation operator 
$\D_\s^W + \D_\s^\b + \D_\s^A$, where
\begin{equation}
\D_\s^A = \int d^4 x \sqrt{-g} \left(\s\b_\m^A \frac{\d}{\d A_\m} 
- \partial_\m\s S \frac{\d}{\d A_\m} \right)\ ,
\label{g2}
\end{equation} 
with the notation $\b_\m^A = \r_I D_\m g^I$, where $D_\m g^I = \partial_\m g^I + (A_\m g)^I$.

The anomaly then has additional contributions. The curvature terms are augmented by
\begin{equation}
\tfrac{1}{4} F_{\m\n} \kappa F^{\m\n} + \tfrac{1}{2}F^{\m\n} \zeta_{IJ} D_\m g^I  D_\n g^J \ ,
\label{g3}
\end{equation} 
while $Z_\m$ may include
\begin{equation}
F^{\m\n} \eta_I D_\n g^I \ ,
\label{g4}
\end{equation} 
where $F_{\m\n}$ is the field strength associated with $A_\m$, which can be viewed as
a background gauge field. 

We can eliminate the $S$ contribution to $Z_\m$ using the $G_F$ Ward identity by defining
\begin{equation}
B^I = \b^I - (S g)^I \ , \qquad \qquad P_I = \r_I + \partial_I S \ , \qquad \qquad B_\m^A = \b_\m^A + D_\m S \ .
\label{g5}
\end{equation} 
so that
\begin{equation}
B_\m^A = \r_I D_\m g^I + D_\m S  =  P_I D_\m g^I \ .
\label{g6}
\end{equation} 
Then, the trace anomaly is written as
\begin{equation}
T^\m{}_\m = B^I O_I + \b_c{\BF} - \b_a {\BG} - \b_b {\BH} - \tfrac{1}{4} F_{\m\n} \kappa F^{\m\n}
+ B_\m^A J^\m + D_\m Z^\m \ .
\label{g7}
\end{equation}

The analysis of the consistency conditions then goes through as before.
We find in particular,
\begin{align}
8 \partial_I \b_a &= \chi^g_{IJ} B^J - {\mathcal L}_B \omega_I - \left(P_I g\right)^J \omega_J  \nonumber \\
B^I P_I &= 0 \ .
\label{g8}
\end{align}

Now consider the consequences of these conditions for renormalisation group flows.
Consider first a general flow defined by a vector field $f_I = dg^I/dt$, where $t$ is a parameter along the flow.
The first consistency (\ref{g8}) is
\begin{equation}
8 \frac{d\tilde{B}_a}{dt} = \chi^g_{IJ} f^I B^J + f^I B^J \left(\partial_I \omega_J - \partial_J \omega_I\right)
- \left(P^I g\right)^J f_I \omega_J \ ,
\label{g9}
\end{equation}
where
\begin{equation}
\tilde{B}_a = \b_a + \tfrac{1}{8} B^I \omega_I \ .
\label{g10}
\end{equation}
Two special cases are of key interest. First, for the RG flow itself with $f_I = B_I$, and using $B^I P_I = 0$, we have
\begin{equation}
B^I \partial_I \tilde{B}_a = \tfrac{1}{8} \chi^g_{IJ} B^I B^J \ ,
\label{g11}
\end{equation}
generalising (\ref{f22}). If $\chi^g_{IJ}$ is indeed positive, this shows that $\tilde{B}_a$ is monotonically decreasing
along an IR-directed RG flow generated by the generalised beta function $B^I$.
Second, we can consider the flow $f_I = -(\omega g)_I$ generated by the action of $G_F$ on the couplings.
In this case, it can be shown that $\tilde{B}_a$ is $G_F$ invariant, so is constant along this flow. This is also true 
of $\tilde{\b}_a$.

\subsection{Limit Cycles}\label{sect 7.2}

Four-dimensional quantum field theories exhibiting limit cycles have been found in gauge theories with
scalar fields exhibiting a global $G_F$ symmetry in their free action in \cite{Fortin:2012cq}.
They have the important feature that along the limit cycle, the RG flow (with $\b^I$) is generated by an element
$Q$ of $G_F$. That is, along the cycle, $\b^I = (Q g)^I$.

Although $\b^I$ is non-zero along the limit cycle, the theory is scale invariant, with $T^\m{}_\m = \partial^\m J_\m$.
We need to resolve the question whether it is also conformal invariant, {\it i.e.}~whether $T^\m{}_\m = 0$.

We now prove the following assertions using the formalism above \cite{Fortin:2012hn}:
\begin{enumerate}
\item $S = Q$ on a limit cycle.
\item $S=0$ at a fixed point.
\item the theory is conformal invariant on its limit cycles.
\end{enumerate}
Consider a flow generated by $B^I(g)$ where the coupling $g$ lies on the limit cycle.
Then,
\begin{equation}
B^I = \b^I - (S g)^I = \left( (Q-S) g\right)^I \ .
\label{g12}
\end{equation}
Since $(Q-S) \in G_F$, the flow is of the second type considered above, generated by the action of the symmetry
group $G_F$ on the couplings. Then, since $\tilde{B}_a$ is constant, and assuming $\chi^g_{IJ}$ is positive definite,
we find from (\ref{g11}) that $B^I = 0$.  It then follows from (\ref{g12}) that $S = Q$.

\noindent At a fixed point, where $\b^I = 0$, we have $B^I = - (S g)^I$ from (\ref{g5}), so the $B^I$ flow is generated
by a $G_F$ transformation. Since this means $\tilde{B}_a$ is constant, the same argument as above using 
(\ref{g11}) and positivity of $\chi^g_{IJ}$ shows $(S g)^I = 0$. So we conclude that $S=0$ at a fixed point.

\noindent The essential idea here is that a potential contribution to $T^\m{}_\m$ involving a current $J_\m$ associated 
to a $G_F$ element $P$ can be absorbed into a redefinition of the $B^I$ function (see the definition (\ref{g5})).
But since $B^I = -(P g)^I$ is a $G_F$ flow, by the familiar argument, $\tilde{B}_a$ is constant and $(P g)^I = 0$.
This shows that the current may be removed, leaving $T^\m{}_\m = 0$ on the limit cycle.

The picture that emerges is as follows:

\noindent Limit cycles exist, with $\b^I = (Q g)^I \neq 0$, where $Q\in G_F$. 
Along the cycle, $\tilde{\b}_a$ is constant, since it is $G_F$ invariant.
So $\tilde{\b}_a$ is {\it not} monotonically decreasing -- there is no $\tilde{\b}_a$ theorem in these theories
with global symmetries and limit cycles.

\noindent However, $B^I = 0$ on the cycle and $\tilde{B}_a$ is constant, consistent with the 
$\tilde{B}_a$ theorem (\ref{g11}) with $\chi^g_{IJ}$ positive definite.
On the cycle, the trace anomaly vanishes, $T^\m{}_\m = 0$, and so the theory is not simply scale-invariant
but truly conformal.

In view of this, we should regard the true theory space as ${\mathcal M} = \{g^I\}\, / \, G_F$, 
{\it i.e.}~identifying couplings related by $G_F$ transformations. The relevant flow is then that generated
by the generalised beta functions $B^I$. Such RG flows end at fixed points on the space ${\mathcal M}$,
where the theory is conformal.\footnote{This is reminiscent of the conditions for conformal invariance 
in 2-dim string-related non-linear sigma models. It was shown in \cite{Tseytlin:1986tt, Shore:1986hk} that 
in these models, the condition for the absence of a conformal anomaly involves not the usual beta functions, 
but generalisations which are invariant under spacetime diffeomorphism invariance, which appears as
a `flavour' symmetry of the target space for the sigma model fields defined over the 2-dim worldsheet.}

\section{Summary and Outlook}\label{sect 8}

In this paper, we have reviewed the development of the local renormalisation group from the initial ideas linking
renormalisation with local couplings and anomalous Weyl symmetry to its elegant modern algebraic formulation.
We have presented a detailed account of the renormalisation of Green functions of the energy-momentum tensor 
within the local RG approach, showing how the RG functions characterising these Green functions are related 
to the coefficients of the trace anomaly in curved spacetime. The key r\^ole of the Weyl consistency conditions 
was explained, highlighting the remarkable relations they imply for the RG flow of the trace anomaly coefficients. 

A key stimulus in the development of the local RGE has been the long-standing search for a four-dimensional version
of the $c$-theorem which incorporates all the properties of the original. As we have seen, while this goal is 
tantalisingly close, it cannot yet be considered to be fully achieved. 
Our approach here has been first to review and investigate many different approaches to proving the two-dimensional
Zamolodchikov theorem in order to identify which route offers the best chance of being successfully extended to
higher dimensions. We have especially tried to emphasise the physical understanding of the theorem, particularly
through a detailed analysis of the spectral function approach. In particular, this allowed us to formulate the
Zamolodchikov-like `theorems' involving the coefficients $\b_c$ and $\b_b$ of the Weyl and Ricci terms 
in the four-dimensional trace anomaly. Neither of these, however, has the key monotonicity property of the 
two-dimensional theorem and so although they encode detailed information about the structure of the spectral 
functions characterising the two-point Green functions of the energy-momentum tensor, their usefulness in constraining 
the nature of RG flows appears limited.

This leaves the coefficient of the Euler-Gauss-Bonnet density $E_4$ in the trace anomaly as the remaining candidate -- the 
would-be $a$-theorem originally conjectured by Cardy \cite{Cardy:1988cwa}. As we have seen, the difficulty here is 
that the second derivative of the integral of $E_4$ with respect to the metric vanishes when evaluated on flat space, 
so the corresponding RG coefficient $\b_a$ does not enter into the two-point Green functions of $T_{\m\n}$.
This is despite the remarkable observation of Osborn \cite{Osborn:1989td} that $\b_a$ satisfies a flow equation 
of precisely the Zamolodchikov form as a consequence of the Weyl consistency conditions.
One option, of studying higher-point Green functions and exploiting the fact that the higher derivatives of the
integral of $E_4$ do not vanish, so that $\b_a$ is then involved, appears at first sight to lose any chance of establishing
the necessary positivity property to constrain its RG flow. 

The alternative, of sticking to two-point Green functions but working on a constant-curvature background where the
Euler density is non-vanishing, was explored in depth in \cite{Osborn:1999az}, 
following earlier work by \cite{Forte:1998dx}, and summarised here in section \ref{sect 9}.
The relation of the Weyl consistency conditions to the two-point Green functions was derived and the
generalisation to curved spacetimes of the analysis leading to the $c$-theorem was pursued in detail.
The conclusion, however, is that it is not possible to establish the
$\b_a$ theorem this way, essentially because the theory then contains {\it two} scales, the RG scale $\m$ and the 
curvature $R$, which changes the relation between the scale dependence and RG flow (see section 14 of 
\cite{Osborn:1999az} for a further discussion of the issues involved). Similar physical considerations obstruct the existence 
of a Zamolodchikov-type theorem in finite temperature QFTs or theories with a non-vanishing chemical potential.

At this point, it is interesting to consider the potential implications of even a weak $a$-theorem for symmetry breaking 
and the spectrum of QFTs (see, for example, \cite{Forte:1998jy} . Given the strong dependence of 
$\b_a \sim (n_s + 11n_f + 62n_v)$ on $n_v$, the clear implication is that as we flow into the IR, the theory 
will try to minimise the number of massless vectors in the spectrum in favour of scalar particles. 
Note that we have shown that $\b_c \sim (n_s + 6n_f + 12n_v)$, with its weaker
$n_v$-dependence, need not reduce in the IR. This accords with the situation in QCD, for example, where the gluons 
and quarks in the asymptotically free UV spectrum are traded for Goldstone bosons in the IR as a consequence of 
chiral symmetry breaking. This illustrates how four-dimensional Zamolodchikov-type theorems can in principle 
give important constraints on phase transitions. However, we should note that there exist examples of theories 
where gauge symmetries are broken at high temperature and restored at low temperature \cite{Weinberg:1974hy}
in contrast to the more familiar situation in QCD and electroweak theories. This of course accords with our
observation that we would not expect an $a$-theorem in theories at finite temperature.

A similar argument can be made for QFTs in curved spacetime. It was first shown in \cite{Shore:1979as} that in 
spontaneously broken gauge theories in spacetimes in spacetimes of constant curvature, the symmetry is 
restored when the curvature $R$ exceeds a critical value. This is the normal situation. However, it is 
possible to contrive models, analogous to \cite{Weinberg:1974hy}, in curved spacetime where gauge symmetries 
are instead broken for large $R$ and restored for small curvature (Shore, unpublished), contradicting a would-be 
$a$-theorem. This supports the conclusion of \cite{Osborn:1999az} that the $a$-theorem cannot  be established 
by consideration of the two-point Green functions of $T_{\m\n}$ in curved spacetime.

Returning now to flat space QFTs, the breakthrough in establishing the weak $a$-theorem 
\begin{equation}
\b_a^{UV} - \b_a^{IR} = \frac{1}{\pi} \int_0^\infty \frac{ds}{s^3}\, {\rm Im}\,{\mathcal A}(s) ~ \ge ~ 0 \ ,
\label{h1}
\end{equation}
in \cite{Komargodski:2011vj,Komargodski:2011xv,Luty:2012ww}
came from the recognition of two key points:~first, that the RG coefficient $\b_a$ does enter the four-point 
Green functions of the energy-momentum tensor trace $T^\m{}_\m$, as determined by the local RGE;
and second, that there is indeed a positivity principle associated with four-point functions arising from the
optical theorem constraint ${\rm Im}\,{\mathcal A}(s) \ge 0$.
Finally, the recognition that the dependence on $\b_a$ can be isolated by choosing a particular kinematical
configuration (null external momenta) allows the two-dimensional dispersion relation
derivation of the weak $c$-theorem presented here in section \ref{sect 4.4} to be extended to the $a$-theorem 
in four dimensions. This remarkable advance seems to be convincingly established, though 
we have been careful to highlight the assumptions made in the derivation in section \ref{sect 10} , 
most notably that of hermitian analyticity of the amplitude ${\mathcal A}(s)$. 
It still remains, however, to fully establish the strong $a$-theorem, which entails identifying a corresponding
non-perturbative, monotonic $C$-function all along the RG trajectory not just in the vicinity of fixed points,
together with a similarly well-defined positive definite metric $G_{IJ}$. This should also exhibit the same close
relation to the Osborn identity
\begin{equation}
\b^I\partial_I \tilde{\b_a} = \tfrac{1}{8} \chi^g_{IJ} \b^I \b^J \ ,
\label{h2}
\end{equation}
as for the two-dimensional $c$-theorem.

The analysis of \cite{Komargodski:2011vj,Komargodski:2011xv,Luty:2012ww}
makes extensive use of a `dilaton' effective action. In this approach, the dilaton field $\t(x)$ is introduced as a
source for the energy-momentum tensor trace $T^\m{}_\m$, so that its effective action is formally
\begin{equation}
\Gamma[\t] = \sum_{n=0}^\infty \frac{1}{n{\rm !}} \int d^4 x_n ~ \ldots ~ \int d^4 x_1 ~ \t(x_n) \, \ldots \, \t(x_1) ~
\frac{\d^n W}{\d\t(x_1)\, \ldots \, \d\t(x_n)} \ .
\label{h3}
\end{equation}
See, for example, \cite{Baume:2013ika,Baume:2014rla} for a particularly clear description of this formalism.
The dilaton effective action therefore serves essentially as a book-keeping device for the anomalous Ward identities
and local RGEs controlling the $n$-point Green functions of $T^\m{}_\m$. Note that any further dynamics
attributed to the dilatons in $\Gamma[\t]$ constitutes an assumption about the properties of the original theory
which is not justified by anomalous Weyl symmetry or renormalisation group considerations. 
In particular, there is no requirement that the dilaton should be a physical particle. This is why we did not need 
to introduce dilatons at all in order to derive the weak $a$-theorem and related results here.
Nevertheless, the dilaton effective action is a very convenient and elegant way of encoding all this
information about the Green functions of $T^\m{}_\m$ and, as demonstrated by 
\cite{Komargodski:2011vj,Komargodski:2011xv,Luty:2012ww,Baume:2014rla}
has also proved invaluable in motivating new directions of investigation.

As a final illustration of the power of the local RGE, we showed how it could be applied to give a clear understanding
into how the existence of limit cycles in QFTs with a $G_F$ global symmetry could be reconciled with monotonicity
of the RG flow. Central to this was the generalisation of (\ref{h2}) to read
\begin{equation}
B^I \partial_I \tilde{B}_a ~=~ \tfrac{1}{8}\, \chi^g_{IJ}\, B^I B^J \ ,
\label{h4}
\end{equation}
where the generalised beta functions $B^I$ vanish on the limit cycle while the $G_F$-invariant RG function 
$\tilde{B}_a = \b_a + \tfrac{1}{8} B^I \omega_I$ is constant. 

The idea of of defining RG functions which are invariant under a symmetry of the theory, whether it is 
diffeomorphisms as in the original string-inspired sigma models \cite{Tseytlin:1986tt, Shore:1986hk}
or global internal symmetries \cite{Jack:1990eb,Fortin:2012hn,Jack:2013sha,Baume:2014rla} 
gives an important insight into the nature of the space of couplings ${\mathcal M}$, the `theory space' of QFTs,
on which the RG flows act. Here, we conclude that ${\mathcal M} = \{g^I\}\,/\,G_F$, showing that the true coupling space 
is one where couplings related by $G_F$ transformations are identified. This shows very clearly how in the
space $\{g^I\}$, a RG flow generated by the usual beta functions $\b^I$ ending on a limit cycle where $\b^I$ 
is non-zero is viewed in the true space ${\mathcal M}$ as a flow ending on a fixed point with $B^I = 0$, 
where the theory is conformal. This is a beautiful illustration of the potential of the local RG and 
anomalous Weyl symmetry to give important insight into the fundamental structure of QFT.

\vskip0.7cm
\noindent{\bf Acknowledgments}

\noindent This review is based on lectures originally given at the workshop on ``{\it Theoretical Methods in 
Particle Physics}'', Higgs Centre for Theoretical Physics, University of Edinburgh, Sept-Oct 2013.
I would like to thank the organisers, Luigi Del Debbio and Roman Zwicky for the opportunity to present this material
and to all the participants for creating such a critical and stimulating environment. 
I am grateful to Ian Jack and Hugh Osborn for illuminating discussions and correspondence over many years
and more recently to Jean-Francois Fortin, Boaz Keren-Zur and Roman Zwicky for helpful conversations.
This work was supported in part by STFC grant ST/L000369/1.

\appendix

\section{Counterterms for renormalised Green functions}\label{sect A}

In this appendix, we give a self-contained derivation of the
counterterms arising in the Green functions considered in section \ref{sect 2}.
To begin, recall the results (\ref{b13}), (\ref{b15}) for $T^\m{}_\m$
and $O_I$, which are determined from the first derivatives of the
action, {\it i.e.}
\begin{equation}
T^\m{}_\m ~=~ \frac{1}{\sqrt{-g}} \frac{\d S}{\d \s} ~=~ \e \left(-k_I g_{B}^I O_{BI} + 
\tfrac{1}{2(n-1)} c_B R - \L_B \right)
+ D^2 c_B + 2 \m^{-\e} \L \ ,
\label{AA1}
\end{equation}
and
\begin{equation}
O_I ~=~\frac{1}{\sqrt{-g}} \frac{\d S}{\d g^I} ~=~
Z_I{}^J O_{BJ} + \m^{-\e}\left( \tfrac{1}{2(n-1)} \partial_I L_c\, R
- \tfrac{1}{2} \partial_I A_{JK} \partial g^J \partial g^K
+ D_\m(A_{IJ}\partial^\m g^J) \right)   \ .
\label{AA2}
\end{equation}
We also need the formulae for transformations under Weyl rescalings
$\d g^{\m\n} = 2\s g^{\m\n}$, {\it viz.} $\d \sqrt{-g} = -n \s\sqrt{-g}$
and
\begin{equation}
\d R = 2\s R + 2(n-1) D^2\s \ , ~~~~~~
\d D^2 = 2\s D^2 - (n-2) \partial_\m\s \partial^\m  \ .
\label{AA3}
\end{equation}
First, we recover the expression (\ref{b17}) for $T^\m{}_\m$ written
entirely in terms of renormalised quantities. The definition of the
beta functions in section 2 gives
\begin{multline}
\e \left(-k_I g_{B}^I O_{BI} + 
\tfrac{1}{2(n-1)} c_B R - \L_B \right) ~=~
\hat{\b}^I Z_I{}^J O_{BJ} ~+~ \m^{-\e} \left(\hat{\b}_c +
  \hat{\b}^I \partial_I L_c\right) \tfrac{1}{2(n-1)} R \\
~-~ \m^{-\e} \left(\hat{\b}_{\L} + \tfrac{1}{2} {\cal L}_{\hat{\b}}
A_{IJ} \partial_\m g^I \partial^\m g^J \right) \ .
\label{AA4}
\end{multline}
Substituting into (\ref{AA1}) and using the definition (\ref{AA2}) of
the renormalised operators $O_I$, this gives
\begin{align}
T^\m{}_\m ~=~ &\hat{\b}^I O_I + \m^{-\e} \left(\tfrac{1}{2(n-1)}
  \hat{\b}_c R  - \hat{\b}_\L \right) 
+ \m^{-\e} (D^2 c + 2 \L) \nonumber \\
&+~ \m^{-\e} D_\m \left(\partial^\m L_c - \hat{\b}^I
  A_{IJ} \partial^\m g^J\right) \nonumber\\
&+~ \m^{-\e} \left[ -\tfrac{1}{2} \left({\cal L}_{\hat{\b}}
  A_{IJ}\right) \partial_\m g^I \partial^\m g^J 
+ \tfrac{1}{2} \left(\hat{\b}^K \partial_K A_{IJ} \right) \partial_\m
g^I \partial^\m g^J + \partial_\m \hat{\b}^I A_{IJ} \partial^\m g^J \right]
\label{AA5}
\end{align}
Using the definition (\ref{b10}) of the Lie derivative, and writing
$\partial_\m \hat{\b}^I = \partial_K \hat{\b}^I \partial_\m g^K$,
we see that the term in the square bracket vanishes. Then, finally,
letting $Z_\m = w_I \partial_\m g^I = \partial_\m L_c - 
\hat{\b}^I A_{IJ} \partial_\m g^J$, we recover the expression
(\ref{b17}) for the trace anomaly.

Now consider the renormalisation of the two-point Green functions
(\ref{b30}) - (\ref{b32}). Here, we need the second variations of the
action w.r.t.~$\s$ and $g^I$. First, taking a Weyl variation of
(\ref{AA1}) for $T^\m{}_\m(y)$ w.r.t.~$\s(x)$, we have
\begin{align}
\frac{1}{\sqrt{-g(x)}} \frac{1}{\sqrt{-g(y)}} \,
\frac{\d^2 S}{\d\s(x) \d\s(y)} ~&=~
\e\langle T^\m{}_\m \rangle \d(x,y) \,+\,
\e\left(c_B D^2 + \partial_\m c_B \,\partial^\m\right) \d(x,y) \nonumber\\
&\hskip4cm ~~~+\,2(\e - 2) \m^{-\e} \Lambda\, \d(x,y) \nonumber\\ 
&\rightarrow ~ \m^{-\e} \e\, L_c \,D^2 \d(x,y) \ ,
\label{AA6}
\end{align}
evaluating at the physical couplings $c=\Lambda = \partial_\m g^I = 0$
and taking the $\e \rightarrow 0$ limit.

Next, varying (\ref{AA2}) for $O_I(y)$ w.r.t.~$\s(x)$, and noting that the bare operator 
$O_{BI}$ transforms as $\d O_{BI} = \left(2 - (p+1)\e\right) \s O_{BI}$, we find
\begin{align}
\frac{1}{\sqrt{-g(x)}} \frac{1}{\sqrt{-g(y)}} \,
\frac{\d^2 S}{\d\s(x) \d g^I(y)} ~&=~
(\e-2) O_I \d(x,y) \,+\, \left(2 - (p+1)\e\right) Z_I{}^J O_{BJ} \d(x,y) \nonumber\\
&+\, 
\m^{-\e}\left[\tfrac{1}{2(n-1)} \partial_I L_c\,\left(2R + 2(n-1)
    D^2\right)\,\d(x,y) \,+\, O(\partial_\m g^I) \right] \nonumber \\
&\rightarrow ~ \m^{-\e} \,\partial_I L_c\,D^2\d(x,y) \ .
\label{AA7}
\end{align}
 The counterterm for the two-point function of the operator $O_I$ 
is a little more subtle. Here, we need
\begin{align}
\frac{1}{\sqrt{-g(x)}} \frac{1}{\sqrt{-g(y)}} \,
\frac{\d^2 S}{\d g^I(x) \d g^J(y)} ~&=~
\left(\partial_I Z_J{}^K\right) O_{BK} \d(x,y) \nonumber \\
& ~~ +~ \m^{-\e} \left[\tfrac{1}{2(n-1)} \partial_I \partial_J L_c \,
  R\,\d(x,y) \,+\, A_{IJ}\,D^2\d(x,y)\right] \ ,
\label{AA8}
\end{align}
where $\partial_I Z_J{}^K = \m^{k_I\e} \partial_I \partial_J L^K$, but
there is no further obvious simplification. In fact, however, we only
need this relation contracted with $\hat{\b}^J$ for the applications
here. In that case, we may use the relation
\begin{align}
\hat{\b}^J\partial_I Z_J{}^K
~&=~  \partial_I\left(\hat{\b}^J\partial_J g_B^K\right) 
-\left(\partial_I \hat{\b}^J\right) Z_J{}^K \nonumber\\
&=~ \left(-k_I \e \d_I{}^J \,-\, \partial_I \hat{\b}^J \right) Z_J{}^K\nonumber\\
&=~ -\left(\partial_I \b^J\right) Z_j{}^K \ ,
\label{AA9}
\end{align}
to show
\begin{align}
\frac{1}{\sqrt{-g(x)}} \frac{1}{\sqrt{-g(y)}} \,
\hat{\b}^J\frac{\d^2 S}{\d g^I(x) \d g^J(y)} ~&=~
-\partial_I \hat{\b}^J\,O_J\,\d(x,y) \,+\,
\m^{-\e} \,A_{IJ}\hat{\b^J}\,D^2\d(x,y)  \nonumber\\
&~~+~\m^{-\e} \left[\tfrac{1}{2(n-1)}R\, \left(
\partial_I\b^J \partial_JL_c \,+\,\hat{\b}^J \partial_J \partial_I L_c
\right)\,\d(x,y)\right] \ .
\label{AA10}
\end{align}
A final simplification then gives
\begin{align}
\frac{1}{\sqrt{-g(x)}} \frac{1}{\sqrt{-g(y)}} \,
\hat{\b}^J\frac{\d^2 S}{\d g^I(x) \d g^J(y)} ~&=~
-\partial_I \hat{\b}^J\,O_J\,\d(x,y) \nonumber\\
&~~+~\m^{-\e} \left[-\tfrac{1}{2(n-1)}R\, \partial_I\b_c\,\d(x,y)
\,+\, A_{IJ}\hat{\b}^J\,D^2 \d(x,y) \right] \ ,
\label{AA11}
\end{align}
as shown in (\ref{b35}). Notice the appearance here of 
$\hat{\gamma}_I{}^J = \partial_I \hat{\b}^J$ as the anomalous
dimension matrix for the operators $O_I$. This follows immediately
from its definition 
$\hat{\gamma}_I{}^J = - \m\frac{d}{d\m}Z_I{}^K \left(Z^{-1}\right)_K{}^J$
using (\ref{b3a}) and (\ref{b16}).

We can make one final consistency check of this whole formalism
by evaluating the 
relation (\ref{AA7}) taking the derivatives in reverse order.
From (\ref{AA1}), this gives
\begin{multline}
\frac{1}{\sqrt{-g(x)}} \frac{1}{\sqrt{-g(y)}} \,\left[
\frac{\d^2 S}{\d g^I(x) \d \s(y)}
\,-\, \hat{\b}^J  \frac{\d^2 S}{\d g^I(x) \d g^J(y)} \right]~=~ 
\partial_I \hat{\b}^J O_J\,\d(x,y) \\
+\, \m^{-\e} \left[\tfrac{1}{2(n-1)} R \partial_I \b_c \d(x,y)
\,+\, w_I D^2\d(x,y)\right] ~+~ O(c,\Lambda,\partial_\m g^I) \ .
\label{AA12}
\end{multline}
Substituting with ({\ref{AA11}), we then find
\begin{align}
\frac{1}{\sqrt{-g(x)}} \frac{1}{\sqrt{-g(y)}} \,
\frac{\d^2 S}{\d g^I(x) \d \s(y)} ~&=~ 
\m^{-\e}\left[w_I \,+\, A_{IJ} \hat{\b}^J \right] D^2 \d(x,y)
\nonumber\\
&=~ \m^{-\e} \,\partial_I L_c \,D^2\d(x,y) \ ,
\label{AA13}
\end{align}
recovering the identity (\ref{AA7}) as required.

\section{RGEs for two-point Green functions}\label{sect B}

In section \ref{sect 3.1} we found the renormalisation group equations for two-point Green functions
by successive differentiation of the fundamental RGE (\ref{c1}) for the generating functional $W$.
As an exercise in renormalisation theory, and a consistency check, we should be able to reproduce these
results from the expressions in section \ref{sect 2.3} giving the renormalised two-point functions in terms
of the bare functions. Note though that the operators in these bare two-point functions
are themselves the renormalised ones, so will contribute to the full RGEs. This exercise turns out to be
remarkably intricate compared with the simplicity of section \ref{sect 2.3}, but is nevertheless instructive.

First, note that since here we are acting on bare functions, we need to use the RG operator including
terms of $O(\e)$, {\it i.e.}
\begin{equation}
\hat{\mathcal D} = \m \frac{\partial}{\partial\m} + \hat{\b}^I \frac{\partial}{\partial g^I} \ .
\label{BB1}
\end{equation} 
Now recall the relation (\ref{b33}) for the renormalised two-point function of energy-momentum tensor traces:
\begin{multline} 
i\langle T^\m{}_\m(x)~T^\r{}_\r(y)\rangle  ~-~
i\langle T^\m{}_\m(x)~T^\r{}_\r(y)\rangle_B  ~+~
(2-\e) \langle T^\m{}_\m(x)\rangle \,\d(x,y) \\ 
~=~   \e \m^{-\e}\, L_c \,D^2\d(x,y) 
\label{BB2} 
\end{multline}
Given that $T^\m{}_\m$ is RG invariant, the bare two-point function does not contribute and we have
\begin{align}
\hat{\mathcal D}~ i \langle T^\m{}_\m(x) ~ T^\r{}_\r(y)\rangle &= \e \hat{\mathcal D} \left(\m^{-\e} L_c\right) D^2 \d(x,y) 
=\e \m^{-\e}\left(-\e L_c + \hat{\b}^I \partial_I L_c\right) D^2 \d(x,y)  \nonumber \\
&= -\e \m^{-\e} \b_c D^2 \d(x,y) \ .
\label{BB3}
\end{align}
Since $\b_c$ is finite, the rhs vanishes at $n=2$ and so we recover
\begin{equation}
{\mathcal D}~ i \langle T^\m{}_\m(x) ~ T^\r{}_\r(y) \rangle = 0 \ .
\label{BB4}
\end{equation}

Next, consider (\ref{b34}),
\begin{equation}
i\langle T^\m{}_\m(x)~O_I(y)\rangle  ~-~
i\langle T^\m{}_\m(x)~O_I(y)\rangle_B   ~=~ 
\m^{-\e} \,\partial_I L_c\, D^2 \d(x,y) \ .
\label{BB5}
\end{equation}
Using the RG equation for $O^I$ from (\ref{c9}), and discarding disconnected terms, we find
\begin{equation}
\hat{\mathcal D}~i\langle T^\m{}_\m(x) ~ O^I(y)\rangle + i \hat{\c}_I{}^J \langle T^\m{}_\m(x) ~ O_J(y)\rangle_B 
=\hat {\mathcal D}\left(\m^{-\e} \partial_I L_c\right) D^2 \d(x,y) \ ,
\label{BB6}
\end{equation}
and substituting for the bare function from (\ref{BB5}),
\begin{multline}
{\mathcal D}~i\langle T^\m{}_\m(x) ~ O^I(y)\rangle + i \c_I{}^J \langle T^\m{}_\m(x) ~ O_J(y)\rangle \\
= \m^{-\e} \hat{\c}_I{}^J \partial_J L_c ~D^2 \d(x,y) + \m^{-\e} \left(-\partial_I\b_c - \hat{\c}_I{}^J \partial_J L_c\right)
D^2 \d(x,y) \ ,
\label{BB7}
\end{multline}
where we have used (\ref{c4}) and (\ref{BB3}). The divergent terms involving $\partial_J L_c$ cancel,
leaving finally
\begin{equation}
{\mathcal D}~i\langle T^\m{}_\m(x) ~ O^I(y)\rangle + i \c_I{}^J \langle T^\m{}_\m(x) ~ O_J(y)\rangle
= -\partial_I \b_c D^2 \d(x,y) \ ,
\label{BB8}
\end{equation}
in agreement with (\ref{c12}).

The remaining two-point function $\langle O_I ~ O_J\rangle$ is considerably more complicated.
From (\ref{b32}) and (\ref{AA8}), we have 
\begin{multline}
i\langle O_I(x)~O_J(y)\rangle -
i\langle O_I(x)~O_J(y)\rangle_B  = K_{IJ}{}^K O_K \d(x,y) \\
+ \m^{-\e} \left(-K_{IJ}{}^K \partial_K L_c + \partial_I \partial_J L_c\right) \tfrac{1}{2(n-1)} R \d(x,y) 
+ \m^{-\e} A_{IJ} D^2 \d(x,y) ~+~ O(\partial g^I) \ ,
\label{BB9}
\end{multline}
where we have defined $K_{IJ}{}^K = \left(\partial_I Z_J{}^L\right) (Z^{-1})_L{}^K$. 
Note that since $\partial_I Z_J{}^K = \m^{k_I \e}\partial_I \partial_J L^K$, we can readily show 
$\hat{\b}^I K_{IJ}{}^K = - \c_J{}^K$ (compare (\ref{AA9}). 
Now, after some calculation, we can show from (\ref{BB9}) that
\begin{align}
{\mathcal D}~i\langle O_I(x) ~ O_J(y)\rangle &+ \c_I{}^K i\langle O_K(x) ~ O_J(y)\rangle
+ \c_J{}^K i\langle O_I(x) ~ O_K(y)\rangle \nonumber \\
&= \left[{\mathcal D}K_{IJ}{}^K - K_{IJ}{}^L \hat\c_L{}^K + \hat\c_I{}^L K_{LJ}{}^K + \hat\c_J{}^L K_{IL}{}^K\right]
\langle O_L\rangle \d(x,y) \nonumber \\
&~~+ \m^{-\e} \left[{\mathcal D}F_{IJ} +\hat\c_I{}^K F_{KJ} + \hat\c_J{}^K F_{IK} - K_{IJ}{}^K \partial_K\hat\b_c\right] 
\tfrac{1}{2(n-1)} R \d(x,y) \nonumber \\
&~~+\m^{-\e} \left[{\mathcal D}A_{IJ} + \hat\c_I{}^L A_{LJ} + \hat\c_J{}^L A_{IL}  \right] D^2 \d(x,y) \ ,
\label{BB10}
\end{align}
where here $F_{IJ}= -K_{IJ}{}^K \partial_K L_c + \partial_I \partial_J L_c$. 
The expressions in square brackets are evaluated by noting that the combinations of ${\mathcal D}$ and the
$\c$ functions can be written (see section 2.2) in terms of the Lie derivative as the operator
$-\e + {\mathcal L}_{\hat\b}$. A tricky calculation using the definitions in section 2.1 
then shows that these expressions are finite and are
respectively $-\partial_I \partial_J \b^K$, $-\partial_I \partial_J \b_c$ and $-\chi_{IJ}$.
Finally therefore, we find the RGE:
\begin{multline}
{\mathcal D}~i\langle O_I(x) ~ O_J(y)\rangle + \c_I{}^K i\langle O_K(x) ~ O_J(y)\rangle
+ \c_J{}^K i\langle O_I(x) ~ O_K(y)\rangle + \partial_I \partial_J \b^K \langle O_K\rangle \d(x,y) \\
= -\tfrac{1}{2}\partial_I \partial_J \b_c~R \d(x,y) - \chi_{IJ} D^2 \d(x,y) \ ,
\label{BB11}
\end{multline}
reproducing the result previously found in (\ref{c14}).

\section{Zamolodchikov functions in $n$ dimensions}\label{sect C}

Here, we give the details of the generalisation to four dimensions of the construction of the Zamolodchikov 
functions $F$, $G$, $H$ and $C$ in terms of spectral functions and Bessel functions described for two dimensions 
in section \ref{sect 4.1}. We also give further details of the identification with RG functions.
This augments the discussion of the four-dimensional $\b_c$ and $\b_b$ theorems in section \ref{sect 6}.
For completeness, we quote formulae for general $n$ dimensions where appropriate.
The presentation here follows \cite{Shore:1990wq,Shore:1990ng}.

The starting point is (\ref{e11}) for the two-point function of $T_{\m\n}$, that is
\begin{equation}
i\langle T_{\m\n}(x) ~ T_{\r\s}(0)\rangle = \Pi^{(0)}_{\m\n\r\s} \, \Omega^{(0)} ~+~ \Pi^{(2)}_{\m\n\r\s}\, \Omega^{(2)} \ ,
\label{CC1}
\end{equation}
where the projection operators are given in (\ref{e12}).
The spectral function representation, given for four dimensions in (\ref{f2}), is
\begin{equation}
\Omega^{(s)} = \frac{i}{(2\pi)^{n/2}} \int d\l^2\, \l^{n-2} \,\r^{(s)}(\l^2) \,(\l|x|)^{1-n/2} K_{n/2 - 1}(\l|x|) \ ,
\label{CC2}
\end{equation}
where we have used the $n$-dimensional expression for the Feynman propagator given in section \ref{sect 4.1}
and taken the points in the Green function to be spacelike separated.

Carrying out the differentiation in (\ref{CC1}) we find
\begin{align}
&\langle T_{\m\n}(x) ~ T_{\r\s}(0)\rangle  \nonumber \\
&= \frac{1}{(n-1)}\,\frac{1}{(2\pi)^{n/2} } \int_0^\infty d\l^2\,\l^{n+2}\,\r^{(0)}(\l^2)
\left[A^{(0)}(\l|x|) \eta_{\m\n}\eta_{\r\s} +\ldots +E^{(0)}(\l|x|) \frac{1}{x^4} x_\m x_\n x_\r x_\s \right] \nonumber\\
&+ \frac{(n-2)}{(n-1)}\,\frac{1}{(2\pi)^{n/2} } \int_0^\infty d\l^2\,\l^{n+2}\,\r^{(2)}(\l^2)
\left[A^{(2)}(\l|x|) \eta_{\m\n}\eta_{\r\s} +\ldots + E^{(2)}(\l|x|) \frac{1}{x^4} x_\m x_\n x_\r x_\s \right] 
\label{CC3}
\end{align}

Contracting indices in the different ways indicated in section \ref{sect 5.2}, we can isolate the spin 0 and spin 2
contributions. For the spin 0 spectral function,
\begin{equation}
\langle T_{\m\n}(x) ~ T^\r{}_\r(0)\rangle = -\frac{1}{(2\pi)^{n/2} } \int_0^\infty d\l^2\,\l^{n+2}\,\r^{(0)}(\l^2)
\left[P^{(0)}(\l|x|) \eta_{\m\n} +Q^{(0)}(\l|x|) \frac{1}{x^2} x_\m x_\n  \right]  \ ,
\label{CC4}
\end{equation}
and
\begin{equation}
\langle T^\m{}_\m(x) ~ T^\r{}_\r(0)\rangle = (n-1) \frac{1}{(2\pi)^{n/2} } \int_0^\infty d\l^2\,\l^{n+2}\,\r^{(0)}(\l^2)\,
R^{(0)}(\l|x|) \ ,
\label{CC5}
\end{equation}
while for the spin 2 spectral function,
\begin{multline}
\langle T_\m{}^\r(x) ~ T_{\n\r}\rangle - \frac{1}{(n-1)} i\langle T_{\m\n}(x) ~ T^\r{}_\r(0)\rangle \\
= -\frac{1}{2}\frac{(n+1)(n-2)}{(n-1)} \frac{1}{(2\pi)^{n/2} } \int_0^\infty d\l^2\,\l^{n+2}\,\r^{(2)}(\l^2)
\left[P^{(2)}(\l|x|) \eta_{\m\n} +Q^{(2)}(\l|x|) \frac{1}{x^2} x_\m x_\n  \right]  \ ,
\label{CC6}
\end{multline}
and
\begin{multline}
\langle T^{\m\n}(x) ~ T_{\m\n}(0)\rangle - \frac{1}{(n-1)}\,i\langle T^\m{}_\m(x) ~ T^\r{}_\r(0)\rangle \\
= \frac{1}{2} (n+1)(n-2) \frac{1}{(2\pi)^{n/2} } \int_0^\infty d\l^2\,\l^{n+2}\,\r^{(2)}(\l^2)\, R^{(2)}(\l|x|) \ .
\label{CC7}
\end{multline}
The coefficient functions are evaluated using the Bessel function recursion relations (\ref{d8}) and we find
for both $s=0$ and $2$,
\begin{align}
E^{(s)}(\l|x|) &= (\l|x|)^{1-n/2} \,K_{n/2 + 3}(\l|x|) \nonumber \\
Q^{(s)}(\l|x|) &= (\l|x|)^{1-n/2} \,K_{n/2 + 1}(\l|x|) \nonumber \\
R^{(s)}(\l|x|) &= (\l|x|)^{1-n/2} \,K_{n/2 -1}(\l|x|) \ .
\label{CC8}
\end{align}
which can be compared with (\ref{d9}) in two dimensions.

The generalised Zamolodchikov functions are then defined from these Green functions, rescaled by a factor 
$x^{2n}$ to produce dimensionless functions and remove contact terms, and we find
\begin{align}
F^{s)} = \frac{1}{(2\pi)^{n/2}} \,\int d\l^2\, \l^{2-n} \r^{(s)}(\l^2) \,(\l|x|)^{2n} E^{(s)}(\l|x|) \nonumber \\
H^{s)} = \frac{1}{(2\pi)^{n/2}} \,\int d\l^2\, \l^{2-n} \r^{(s)}(\l^2) \,(\l|x|)^{2n} Q^{(s)}(\l|x|) \nonumber \\
G^{s)} = \frac{1}{(2\pi)^{n/2}} \,\int d\l^2\, \l^{2-n} \r^{(s)}(\l^2) \,(\l|x|)^{2n} R^{(s)}(\l|x|)  \ .
\label{CC9}
\end{align}
As explained in section \ref{sect 6.1}, we now search for a combination $C^{(s)}$ of these functions which reproduces 
as closely as possible the properties of the Zamolodchikov $C$ function (\ref{d11}) in two dimensions.
With 
\begin{equation}
C^{(s)}  = F^{(s)} + \left(\frac{n}{2} + 1\right)H^{(s)} - \left(\frac{n}{2}+2\right) G^{(s)} \ ,
\label{CC10}
\end{equation}
we have
\begin{equation}
|x| \frac{\partial C^{(s)}}{\partial |x|} = (n-2) C^{(s)} - \tfrac{1}{2}(n+2)(n+4) G^{(s)} \ ,
\label{CC11}
\end{equation}
as quoted in (\ref{f3}) and (\ref{f4}).
Explicitly,
\begin{multline}
C^{(s)} = \frac{1}{(2\pi)^{n/2}} \, \int d\l^2 \, \l^{2-n} \,\r^{(s)}(\l^2) \, (\l|x|)^{3n/2 + 1} \\
\times \, \left[K_{n/2 +3}(\l|x|) + \left(\frac{n}{2}+1\right) K_{n/2 +1}(\l|x|) 
- \left(\frac{n}{2}+2\right) K_{n/2 -1}(\l|x|) \right] \ ,
\label{CC12}
\end{multline}
and
\begin{equation}
G^{(s)} = \frac{1}{(2\pi)^{n/2}} \, \int d\l^2 \, \l^{2-n} \,\r^{(s)}(\l^2) \, (\l|x|)^{3n/2 + 1} \,K_{n/2 -1}(\l|x|) \ .
\label{CC13}
\end{equation}

As discussed in sections \ref{sect 4.2} and \ref{sect 6.2} (see \cite{Shore:1990wq,Shore:1990ng}), 
we can also express the Zamolodchikov functions 
in terms of the Fourier transforms $\tilde{\Omega}^{(s)}_{TT}$ and $\tilde{\Omega}_{IJ}$ defined by
\begin{align}
\Omega^{(s)} = a^{(s)} \int \frac{d^n k}{(2\pi)^n}\, e^{ik.x}\, k^{n-4}\, \tilde{\Omega}^{(s)}_{TT}(t) \ ,
\label{CC14}
\end{align}
and
\begin{equation}
\Omega_{IJ} = \int \frac{d^n k}{(2\pi)^n}\, e^{ik.x}\, k^{n-4}\, \tilde{\Omega}_{IJ} \ ,
\label{CC140}
\end{equation}
where $t = \tfrac{1}{2} \log(-\m^2 x^2)$ and we have scaled so that $\tilde{\Omega}^{(s)}_{TT}(t)$ 
and $\tilde{\Omega}_{IJ}$ are dimensionless.
The numerical coefficients are $a^{(0)} = \frac{1}{n-1}$ and $a^{(2)} = \frac{2}{(n+1)(n-2)}$.

The RGEs for $\tilde{\Omega}^{(s)}_{TT}$ and $\tilde{\Omega}_{IJ}$ follow from (\ref{f13}).
From the definition (\ref{CC14}), in four dimensions we find
\begin{equation}
{\mathcal D} \tilde{\Omega}^{(0)}_{TT} =  8 \b_b \ , \qquad \qquad
{\mathcal D} \tilde{\Omega}^{(2)}_{TT} = - 20 \b_c \ ,
\label{CC17}
\end{equation}
while
\begin{equation}
{\mathcal D} \tilde{\Omega}_{IJ} + \c_I{}^K \tilde{\Omega}_{KJ} + \c_j{}^K \tilde{\Omega}_{IK} 
= - \chi^a_{IJ} \ .
\label{CC18}
\end{equation}
We can then solve the RGEs for $C^{(s)}$, $G^{(s)}$ and $G_{IJ}$ in exactly the way described in detail
in section \ref{sect 4.2}, with the results discussed in section \ref{sect 6.2}.

\end{document}